\documentclass[11pt]{article}
\usepackage[latin9]{inputenc}
\usepackage{amsfonts}
\usepackage{amsmath}
\usepackage{amssymb}
\usepackage{indentfirst}
\usepackage{graphicx}
\usepackage{hyperref}
\usepackage{cite}

\numberwithin{equation}{section}
\begin{document}

\begin{titlepage}
\vspace{3cm}
\baselineskip=24pt

\begin{center}
\textbf{\LARGE{On the supersymmetric extension of asymptotic symmetries in three spacetime dimensions}}
\par\end{center}{\LARGE \par}

\begin{center}
	\vspace{1cm}
	\textbf{Ricardo Caroca}$^{\ast}$,
	\textbf{Patrick Concha}$^{\ast}$,
    \textbf{Octavio Fierro}$^{\ast}$,
	\textbf{Evelyn Rodríguez}$^{\dag}$,
	\small
	\\[5mm]
    $^{\ast}$\textit{Departamento de Matemática y Física Aplicadas, }\\
	\textit{ Universidad Católica de la Santísima Concepción, }\\
\textit{ Alonso de Ribera 2850, Concepción, Chile.}
	\\[2mm]

	$^{\dag}$\textit{Departamento de Ciencias, Facultad de Artes Liberales,} \\
	\textit{Universidad Adolfo Ibáñez, Viña del Mar-Chile.} \\[5mm]
	\footnotesize
	\texttt{rcaroca@ucsc.cl},
	\texttt{patrick.concha@ucsc.cl},
    \texttt{ofierro@ucsc.cl},
	\texttt{evelyn.rodriguez@edu.uai.cl},
	\par\end{center}
\vskip 26pt
\begin{abstract}
\noindent In this work we obtain known and new supersymmetric extensions of diverse asymptotic symmetries defined in three spacetime dimensions by considering the semigroup expansion method. The super-$BMS_3$, the superconformal algebra and new infinite-dimensional superalgebras are obtained by expanding the super-Virasoro algebra. The new superalgebras obtained are supersymmetric extensions of the asymptotic algebras of the Maxwell and the $\mathfrak{so}(2,2)\oplus\mathfrak{so}(2,1)$ gravity theories. We extend our results to the $\mathcal{N}=2$ and $\mathcal{N}=4$ cases and find that R-symmetry generators are required. We also show that the new infinite-dimensional structures are related through a flat limit $\ell \rightarrow \infty$.

\end{abstract}
\end{titlepage}\newpage {} {\baselineskip=12pt {} \tableofcontents}

\section{Introduction}

In recent years, infinite-dimensional symmetries have received a growing
interest in the study of fluid mechanics, string theory, two-dimensional
field theory, soliton theory and gravity theory among others. In particular,
the infinite-dimensional (super)symmetries of the Virasoro type result to
describe the boundary dynamics of three-dimensional (super)gravity theories.
In particular, three-dimensional theories are worth to study and are
interesting toy models since they could be useful to approach and understand
open issues in higher-dimensional cases.

At the bosonic level, the asymptotic symmetry of a three-dimensional gravity
in presence of a negative cosmological constant corresponds to two copies of
the Virasoro algebra \cite{BH}. Such structure is obtained by considering
suitable boundary conditions. In the vanishing-cosmological constant case,
the symmetry of asymptotically flat spacetimes at null infinity is described
by the $BMS_{3}$ algebra \cite{ABS, BC, BT1} which is the three-dimensional
version of the $BMS$ algebra introduced more than a half century ago \cite%
{BBM, Sachs}. Extensions of the $BMS_{3}$ symmetry have been subsequently
studied in \cite{GMPT, ABFGR, GP, MPTT, FMT1, BJMN, DR, SA, PSSJ, SSJ}.

More recently, an extended and deformed $BMS_{3}$ algebra (which we have
called deformed $\widetilde{BMS}_{3}$ algebra) appears as the asymptotic
symmetry of a three-dimensional gravity theory invariant under the so-called
Maxwell algebra \cite{CMMRSV}. The Maxwell symmetry has been presented in
\cite{BCR, Schrader, GK} in order to describe the presence of a constant
electromagnetic field background in Minkowski space. Such symmetry has then
been generalized by diverse authors with different applications \cite{AKL,
DKGS, AKL2, CPRS1, CPRS2, CPRS3, SSV, HR, CK, AFGHZ, GKP, KSC, SR}.
Subsequently, a semi-simple enlargement of the $BMS_{3}$ algebra has been
introduced in \cite{CMRSV} corresponding to the asymptotic symmetry of a
gravity theory invariant under the so-called AdS-Lorentz algebra which can
be seen as a semi-simple enlargement of the Poincaré algebra. The
AdS-Lorentz symmetry has been introduced in \cite{Sorokas, GKL, DFIMRSV, SS}
and has lead to diverse applications in the context of Lovelock gravity \cite%
{CDIMR, CMR, CR3} and non-relativistic gravity theory \cite{CR4}. An
interesting feature of the enlarged $BMS_{3}$ algebra obtained in \cite%
{CMRSV} is the connection to the deformed $\widetilde{BMS}_{3}$ algebra
through the flat limit $\ell \rightarrow \infty $.

A minimal supersymmetric extension of the $BMS_{3}$ algebra has been shown
to describe the asymptotic structure of the $\mathcal{N}=1$ supergravity in
three spacetime dimensions considering suitable boundary conditions \cite%
{BDMT}. The extensions to $\mathcal{N}=2$ \cite{LM, FMT, BBNN}, $\mathcal{N}%
=4$ \cite{BLN} and $\mathcal{N}=8$ \cite{BBLN} have later been explored by
diverse authors. On the other hand, the supersymmetric extensions of the
so-called deformed $\widetilde{BMS}_{3}$ algebra and the semi-simple
enlargement of the $BMS_{3}$ algebra remain unknown. Interestingly, the $%
BMS_{3}$, the deformed $\widetilde{BMS}_{3}$ and the enlarged $BMS_{3}$
algebras can alternatively be recovered by applying a semigroup expansion
\cite{Sexp} ($S$-expansion) to the Virasoro algebra \cite{CCRS}.
Furthermore, the super-$BMS_{3}$ algebra and its $\mathcal{N}$-extended
versions have been recently obtained applying the $S$-expansion to the
super-Virasoro algebra \cite{CCFR}.

In this paper, we extend the approach of \cite{CCRS, CCFR} to others
asymptotic symmetries whose supersymmetric extensions are unknown. In
particular, we apply the $S$-expansion to the super-Virasoro algebra in
order to introduce novel supersymmetric extensions of known asymptotic
symmetries. The new infinite-dimensional superalgebras obtained correspond
to the supersymmetric extensions of the deformed $\widetilde{BMS}_{3}$
algebra and the enlarged $BMS_{3}$ algebra. Interestingly, they can be seen
as the infinite-dimensional lifts of the Maxwell and AdS-Lorentz
superalgebra introduced in \cite{BGKL} and \cite{CPR}, respectively.
Furthermore, as their respective finite subalgebras, they are related
through a flat limit $\ell \rightarrow \infty $. We extend our results to
the $\mathcal{N}=2$ and $\mathcal{N}=4$ cases and show that the new $%
\mathcal{N}$-extended infinite-dimensional superalgebras require the
presence of R-symmetry generators.

The paper is organized as follows: In section 2 we give a brief review of
the $S$-expansion procedure and the super-Virasoro algebra. Section 3 and 4
contain our main results. In section 3, we show that known asymptotic
supersymmetries can alternatively be recovered using the $S$-expansion
method. In section 4, we present novel supersymmetric extensions of the
deformed and enlarged $BMS_{3}$ algebras applying different semigroups to
the super-Virasoro algebra. The extensions to $\mathcal{N}=2$ and $\mathcal{N%
}=4$ are also considered. Section 5 is devoted to discussion and possible
developments.

\section{The Semigroup expansion method and Super-Virasoro algebra}

The Lie algebra expansion method was first introduced in \cite{HS} and
subsequently developed in \cite{AIPV, AIPV2, AIPV3} in the context of
three-dimensional Chern-Simons (CS) supergravity and M-theory superalgebra.
A generalization of the expansion procedure using semigroups was later
presented in \cite{Sexp}. Such Abelian semigroup expansion method consists
in combining the elements of a semigroup $S$ with the structure constants of
a Lie algebra $\mathfrak{g}$ in order to obtain a new expanded Lie algebra $%
\mathfrak{G}=S\times \mathfrak{g}$. The $S$-expanded Lie algebra satisfies%
\begin{equation}
\left[ T_{\left( A,\alpha \right) },T_{\left( B,\beta \right) }\right]
=K_{\alpha \beta }^{\text{ \ }\gamma }C_{AB}^{\text{ \ }C}T_{\left( C,\gamma
\right) }\,,
\end{equation}%
where $T_{\left( A,\alpha \right) }=\lambda _{\alpha }T_{A}$ are the
generators of the expanded algebra $\mathfrak{G}$ defined in terms of the
generators $T_{A}$ of the original algebra $\mathfrak{g}$ and in terms of
the elements $\lambda _{\alpha }$ of the semigroup $S$. Here $C_{AB}^{\text{
\ }C}$ are the structure constants of the original algebra $\mathfrak{g}$
while $K_{\alpha \beta }^{\text{ \ }\gamma }$ is the so-called 2-selector
defined by%
\begin{equation}
K_{\alpha \beta }^{\text{ \ }\gamma }=\left\{
\begin{array}{c}
1\text{\qquad when }\lambda _{\alpha }\lambda _{\beta }=\lambda _{\gamma }
\\
0\qquad \text{otherwise. \ \ \ \ \ \ \ \ \ }%
\end{array}%
\right.
\end{equation}%
Then, the $S$-expanded algebra $\mathfrak{G}=S\times \mathfrak{g\,\ }$is a
Lie algebra with structure constants%
\begin{equation}
C_{\left( A,\alpha \right) \left( B,\beta \right) }^{\text{ \ \ \ \ \ \ \ }%
\left( C,\gamma \right) }=K_{\alpha \beta }^{\text{ \ }\gamma }C_{AB}^{\text{
\ }C}\,.
\end{equation}

Interestingly, it is possible to extract a smaller algebra of the expanded
one when the semigroup has a zero element $0_{S}\in S$. In particular, the
algebra obtained by imposing the condition $0_{S}T_{A}=0$ on the expanded
algebra $\mathfrak{G}$ is called $0_{S}$-reduced algebra of $\mathfrak{G}$.

There is an alternative procedure to extract smaller algebra which requires
to consider a decomposition of the original algebra $\mathfrak{g}$ and of
the semigroup $S$. Let $\mathfrak{g}=\bigoplus\nolimits_{p\in I}V_{p}$ be a
subspace decomposition of the original algebra $\mathfrak{g}$, where $I$
denotes a set of indices. Then for each $p,q\in I$ it is possible to define $%
i_{\left( p,q\right) }\subset I$ such that%
\begin{equation}
\left[ V_{p},V_{q}\right] \subset \bigoplus\limits_{r\in i_{\left(
p,q\right) }}V_{r}\,.
\end{equation}%
On the other hand, let us consider a subset decomposition of the semigroup $%
S=\bigcup\nolimits_{p\in I}S_{p}$ such that%
\begin{equation}
S_{p}\cdot S_{q}\subset \bigcap\limits_{r\in i_{\left( p,q\right) }}S_{r}\,.
\end{equation}%
When such subset decomposition exists, we say that it is in resonance with
the subspace decomposition of the algebra $\mathfrak{g}$. In particular, the
subalgebra%
\begin{equation}
\mathfrak{G}_{R}=\bigoplus\limits_{p\in I}S_{p}\times V_{p}\,,
\end{equation}%
is called the resonant subalgebra of $\mathfrak{G}$. Further studies of the $%
S$-expansion method have been developed by diverse authors and can be found
in \cite{Caroca2010a, Caroca2010b, Caroca2011, CKMN, AMNT, ACCSP, ILPR,
IKMN, IKMN2}.

It is interesting to notice that non-trivial (anti-)commutation relations
can be obtained by choosing suitable semigroups $S$ and a pertinent Lie
(super)algebra as the original (super)algebra $\mathfrak{g}$. Here we shall
consider a particular infinite-dimensional superalgebra as our starting
point of our construction. Let $\mathfrak{g}$ be the super-Virasoro algebra,
which we shall denote as $\mathfrak{svir}$ and whose generators satisfy the
following (anti-)commutation relations:%
\begin{equation}
\begin{array}{lcl}
\left[ \ell _{m},\ell _{n}\right] & = & \left( m-n\right) \ell _{m+n}+\frac{c%
}{12}\,m\left( m^{2}-1\right) \delta _{m+n,0}\,, \\[6pt]
\left[ \ell _{m},\mathcal{Q}_{r}\right] & = & \left( \frac{m}{2}-r\right)
\mathcal{Q}_{m+r}\,, \\[6pt]
\left\{ \mathcal{Q}_{r},\mathcal{Q}_{s}\right\} & = & \ell _{r+s}+\frac{c}{6}%
\,\left( r^{2}-\frac{1}{4}\right) \delta _{r+s,0}\,,%
\end{array}
\label{svir}
\end{equation}%
where $c$ is a central extension. Such infinite-dimensional algebra can be
decomposed in subspaces as%
\begin{equation}
\mathfrak{svir}=V_{0}\oplus V_{1}\,,
\end{equation}%
where $V_{0}$ is spanned by the bosonic generators, while $V_{1}$ is the
fermionic subspace. One can note that the subspaces satisfy a graded Lie
algebra,%
\begin{eqnarray}
\left[ V_{0},V_{0}\right] &\subset &V_{0}\,,  \notag \\
\left[ V_{0},V_{1}\right] &\subset &V_{1}\,,  \label{subspace} \\
\left[ V_{1},V_{1}\right] &\subset &V_{0}\,.  \notag
\end{eqnarray}%
Let us note that there is a finite subalgebra spanned by the generators $%
\ell _{0}$, $\ell _{1}$, $\ell _{-1}$, $Q_{\pm \frac{1}{2}}$ which are
related to the super-Lorentz generators through the following change of
basis:%
\begin{equation}
\begin{tabular}{lllll}
$\ell _{-1}=-\sqrt{2}M_{0}$ & , & $\ell _{1}=\sqrt{2}M_{1}$ & , & $\ell
_{0}=M_{2}\,,$ \\
$\mathcal{Q}_{-\frac{1}{2}}=\sqrt{2}\tilde{Q}_{+}$ & , & $\mathcal{Q}_{\frac{%
1}{2}}=\sqrt{2}\tilde{Q}_{-}\,.$ &  &
\end{tabular}%
\end{equation}%
The supersymmetric extension of the Lorentz algebra has been introduced in
\cite{LuNo} and can be used to construct an exotic supersymmetric CS action
\cite{CCFR}.

An $S$-expanded super-Virasoro algebra can be obtained by considering a
semigroup $S=\left\{ \lambda _{i}\right\} $ such that the new algebra is
given by the direct product $S\times \mathfrak{sv}\mathfrak{ir}$. The
expanded generators are given in terms of the super-Virasoro ones as%
\begin{eqnarray}
\ell _{\left( m,\alpha\right) } &=&\lambda _{\alpha}\ell _{m}\,,  \notag \\
Q_{\left( r,\alpha \right) } &=&\lambda _{\alpha }Q_{r}\,,  \label{expgen}
\end{eqnarray}%
and satisfy the (anti-)commutation relations%
\begin{eqnarray*}
\left[ \ell _{\left( m,\alpha \right) },\ell _{\left( n,\beta \right) }%
\right] &=&\left( m-n\right) K_{\alpha \beta }^{\gamma }\ell _{\left(
m+n,\gamma \right) }+\frac{c_{\alpha \beta }}{12}m\left( m^{2}-1\right)
\delta _{m+n,0}\,, \\
\left[ \ell _{\left( m,\alpha \right) },\mathcal{Q}_{\left( r,\beta \right) }%
\right] &=&\left( \frac{m}{2}-r\right) K_{\alpha \beta }^{\gamma }\mathcal{Q}%
_{\left( m+r,\gamma \right) }\,, \\
\left\{ \mathcal{Q}_{\left( r,\alpha \right) },\mathcal{Q}_{\left( s,\beta
\right) }\right\} &=&K_{\alpha \beta }^{\gamma }\ell _{\left( r+s,\gamma
\right) }+\frac{c_{\alpha \beta }}{6}\,\left( r^{2}-\frac{1}{4}\right)
\delta _{r+s,0}\,.
\end{eqnarray*}%
Here, $c_{\alpha \beta }$ denotes a set of central charges given by%
\begin{equation}
c_{\alpha \beta }=cK_{\alpha \beta }^{\gamma }\lambda _{\gamma }\,.
\label{expcen}
\end{equation}%
Interestingly, the $S$-expanded super-Virasoro algebra has a finite
subalgebra which results to be the direct product $S\times \mathcal{SL}$
where $\mathcal{SL}$ is the finite super-Lorentz subalgebra of the original
super-Virasoro algebra. Thus, the semigroup $S$ which allows us to obtain a
new finite (super)algebra also reproduces its infinite-dimensional version.
Such particularity has first been observed at the bosonic level in \cite%
{CCRS} and subsequently noticed at the supersymmetric level in \cite{CCFR}.

In what follows, we shall first recover known asymptotic supersymmetries
using the semigroup expansion method. We then extend our methodology to
obtain new supersymmetric extension of particular asymptotic symmetries.
Naturally, in order to get $\mathcal{N}$-extended superalgebras we shall
require to consider an $\mathcal{N}$-extended super-Virasoro algebra as the
original superalgebra.

\section{Known examples}

It is well known that in asymptotically flat spacetimes, the $BMS_{3}$
algebra describes the asymptotic symmetry of General Relativity in three
spacetime dimensions \cite{ABS, BC, BT1}. At the supersymmetric level, the
super-$BMS_{3}$ algebra appears as the asymptotic symmetry of a
three-dimensional minimal supergravity for a suitable set of asymptotically
flat boundary conditions \cite{BDMT}. The $\mathcal{N}$-extension of the $%
BMS_{3}$ algebra has subsequently been explored by diverse authors in \cite%
{LM, FMT, BBNN, BLN, BBLN}. In particular, the $\mathcal{N}$-extended super-$%
BMS_{3}$ algebras can alternatively be obtained by performing a suitable
contraction of the $\mathcal{N}$-extended superconformal algebras \cite%
{BJLMN}.

Recently, it has been shown in \cite{CCFR} that the $\mathcal{N}$-extended
super-$BMS_{3}$ algebra can also be recovered through the semigroup
expansion method considering an $\mathcal{N}$-extended super-Virasoro
algebra as the original superalgebra. Here we briefly review the obtention
of the super-$BMS_{3}$ algebra following \cite{CCFR}. Then, we show that the
superconformal algebra can also be obtained considering a different
semigroup.

\subsection{Super-$BMS_{3}$ algebra}

As was shown in \cite{CCFR}, the super-$BMS_{3}$ algebra appears as an $S$%
-expansion of the super-Virasoro algebra (\ref{svir}). This can be done by
considering $S_{E}^{\left( 2\right) }=\left\{ \lambda _{0},\lambda
_{1},\lambda _{2},\lambda _{3}\right\} $ as the relevant Abelian semigroup
whose elements satisfy%
\begin{equation}
\begin{tabular}{l|llll}
$\lambda _{3}$ & $\lambda _{3}$ & $\lambda _{3}$ & $\lambda _{3}$ & $\lambda
_{3}$ \\
$\lambda _{2}$ & $\lambda _{2}$ & $\lambda _{3}$ & $\lambda _{3}$ & $\lambda
_{3}$ \\
$\lambda _{1}$ & $\lambda _{1}$ & $\lambda _{2}$ & $\lambda _{3}$ & $\lambda
_{3}$ \\
$\lambda _{0}$ & $\lambda _{0}$ & $\lambda _{1}$ & $\lambda _{2}$ & $\lambda
_{3}$ \\ \hline
& $\lambda _{0}$ & $\lambda _{1}$ & $\lambda _{2}$ & $\lambda _{3}$%
\end{tabular}
\label{ml}
\end{equation}%
where $\lambda _{3}=0_{S}$ is the zero element of the semigroup. A
particular subset decomposition $S_{E}^{\left( 2\right) }=S_{0}\cup S_{1}$,
with%
\begin{eqnarray}
S_{0} &=&\left\{ \lambda _{0},\lambda _{2},\lambda _{3}\right\} \,,  \notag
\\
S_{1} &=&\left\{ \lambda _{1},\lambda _{3}\right\} \,,
\end{eqnarray}%
is said to be resonant since it satisfies the same structure than the
subspaces (\ref{subspace}),%
\begin{eqnarray}
S_{0}\cdot S_{0} &\subset &S_{0}\,,  \notag \\
S_{0}\cdot S_{1} &\subset &S_{1}\,, \\
S_{1}\cdot S_{1} &\subset &S_{0}\,.  \notag
\end{eqnarray}%
The minimal super-$BMS_{3}$ algebra is obtained after extracting a resonant
subalgebra of $S_{E}^{\left( 2\right) }\times \mathfrak{svir}$ and
performing a $0_{S}$-reduction. In particular, the super-$BMS_{3}$
generators are related to the super-Virasoro ones through%
\begin{equation}
\begin{tabular}{ll}
$\mathcal{J}_{m}=\lambda _{0}\ell _{m}\,,$ & $c_{1}=\lambda _{0}c\,,$ \\
$\mathcal{P}_{m}=\lambda _{2}\ell _{m}\,,$ & $c_{2}=\lambda _{2}c\,,$ \\
$\mathcal{G}_{r}=\lambda _{1}\mathcal{Q}_{r}\,.$ &
\end{tabular}%
\end{equation}%
Then, using the multiplication law of the semigroup (\ref{ml}) and the
(anti-)commutation relations of the super-Virasoro algebra (\ref{svir}), one
finds the (anti-)commutators of the minimal super-$BMS_{3}$ algebra with two
central charges \cite{BDMT, BDMT2}:%
\begin{equation}
\begin{array}{lcl}
\left[ \mathcal{J}_{m},\mathcal{J}_{n}\right] & = & \left( m-n\right)
\mathcal{J}_{m+n}+\dfrac{c_{1}}{12}\left( m^{3}-m\right) \delta _{m+n,0}\,,
\\[6pt]
\left[ \mathcal{J}_{m},\mathcal{P}_{n}\right] & = & \left( m-n\right)
\mathcal{P}_{m+n}+\dfrac{c_{2}}{12}\left( m^{3}-m\right) \delta _{m+n,0}\,,
\\[6pt]
\left[ \mathcal{J}_{m},\mathcal{G}_{r}\right] & = & \left( \frac{m}{2}%
-r\right) \mathcal{G}_{m+r}\,, \\[6pt]
\left\{ \mathcal{G}_{r},\mathcal{G}_{s}\right\} & = & \mathcal{P}_{r+s}+%
\frac{c_{2}}{6}\,\left( r^{2}-\frac{1}{4}\right) \delta _{r+s,0}\,.%
\end{array}
\label{sbms3}
\end{equation}%
The central charges $c_{1}$ and $c_{2}$ are related to the $\mathcal{N}=1$
CS Poincaré supergravity action with%
\begin{equation}
c_{1}=12k\alpha _{0}\,,\qquad c_{2}=12k\alpha _{1}\,,
\end{equation}%
where $\alpha _{0}$ and $\alpha _{1}$ are the respective coupling constants
of the exotic CS term and the Einstein-Hilbert term, respectively. It is
interesting to note that the super-$BMS_{3}$ algebra (\ref{sbms3}) is
isomorphic to the supersymmetric extension of the two-dimensional Galilean
conformal algebra introduced in \cite{BM, M}.

One can notice that the super-$BMS_{3}$ algebra has a finite subalgebra
spanned by $\mathcal{J}_{0}$, $\mathcal{J}_{1}$, $\mathcal{J}_{-1}$, $%
\mathcal{P}_{0}$, $\mathcal{P}_{1}$, $\mathcal{P}_{-1}$ and $\mathcal{G}_{%
\frac{1}{2}}$, $\mathcal{G}_{-\frac{1}{2}}$ which are related to the Poincaré
superalgebra through the following change of basis:%
\begin{equation}
\begin{tabular}{lllll}
$\mathcal{J}_{-1}=-\sqrt{2}J_{0}$ & , & $\mathcal{J}_{1}=\sqrt{2}J_{1}$ & ,
& $\mathcal{J}_{0}=J_{2}\,,$ \\
$\mathcal{P}_{-1}=-\sqrt{2}P_{0}$ & , & $\mathcal{P}_{1}=\sqrt{2}P_{1}$ & ,
& $\mathcal{P}_{0}=P_{2}\,,$ \\
$\mathcal{G}_{-\frac{1}{2}}=\sqrt{2}Q_{+}$ & , & $\mathcal{G}_{\frac{1}{2}}=%
\sqrt{2}Q_{-}$ & , &
\end{tabular}%
\end{equation}%
and using a non-diagonal Minkowski metric%
\begin{equation}
\eta _{ab}=\left(
\begin{array}{ccc}
0 & 1 & 0 \\
1 & 0 & 0 \\
0 & 0 & 1%
\end{array}%
\right) \,.
\end{equation}%
Interestingly, such subalgebra can also be obtained by considering the $%
S_{E}^{\left( 2\right) }$-expansion of the super-Lorentz subalgebra of the
super-Virasoro. Indeed, as was shown in \cite{CCFR}, the super Poincaré
structure appears as an $S_{E}^{\left( 2\right) }$-expansion of the
super-Lorentz algebra. Such particularity was also observed at the bosonic
level for the $BMS_{3}$ algebra in \cite{CCRS}. An alternative procedure to
derive the $BMS_{3}$ algebra using an algebraic operation can be found in
\cite{KRR}.

The generalizations to $\mathcal{N}=2$ and $\mathcal{N}=4$ using the
semigroup expansion method can be found in \cite{CCFR}.

\subsection{Superconformal algebra}

The superconformal algebra with $\left( 1,0\right) $ supersymmetry can be
obtained by considering a particular $S$-expansion of the super-Virasoro
algebra (\ref{svir}). In fact, let $S_{L}^{\left( 1\right) }=\left\{ \lambda
_{0},\lambda _{1},\lambda _{2}\right\} $ be the relevant semigroup whose
elements satisfy the following multiplication law%
\begin{equation}
\begin{tabular}{l|lll}
$\lambda _{2}$ & $\lambda _{2}$ & $\lambda _{2}$ & $\lambda _{2}$ \\
$\lambda _{1}$ & $\lambda _{2}$ & $\lambda _{1}$ & $\lambda _{2}$ \\
$\lambda _{0}$ & $\lambda _{0}$ & $\lambda _{2}$ & $\lambda _{2}$ \\ \hline
& $\lambda _{0}$ & $\lambda _{1}$ & $\lambda _{2}$%
\end{tabular}
\label{ml2}
\end{equation}%
with $\lambda _{2}=0_{S}$ being the zero element of the semigroup. Let us
consider now a resonant decomposition of the semigroup%
\begin{eqnarray}
S_{0} &=&\left\{ \lambda _{0},\lambda _{1},\lambda _{2}\right\} \,,  \notag
\\
S_{1} &=&\left\{ \lambda _{1},\lambda _{2}\right\} \,,
\end{eqnarray}%
which satisfies%
\begin{eqnarray}
S_{0}\cdot S_{0} &\subset &S_{0}\,,  \notag \\
S_{0}\cdot S_{1} &\subset &S_{1}\,, \\
S_{1}\cdot S_{1} &\subset &S_{0}\,.  \notag
\end{eqnarray}%
Then, the resonant subalgebra is given by%
\begin{equation}
W_{R}=W_{0}\oplus W_{1}=S_{0}\times V_{0}\,\oplus S_{1}\times V_{1}\,,
\end{equation}%
with $V_{0}$ and $V_{1}$ being the subspaces of the super-Virasoro algebra.
The superconformal algebra is obtained after performing a $0_{S}$-reduction
whose generators are related to the super-Virasoro ones through%
\begin{equation}
\begin{tabular}{ll}
$\mathcal{\bar{L}}_{m}=\lambda _{0}\ell _{m}\,,$ & $\bar{c}=\lambda _{0}c\,,$
\\
$\mathcal{L}_{m}=\lambda _{1}\ell _{m}\,,$ & $c=\lambda _{1}c\,,$ \\
$\mathcal{Q}_{r}=\lambda _{1}\mathcal{Q}_{r}\,.$ &
\end{tabular}%
\end{equation}%
In particular, using the multiplication law of the semigroup (\ref{ml2}) and
the (anti-)commutation relations of the super-Virasoro algebra (\ref{svir}),
one finds the (anti-)commutators of the superconformal algebra:%
\begin{eqnarray}
\left[ \mathcal{L}_{m},\mathcal{L}_{n}\right] &=&\left( m-n\right) \mathcal{L%
}_{m+n}+\frac{c}{12}\,m\left( m^{2}-1\right) \delta _{m+n,0}\,,  \notag \\
\left[ \mathcal{\bar{L}}_{m},\mathcal{\bar{L}}_{n}\right] &=&\left(
m-n\right) \mathcal{\bar{L}}_{m+n}+\frac{\bar{c}}{12}\,m\left(
m^{2}-1\right) \delta _{m+n,0}\,,  \notag \\
\left[ \mathcal{L}_{m},\mathcal{Q}_{r}\right] &=&\left( \frac{m}{2}-r\right)
\mathcal{Q}_{m+r}\,,  \label{sca} \\
\left\{ \mathcal{Q}_{r},\mathcal{Q}_{s}\right\} &=&\mathcal{L}_{r+s}+\frac{c%
}{6}\,\left( r^{2}-\frac{1}{4}\right) \delta _{r+s,0}\,.  \notag
\end{eqnarray}%
Such infinite-dimensional superalgebra corresponds to two copies of the
Virasoro algebra, one of which is supersymmetric, and results to be the
corresponding asymptotic symmetry of the three-dimensional supergravity
theory.

Let us note that a flat limit can be performed by considering first a
redefinition of the generators,%
\begin{equation}
\mathcal{J}_{m}=\mathcal{L}_{m}-\mathcal{\bar{L}}_{-m}\,,\quad \mathcal{P}%
_{m}=\frac{1}{\ell }\left( \mathcal{L}_{m}+\mathcal{\bar{L}}_{-m}\right)
\,,\quad \mathcal{G}_{r}=\sqrt{\frac{2}{\ell }}\mathcal{Q}_{r}\,,
\end{equation}%
which allows us to rewrite the superconformal algebra as the
infinite-dimensional lift of the AdS superalgebra%
\begin{eqnarray}
\left[ \mathcal{J}_{m},\mathcal{J}_{n}\right] &=&\left( m-n\right) \mathcal{J%
}_{m+n}+\frac{c_{1}}{12}\,m\left( m^{2}-1\right) \delta _{m+n,0}\,,  \notag
\\
\left[ \mathcal{J}_{m},\mathcal{P}_{n}\right] &=&\left( m-n\right) \mathcal{P%
}_{m+n}+\frac{c_{2}}{12}\,m\left( m^{2}-1\right) \delta _{m+n,0}\,,  \notag
\\
\left[ \mathcal{P}_{m},\mathcal{P}_{n}\right] &=&\frac{1}{\ell ^{2}}\left(
m-n\right) \mathcal{J}_{m+n}+\frac{c_{1}}{12\ell ^{2}}\,m\left(
m^{2}-1\right) \delta _{m+n,0}\,,  \notag \\
\left[ \mathcal{J}_{m},\mathcal{G}_{r}\right] &=&\left( \frac{m}{2}-r\right)
\mathcal{G}_{m+r}\,,  \label{sca2} \\
\left[ \mathcal{P}_{m},\mathcal{G}_{r}\right] &=&\frac{1}{\ell }\left( \frac{%
m}{2}-r\right) \mathcal{G}_{m+r}\,,  \notag \\
\left\{ \mathcal{G}_{r},\mathcal{G}_{s}\right\} &=&\frac{\mathcal{J}_{r+s}}{%
\ell }+\mathcal{P}_{r+s}+\frac{\left( c_{1}/\ell +c_{2}\right) }{6}\,\left(
r^{2}-\frac{1}{4}\right) \delta _{r+s,0}\,.  \notag
\end{eqnarray}%
where we have defined $c_{1}=c-\bar{c}$ and $c_{2}=\frac{1}{\ell }\left( c+%
\bar{c}\right) $. In particular, the central charges $c_{1}$ and $c_{2}$ are
related to the$\mathcal{\,}$minimal CS AdS supergravity action with%
\begin{equation}
c_{1}=12k\alpha _{0}\,,\qquad c_{2}=12k\alpha _{1}\,,
\end{equation}%
where $\alpha _{0}$ and $\alpha _{1}$ are the respective coupling constant
of the exotic Lagrangian term and Einstein-Hilbert term. One can see that
the superconformal algebra in the basis $\left\{ \mathcal{J}_{m},\mathcal{P}%
_{m},\mathcal{G}_{r}\right\} $ reproduces the super-$BMS_{3}$ algebra (\ref%
{sbms3}) in the flat limit $\ell \rightarrow \infty $.

Let us note that the superconformal algebra (\ref{sca2}) contains a finite
subalgebra spanned by $\mathcal{J}_{0}$, $\mathcal{J}_{1}$, $\mathcal{J}%
_{-1} $, $\mathcal{P}_{0}$, $\mathcal{P}_{1}$, $\mathcal{P}_{-1}$ and $%
\mathcal{G}_{\frac{1}{2}}$, $\mathcal{G}_{-\frac{1}{2}}$ which are related
to the AdS superalgebra through the following change of basis:
\begin{equation}
\begin{tabular}{lllll}
$\mathcal{J}_{-1}=-\sqrt{2}J_{0}$ & , & $\mathcal{J}_{1}=\sqrt{2}J_{1}$ & ,
& $\mathcal{J}_{0}=J_{2}\,,$ \\
$\mathcal{P}_{-1}=-\sqrt{2}P_{0}$ & , & $\mathcal{P}_{1}=\sqrt{2}P_{1}$ & ,
& $\mathcal{P}_{0}=P_{2}\,,$ \\
$\mathcal{G}_{-\frac{1}{2}}=\sqrt{2}Q_{+}$ & , & $\mathcal{G}_{\frac{1}{2}}=%
\sqrt{2}Q_{-}\,.$ &  &
\end{tabular}%
\end{equation}%
As it is well known, the generators of the AdS superalgebra can be rewritten
as two copies of the Lorentz algebra, one of which is augmented by
supersymmetry. Such superalgebra, which results to be the finite subalgebra
of the superconformal algebra in the basis $\left\{ \mathcal{L}_{m},\mathcal{%
\bar{L}}_{m},\mathcal{Q}_{r}\right\} $, appears as a $0_{S}$-reduced
resonant expansion of the super Lorentz algebra using the same semigroup $%
S_{L}^{\left( 1\right) }$.

The following diagram summarizes our construction and the relationship between algebras:%
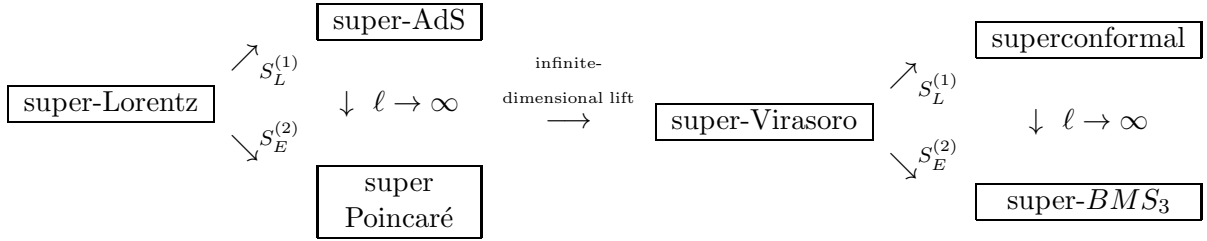
\begin{equation*}
\begin{tabular}{ccc}
\cline{3-3}
&  & \multicolumn{1}{|c|}{super-AdS} \\ \cline{3-3}
& $\nearrow _{S_{L}^{\left( 1\right) }}$ &  \\ \cline{1-1}
\multicolumn{1}{|c}{super-Lorentz} & \multicolumn{1}{|c}{} & $\downarrow $ \
$\ell \rightarrow \infty $ \\ \cline{1-1}
& $\searrow ^{S_{E}^{\left( 2\right) }}$ &  \\ \cline{3-3}
&  & \multicolumn{1}{|c|}{super} \\
&  & \multicolumn{1}{|c|}{Poincaré} \\ \cline{3-3}
\end{tabular}%
\overset{%
\begin{array}{c}
\text{{\tiny infinite-}} \\
\text{{\tiny dimensional lift}}%
\end{array}%
}{\longrightarrow }%
\begin{tabular}{ccc}
\cline{3-3}
&  & \multicolumn{1}{|c|}{superconformal} \\ \cline{3-3}
& $\nearrow _{S_{L}^{\left( 1\right) }}$ &  \\ \cline{1-1}
\multicolumn{1}{|c}{super-Virasoro} & \multicolumn{1}{|c}{} & $\downarrow $
\ $\ell \rightarrow \infty $ \\ \cline{1-1}
& $\searrow ^{S_{E}^{\left( 2\right) }}$ &  \\ \cline{3-3}
&  & \multicolumn{1}{|c|}{super-$BMS_{3}$} \\ \cline{3-3}
\end{tabular}%
\end{equation*}%
Along the paper we shall see that such diagram can be generalized to an
enlarged symmetry. In particular, the election of the semigroups is based on
those used to relate finite superalgebras. As we shall see, new
infinite-dimensional algebras can be obtained using the same semigroups used
at the finite level.

\subsection{$\left( 1,1\right) $ superconformal algebra}

For completeness we extend our method to obtain the $\left( 1,1\right) $
superconformal algebra. This can be done by considering the same semigroup $%
S_{L}^{\left( 1\right) }=\left\{ \lambda _{0},\lambda _{1},\lambda
_{2}\right\} $ but without considering a resonant decomposition as in the
previous case. Indeed, by considering the $S_{L}^{\left( 1\right) }$%
-expansion of the super-Virasoro algebra we have%
\begin{equation}
S_{L}^{\left( 1\right) }\times \mathfrak{svir}=\left\{ \lambda _{0}\ell
_{m},\lambda _{1}\ell _{m},\lambda _{2}\ell _{m},\lambda _{0}\mathcal{Q}%
_{r},\lambda _{1}\mathcal{Q}_{r},\lambda _{2}\mathcal{Q}_{r}\right\} \,,
\end{equation}%
Then, the $\left( 1,1\right) $ superconformal algebra is obtained by
performing the $0_{S}$-reduction with $\lambda _{2}=0_{S}$. In particular,
the $\left( 1,1\right) $ superconformal generators are related to the
super-Virasoro ones through%
\begin{equation}
\begin{tabular}{ll}
$\mathcal{\bar{L}}_{m}=\lambda _{0}\ell _{m}\,,$ & $\bar{c}=\lambda _{0}c\,,$
\\
$\mathcal{L}_{m}=\lambda _{1}\ell _{m}\,,$ & $c=\lambda _{1}c\,,$ \\
$\mathcal{Q}_{r}=\lambda _{1}\mathcal{Q}_{r}\,,$ & $\mathcal{\bar{Q}}%
_{r}=\lambda _{0}\mathcal{Q}_{r}\,.$%
\end{tabular}%
\end{equation}%
The (anti-)commutators of the expanded superalgebra appears using the
multiplication law of the semigroup (\ref{ml2}) and the original commutators
of the super-Virasoro algebra:%
\begin{eqnarray}
\left[ \mathcal{L}_{m},\mathcal{L}_{n}\right] &=&\left( m-n\right) \mathcal{L%
}_{m+n}+\frac{c}{12}\,m\left( m^{2}-1\right) \delta _{m+n,0}\,,  \notag \\
\left[ \mathcal{\bar{L}}_{m},\mathcal{\bar{L}}_{n}\right] &=&\left(
m-n\right) \mathcal{\bar{L}}_{m+n}+\frac{\bar{c}}{12}\,m\left(
m^{2}-1\right) \delta _{m+n,0}\,,  \notag \\
\left[ \mathcal{L}_{m},\mathcal{Q}_{r}\right] &=&\left( \frac{m}{2}-r\right)
\mathcal{Q}_{m+r}\,,\,\left[ \mathcal{\bar{L}}_{m},\mathcal{\bar{Q}}_{r}%
\right] =\left( \frac{m}{2}-r\right) \mathcal{\bar{Q}}_{m+r}\,,  \label{scon}
\\
\left\{ \mathcal{Q}_{r},\mathcal{Q}_{s}\right\} &=&\mathcal{L}_{r+s}+\frac{c%
}{6}\,\left( r^{2}-\frac{1}{4}\right) \delta _{r+s,0}\,,  \notag \\
\left\{ \mathcal{\bar{Q}}_{r},\mathcal{\bar{Q}}_{s}\right\} &=&\mathcal{\bar{%
L}}_{r+s}+\frac{\bar{c}}{6}\,\left( r^{2}-\frac{1}{4}\right) \delta
_{r+s,0}\,,  \notag
\end{eqnarray}%
which corresponds to two copies of the super-Virasoro algebra. Let us note
that such structure can also be derived from the $\mathcal{N}=\left(
2,0\right) $ superconformal algebra \cite{BJLMN}.

Interestingly, the $\mathcal{N}=2$ super-$BMS_{3}$ algebra can be recovered
as a flat limit after a suitable redefinition of the generators. In fact,
the $\left( 1,1\right) $ superconformal algebra can be rewritten as the
infinite-dimensional lift of the $\mathcal{N}=2$ super-AdS algebra%
\begin{eqnarray}
\left[ \mathcal{J}_{m},\mathcal{J}_{n}\right] &=&\left( m-n\right) \mathcal{J%
}_{m+n}+\frac{c_{1}}{12}\,m\left( m^{2}-1\right) \delta _{m+n,0}\,,  \notag
\\
\left[ \mathcal{J}_{m},\mathcal{P}_{n}\right] &=&\left( m-n\right) \mathcal{P%
}_{m+n}+\frac{c_{2}}{12}\,m\left( m^{2}-1\right) \delta _{m+n,0}\,,  \notag
\\
\left[ \mathcal{P}_{m},\mathcal{P}_{n}\right] &=&\frac{1}{\ell ^{2}}\left(
m-n\right) \mathcal{J}_{m+n}+\frac{c_{1}}{12\ell ^{2}}\,m\left(
m^{2}-1\right) \delta _{m+n,0}\,,  \notag \\
\left[ \mathcal{J}_{m},\mathcal{G}_{r}^{i}\right] &=&\left( \frac{m}{2}%
-r\right) \mathcal{G}_{m+r}^{i}\,, \\
\left[ \mathcal{P}_{m},\mathcal{G}_{r}^{i}\right] &=&\frac{1}{\ell }\left(
\frac{m}{2}-r\right) \mathcal{G}_{m+r}^{i}\,,  \notag \\
\left\{ \mathcal{G}_{r}^{i},\mathcal{G}_{s}^{j}\right\} &=&\delta ^{ij}\left[
\frac{\mathcal{J}_{r+s}}{\ell }+\mathcal{P}_{r+s}+\frac{\left( c_{1}/\ell
+c_{2}\right) }{6}\,\left( r^{2}-\frac{1}{4}\right) \delta _{r+s,0}\right]
\,.  \notag
\end{eqnarray}%
where we have considered the following redefinitions of the generators,%
\begin{eqnarray}
\mathcal{J}_{m} &=&\mathcal{L}_{m}-\mathcal{\bar{L}}_{-m}\,,\quad \mathcal{P}%
_{m}=\frac{1}{\ell }\left( \mathcal{L}_{m}+\mathcal{\bar{L}}_{-m}\right) \,,
\notag \\
\mathcal{G}_{r}^{1} &=&\sqrt{\frac{2}{\ell }}\mathcal{Q}_{r}\,,\qquad
\mathcal{G}_{r}^{2}=\sqrt{\frac{2}{\ell }}\mathcal{\bar{Q}}_{-r}\,,
\end{eqnarray}%
with $c_{1}=c-\bar{c}$ and $c_{2}=\frac{1}{\ell }\left( c+\bar{c}\right) $.
Let us note that the $\left( 1,1\right) $ superconformal algebra written in
the basis $\left\{ \mathcal{J}_{m},\mathcal{P}_{m},\mathcal{G}%
_{r}^{i}\right\} $ leads to the $\mathcal{N}=2$ super-$BMS_{3}$ algebra in
the flat limit $\ell \rightarrow \infty $. An alternative way to obtain a $%
\mathcal{N}=2$ super-$BMS_{3}$ algebra in presence of R-symmetry generators $%
\mathcal{R}_{m}$ has been presented in \cite{CCFR} by expanding a $\mathcal{N%
}=2$ super-Virasoro algebra. The $\mathcal{N}=2$ super-$BMS_{3}$ algebra can
also be recovered from the $\mathcal{N}=4$ super-$BMS_{3}$ appearing in \cite%
{BJLMN} after setting some fermionic generators to zero. On the other hand,
a "despotic" contraction of the $\mathcal{N}=\left( 2,2\right) $
superconformal algebra reproduces an inequivalent $\mathcal{N}=2$ super-$%
BMS_{3}$ algebra \cite{LM}.

\section{New examples}

\subsection{Supersymmetric extension of the asymptotic algebra of the
Maxwell gravity theory}

In this section, we explore the supersymmetry extension of a particular
infinite-dimensional algebra introduced in \cite{CCRS}. Such
infinite-dimensional algebra results to describe the asymptotic symmetry of
the three-dimensional Chern-Simons gravity theory invariant under the
Maxwell algebra \cite{CMMRSV}. In particular, as was shown in \cite{CMMRSV},
the new infinite-dimensional algebra is obtained by studying solutions of
the Maxwell CS gravity theory with null boundary in the BMS gauge. Similarly
to the finite subalgebra, given by the Maxwell algebra, the asymptotic
symmetry contains an additional Abelian generator $\mathcal{Z}_{m}$. The
presence of this additional generator modifies the $BMS_{3}$ algebra to an
extension and deformation of such algebra introduced first as an expansion
of the Virasoro algebra in \cite{CCRS}. Appendix A contains a brief review
of the Maxwell gravity theory and its asymptotic symmetry.

The supersymmetric version of the extended and deformed $BMS_{3}$ algebra is
unknown. Although the $\mathcal{N}$-extended CS supergravity theory based on
a $\mathcal{N}$-extended Maxwell superalgebra has been studied in \cite{CPR,
Concha}, the asymptotic structure remains unexplored. In what follows, we
extend the results obtained in \cite{CCRS} to incorporate supersymmetry. In
particular, we present diverse $\mathcal{N}$-extended supersymmetric
versions of the asymptotic symmetry of the Maxwell CS gravity theory \cite%
{CMMRSV} by expanding the $\mathcal{N}$-extended super Virasoro algebra. For
this purpose, we shall consider $S_{E}^{(4)}$ as the relevant semigroup. The
election of the semigroup is not arbitrary and has two origins. First, as
was shown in \cite{CCRS,CPR}, the family $S_{E}^{(n)}$ of semigroups has
been used to obtain the Maxwell (super)algebras from (super) Lorentz
algebra. In particular, the Maxwell algebra belongs to a family of Maxwell
like algebras denoted by $\mathfrak{B}_{k}$ where $k=4$ reproduces the
Maxwell algebra. Such family can be obtained as a $S_{E}^{k-2}$-expansion
\cite{GRCS}. Second, as was discussed at the bosonic level in \cite{CCRS},
the semigroups used to relate finite algebras can also be used to relate
their respective infinite-dimensional enhancements. For instance, as was
shown in \cite{CCRS}, the Maxwell algebra can be obtained as a $S_{E}^{(2)}$%
-expansion of the Lorentz algebra. Interestingly, the same semigroup is used
in \cite{CCRS} to relate the respective infinite-dimensional algebras of the
Maxwell and Lorentz one. Then it seems natural to choose $S_{E}^{(4)}$ as
the relevant semigroup for our task since the Maxwell superalgebra can be
obtained as a $S_{E}^{(4)}$-expansion of the super-Lorentz algebra \cite{CPR}%
.

We conjecture that the new infinite-dimensional superalgebras obtained here
would correspond to the asymptotic symmetries of three-dimensional CS
supergravity theories invariant under the $\mathcal{N}$-extended Maxwell
superalgebra. The explicit obtention of such infinite-dimensional
superalgebras by imposing suitable boundary conditions would be explored in
a future work.\newline

\subsubsection{Minimal deformed super-$\widetilde{BMS}_{3}$ algebra}

Let us consider the $S$-expansion of the super-Virasoro algebra (\ref{svir})
with $S_{E}^{(4)}\mathcal{=}\left\{ \lambda _{0},\lambda _{1},\lambda
_{2},\lambda _{3},\lambda _{4},\lambda _{5}\right\} $ as the relevant finite
semigroup whose element satisfy the following multiplication law%
\begin{equation}
\begin{tabular}{l|llllll}
$\lambda _{5}$ & $\lambda _{5}$ & $\lambda _{5}$ & $\lambda _{5}$ & $\lambda
_{5}$ & $\lambda _{5}$ & $\lambda _{5}$ \\
$\lambda _{4}$ & $\lambda _{4}$ & $\lambda _{5}$ & $\lambda _{5}$ & $\lambda
_{5}$ & $\lambda _{5}$ & $\lambda _{5}$ \\
$\lambda _{3}$ & $\lambda _{3}$ & $\lambda _{4}$ & $\lambda _{5}$ & $\lambda
_{5}$ & $\lambda _{5}$ & $\lambda _{5}$ \\
$\lambda _{2}$ & $\lambda _{2}$ & $\lambda _{3}$ & $\lambda _{4}$ & $\lambda
_{5}$ & $\lambda _{5}$ & $\lambda _{5}$ \\
$\lambda _{1}$ & $\lambda _{1}$ & $\lambda _{2}$ & $\lambda _{3}$ & $\lambda
_{4}$ & $\lambda _{5}$ & $\lambda _{5}$ \\
$\lambda _{0}$ & $\lambda _{0}$ & $\lambda _{1}$ & $\lambda _{2}$ & $\lambda
_{3}$ & $\lambda _{4}$ & $\lambda _{5}$ \\ \hline
& $\lambda _{0}$ & $\lambda _{1}$ & $\lambda _{2}$ & $\lambda _{3}$ & $%
\lambda _{4}$ & $\lambda _{5}$%
\end{tabular}
\label{ml3}
\end{equation}%
with $\lambda _{5}=0_{S}$ being the zero element of the semigroup. Let $%
S_{E}^{\left( 4\right) }=S_{0}\cup S_{1}$, with%
\begin{eqnarray}
S_{0} &=&\left\{ \lambda _{0},\lambda _{2},\lambda _{4},\lambda _{5}\right\}
\,,  \notag \\
S_{1} &=&\left\{ \lambda _{1},\lambda _{3},\lambda _{5}\right\} \,,
\end{eqnarray}%
be the resonant subset decomposition. One can see that such decomposition
has the same structure than the super-Virasoro subspaces (\ref{subspace}),%
\begin{eqnarray}
S_{0}\cdot S_{0} &\subset &S_{0}\,,  \notag \\
S_{0}\cdot S_{1} &\subset &S_{1}\,, \\
S_{1}\cdot S_{1} &\subset &S_{0}\,.  \notag
\end{eqnarray}%
A supersymmetric extension of the deformed $\widetilde{BMS}_{3}$ algebra (%
\ref{dbms3}) is obtained by considering a resonant subalgebra of $%
S_{E}^{\left( 4\right) }\times \mathfrak{svir}$ and performing a $0_{S}$%
-reduction. Denoting the generators (\ref{expgen}) and the central charges (%
\ref{expcen}) of the corresponding $S$-expanded superalgebra as%
\begin{equation}
\begin{tabular}{lll}
$\mathcal{J}_{m}=\lambda _{0}\ell _{m}$ & , & $c_{1}=\lambda _{0}c\,,$ \\
$\ell \,\mathcal{P}_{m}=\lambda _{2}\ell _{m}$ & , & $\ell \,c_{2}=\lambda
_{2}c\,,$ \\
$\ell ^{2}\,\mathcal{Z}_{m}=\lambda _{4}\ell _{m}\,$ & , & $\ell
^{2}\,c_{3}=\lambda _{4}c\,,$ \\
$\ell ^{1/2}\,\mathcal{G}_{r}=\lambda _{1}\mathcal{Q}_{r}$ & , & $\ell
^{3/2}\,\mathcal{H}_{r}=\lambda _{3}\mathcal{Q}_{r}\,,$%
\end{tabular}%
\end{equation}%
the $0_{S}$-reduced and resonant $S_{E}^{\left( 4\right) }$-expanded
superalgebra satisfies the following non-vanishing (anti-)commutation
relations:%
\begin{eqnarray}
\left[ \mathcal{J}_{m},\mathcal{J}_{n}\right] &=&\left( m-n\right) \mathcal{J%
}_{m+n}+\frac{c_{1}}{12}\,m\left( m^{2}-1\right) \delta _{m+n,0}\,,  \notag
\\
\left[ \mathcal{J}_{m},\mathcal{P}_{n}\right] &=&\left( m-n\right) \mathcal{P%
}_{m+n}+\frac{c_{2}}{12}\,m\left( m^{2}-1\right) \delta _{m+n,0}\,,  \notag
\\
\left[ \mathcal{P}_{m},\mathcal{P}_{n}\right] &=&\left( m-n\right) \mathcal{Z%
}_{m+n}+\frac{c_{3}}{12}\,m\left( m^{2}-1\right) \delta _{m+n,0}\,,  \notag
\\
\left[ \mathcal{J}_{m},\mathcal{Z}_{n}\right] &=&\left( m-n\right) \mathcal{Z%
}_{m+n}+\frac{c_{3}}{12}\,m\left( m^{2}-1\right) \delta _{m+n,0}\,,  \notag
\\
\left[ \mathcal{J}_{m},\mathcal{G}_{r}\right] &=&\left( \frac{m}{2}-r\right)
\mathcal{G}_{m+r}\,,\quad \left[ \mathcal{P}_{m},\mathcal{G}_{r}\right]
=\left( \frac{m}{2}-r\right) \mathcal{H}_{m+r}\,,  \label{sdbms3} \\
\left[ \mathcal{J}_{m},\mathcal{H}_{r}\right] &=&\left( \frac{m}{2}-r\right)
\mathcal{H}_{m+r}\,,  \notag \\
\left\{ \mathcal{G}_{r},\mathcal{G}_{s}\right\} &=&\mathcal{P}_{r+s}+\frac{%
c_{2}}{6}\,\left( r^{2}-\frac{1}{4}\right) \delta _{r+s,0}\,,  \notag \\
\left\{ \mathcal{G}_{r},\mathcal{H}_{s}\right\} &=&\mathcal{Z}_{r+s}+\frac{%
c_{3}}{6}\,\left( r^{2}-\frac{1}{4}\right) \delta _{r+s,0}\,.  \notag
\end{eqnarray}%
Such infinite-dimensional superalgebra is the minimal supersymmetric
extension of the so-called deformed $\widetilde{BMS}_{3}$ algebra (\ref%
{dbms3}) and appears by combining the semigroup multiplication law (\ref{ml3}%
) with the original (anti-)commutators of the super-Virasoro algebra (\ref%
{svir}). Such supersymmetric extension is characterized by the introduction
of an additional spinor charge $\mathcal{H}_{r}$ whose presence assures the
Jacobi identity $\left( \mathcal{P},\mathcal{G},\mathcal{G}\right) $.
Analogously to its bosonic version, the deformed super-$\widetilde{BMS}_{3}$
algebra contains a finite subalgebra spanned by the generators $\mathcal{J}%
_{0}$, $\mathcal{J}_{1}$, $\mathcal{J}_{-1}$, $\mathcal{P}_{0}$, $\mathcal{P}%
_{1}$, $\mathcal{P}_{-1}$, $\mathcal{Z}_{0}$, $\mathcal{Z}_{1}$, $\mathcal{Z}%
_{-1}$, $\mathcal{G}_{\frac{1}{2}}$, $\mathcal{G}_{-\frac{1}{2}}$ and $%
\mathcal{H}_{\frac{1}{2}}$, $\mathcal{H}_{-\frac{1}{2}}$ which are related
to the minimal Maxwell superalgebra%
\begin{equation}
\begin{tabular}{lllll}
$\mathcal{J}_{-1}=-\sqrt{2}J_{0}$ & , & $\mathcal{J}_{1}=\sqrt{2}J_{1}$ & ,
& $\mathcal{J}_{0}=J_{2}\,,$ \\
$\mathcal{P}_{-1}=-\sqrt{2}P_{0}$ & , & $\mathcal{P}_{1}=\sqrt{2}P_{1}$ & ,
& $\mathcal{P}_{0}=P_{2}\,,$ \\
$\mathcal{Z}_{-1}=-\sqrt{2}Z_{0}$ & , & $\mathcal{Z}_{1}=\sqrt{2}Z_{1}$ & ,
& $\mathcal{Z}_{0}=Z_{2}\,,$ \\
$\mathcal{G}_{-\frac{1}{2}}=\sqrt{2}Q_{+}$ & , & $\mathcal{G}_{\frac{1}{2}}=%
\sqrt{2}Q_{-}$ & , &  \\
$\mathcal{H}_{-\frac{1}{2}}=\sqrt{2}\Sigma _{+}$ & , & $\mathcal{H}_{\frac{1%
}{2}}=\sqrt{2}\Sigma _{-}$ & . &
\end{tabular}
\label{cob}
\end{equation}%
This means that the minimal deformed super-$\widetilde{BMS}_{3}$ algebra (%
\ref{sdbms3}) obtained here corresponds to an infinite-dimensional lift of
the minimal Maxwell superalgebra in the very same way as the deformed $%
\widetilde{BMS}_{3}$ algebra is an infinite-dimensional lift of the Maxwell
algebra. In particular, the super-Maxwell generators satisfy the following
non-vanishing (anti-)commutators:%
\begin{eqnarray}
\left[ J_{a},J_{b}\right] &=&\epsilon _{abc}J^{c}\,,\qquad \left[ J_{a},P_{b}%
\right] =\epsilon _{abc}P^{c}\,,  \notag \\
\left[ J_{a},Z_{b}\right] &=&\epsilon _{abc}Z^{c}\,,\qquad \left[ P_{a},P_{b}%
\right] =\epsilon _{abc}Z^{c}\,,  \notag \\
\left[ J_{a},Q_{\alpha }\right] &=&\frac{1}{2}\,\left( \Gamma _{a}\right) _{%
\text{ }\alpha }^{\beta }Q_{\beta }\,,\text{ \ \ }  \notag \\
\left[ J_{a},\Sigma _{\alpha }\right] &=&\frac{1}{2}\,\left( \Gamma
_{a}\right) _{\text{ }\alpha }^{\beta }\Sigma _{\beta }\,,\text{ \ \ }
\label{sp1} \\
\left[ P_{a},Q_{\alpha }\right] &=&\frac{1}{2}\,\left( \Gamma _{a}\right) _{%
\text{ }\alpha }^{\beta }\Sigma _{\beta }\,,\text{ }  \notag \\
\left\{ Q_{\alpha },Q_{\beta }\right\} &=&\frac{1}{2}\left( C\Gamma
^{a}\right) _{\alpha \beta }P_{a}\,,  \notag \\
\left\{ Q_{\alpha },\Sigma _{\beta }\right\} &=&\frac{1}{2}\,\left( C\Gamma
^{a}\right) _{\alpha \beta }Z_{a}\,.  \notag
\end{eqnarray}%
The supersymmetrization of the Maxwell algebra is not unique and have been
studied by diverse authors \cite{BGKL2, Lukierski, FL, AILW, AI, CR1, CR2,
CFRS, CFR, PR, Ravera, KC}. However as was discussed in \cite{CPR}, the
superalgebra spanned by $\left\{ J_{a},P_{a},Z_{a},Q_{\alpha },\Sigma
_{\alpha }\right\} $ is the minimal supersymmetric extension of the Maxwell
algebra allowing to define a consistent supergravity action in three
spacetime dimensions. Let us note that the presence of a second Abelian
spinorial charge has already been discussed in the context of superstring
theory \cite{Green} and $D=11$ supergravity \cite{AF}.

As an ending remark, it is worth it to mention that the three-dimensional
minimal Maxwell supergravity theory constructed in \cite{CPR} has directly
been obtained through the semigroup expansion using $S_{E}^{\left( 4\right)
} $ as the relevant finite semigroup. 
As we have discussed in section 2, the semigroup used to obtain a finite
(super)algebra from a finite one can also be used to relate their respective
infinite-dimensional enhancements. Such particular feature is the main
reason behind our election of semigroups.

\subsubsection{$\mathcal{N}=2$ deformed super-$\widetilde{BMS}_{3}$ algebra}

The extension to $\mathcal{N}=\left( 2,0\right) $ of the deformed super-$%
BMS_{3}$ algebra requires to consider the $\mathcal{N}=2$ super-Virasoro
algebra as the starting point. Interestingly, the $\mathcal{N}=2$ deformed
super-$\widetilde{BMS}_{3}$ algebra obtained through the semigroup expansion
procedure corresponds to the supersymmetric extension of the deformed $%
\widetilde{BMS}_{3}$ algebra (\ref{dbms3}) endowed with three $\mathfrak{%
\hat{u}}\left( 1\right) $ current algebra.

The $\mathcal{N}=2$ super-Virasoro algebra, which we shall denote as $%
\mathfrak{svir}_{\left( 2\right) }$, is characterized by the presence of an
R-symmetry generator $\mathcal{R}_{m}$ and its (anti-)commutators are given
by%
\begin{eqnarray}
\left[ \ell _{m},\ell _{n}\right] &=&\left( m-n\right) \ell _{m+n}+\frac{c}{%
12}\,m\left( m^{2}-1\right) \delta _{m+n,0}\,,  \notag \\
\left[ \ell _{m},\mathcal{Q}_{r}^{i}\right] &=&\left( \frac{m}{2}-r\right)
\mathcal{Q}_{m+r}^{i}\,,  \notag \\
\left[ \ell _{m},\mathcal{R}_{n}\right] &=&-n\mathcal{R}_{m+n}\,,  \notag \\
\left[ \mathcal{R}_{m},\mathcal{R}_{n}\right] &=&\frac{c}{3}\,m\delta
_{m+n,0}\,,  \label{n2sv} \\
\left[ \mathcal{Q}_{r}^{i},\mathcal{R}_{m}\right] &=&\epsilon ^{ij}\mathcal{Q%
}_{m+r}^{j}\,,  \notag \\
\left\{ \mathcal{Q}_{r}^{i},\mathcal{Q}_{s}^{j}\right\} &=&\delta ^{ij}\,%
\left[ \ell _{r+s}+\frac{c}{6}\,\left( r^{2}-\frac{1}{4}\right) \delta
_{r+s,0}\,\right] -2\epsilon ^{ij}\left( r-s\right) \mathcal{R}_{r+s}\,,
\notag
\end{eqnarray}%
One can notice that the $\mathcal{N}=2$ super-Virasoro algebra can be
decomposed in a bosonic subspace $V_{0}=\left\{ \ell _{m},\mathcal{R}%
_{m},c\right\} $ and a fermionic subspace $V_{1}=\left\{ \mathcal{Q}%
_{r}^{i}\right\} $ such that%
\begin{equation}
\mathfrak{svir}_{\left( 2\right) }=V_{0}\oplus V_{1}\,,  \label{ssubspace}
\end{equation}%
where they satisfy a graded Lie algebra (\ref{subspace}).

On the other hand, let us consider $S_{E}^{\left( 4\right) }=\left\{ \lambda
_{0},\lambda _{1},\lambda _{2},\lambda _{3},\lambda _{4},\lambda
_{5}\right\} $ as the relevant finite semigroup whose elements satisfy the
multiplication law (\ref{ml3}) and $\lambda _{5}=0_{S}$ being the zero
element of the semigroup. Let $S_{E}^{\left( 4\right) }=S_{0}\cup S_{1}$ with%
\begin{eqnarray}
S_{0} &=&\left\{ \lambda _{0},\lambda _{2},\lambda _{4},\lambda _{5}\right\}
\,,  \notag \\
S_{1} &=&\left\{ \lambda _{1},\lambda _{3},\lambda _{5}\right\} \,,
\label{resonant}
\end{eqnarray}%
be the subset decomposition which is resonant with the subspace
decomposition (\ref{ssubspace}). Then, an $S$-expanded superalgebra
generated by the set $\left\{ \mathcal{J}_{m},\mathcal{P}_{m},\mathcal{Z}%
_{m},\mathtt{T}_{m},\mathtt{B}_{m},\mathtt{Z}_{m},\mathcal{G}_{r}^{i},%
\mathcal{H}_{r}^{i},c_{1},c_{2},c_{3}\right\} $ can be obtained by
considering a $0_{S}$-reduced resonant subalgebra of $S_{E}^{\left( 4\right)
}\times \mathfrak{svir}_{\left( 2\right) }$. In particular, the expanded
generators and central charges are related to the $\mathcal{N}=2$
super-Virasoro ones through%
\begin{equation}
\begin{tabular}{lll}
$\mathcal{J}_{m}=\lambda _{0}\ell _{m}$ & , & $c_{1}=\lambda _{0}c\,,$ \\
$\ell \,\mathcal{P}_{m}=\lambda _{2}\ell _{m}$ & , & $\ell \,c_{2}=\lambda
_{2}c\,,$ \\
$\ell ^{2}\,\mathcal{Z}_{m}=\lambda _{4}\ell _{m}\,$ & , & $\ell
^{2}\,c_{3}=\lambda _{4}c\,,$ \\
\texttt{T}$_{m}=\lambda _{0}\mathcal{R}_{m}\,$ & , & $\ell \,$\texttt{B}$%
_{m}=\lambda _{2}\mathcal{R}_{m}\,,$ \\
$\ell ^{2}\,$\texttt{Z}$_{m}=\lambda _{4}\mathcal{R}_{m}\,,$ & , &  \\
$\ell ^{1/2}\,\mathcal{G}_{r}=\lambda _{1}\mathcal{Q}_{r}$ & , & $\ell
^{3/2}\,\mathcal{H}_{r}=\lambda _{3}\mathcal{Q}_{r}\,.$%
\end{tabular}%
\end{equation}%
Such generators satisfy an $\mathcal{N}=2$ \ supersymmetric extension of the
deformed $\widetilde{BMS}_{3}$ algebra whose (anti-)commutation relations
are directly obtained by combining the multiplication law of the semigroup (%
\ref{ml3}) and the original (anti-)commutators of the $\mathcal{N}=2$
super-Virasoro algebra (\ref{n2sv}). In fact, the (anti-)commutators of the $%
S$-expanded superalgebra are given by%
\begin{eqnarray}
\left[ \mathcal{J}_{m},\mathcal{J}_{n}\right] &=&\left( m-n\right) \mathcal{J%
}_{m+n}+\frac{c_{1}}{12}\,m\left( m^{2}-1\right) \delta _{m+n,0}\,,  \notag
\\
\left[ \mathcal{J}_{m},\mathcal{P}_{n}\right] &=&\left( m-n\right) \mathcal{P%
}_{m+n}+\frac{c_{2}}{12}\,m\left( m^{2}-1\right) \delta _{m+n,0}\,,  \notag
\\
\left[ \mathcal{P}_{m},\mathcal{P}_{n}\right] &=&\left( m-n\right) \mathcal{Z%
}_{m+n}+\frac{c_{3}}{12}\,m\left( m^{2}-1\right) \delta _{m+n,0}\,,
\label{BosDefBMS3} \\
\left[ \mathcal{J}_{m},\mathcal{Z}_{n}\right] &=&\left( m-n\right) \mathcal{Z%
}_{m+n}+\frac{c_{3}}{12}\,m\left( m^{2}-1\right) \delta _{m+n,0}\,,  \notag
\end{eqnarray}%
\begin{eqnarray}
\left[ \mathcal{J}_{m},\mathtt{T}_{n}\right] &=&-n\mathtt{T}_{m+n}\,,\qquad %
\left[ \mathcal{P}_{m},\mathtt{T}_{n}\right] =-n\mathtt{B}_{m+n}\,,  \notag
\\
\left[ \mathcal{Z}_{m},\mathtt{T}_{n}\right] &=&-n\mathtt{Z}_{m+n}\,,\qquad %
\left[ \mathcal{J}_{m},\mathtt{B}_{n}\right] =-n\mathtt{B}_{m+n}\,,  \notag
\\
\left[ \mathcal{P}_{m},\mathtt{B}_{n}\right] &=&-n\mathtt{Z}_{m+n}\,,\qquad %
\left[ \mathcal{J}_{m},\mathtt{Z}_{n}\right] =-n\mathtt{Z}_{m+n}\,,
\label{bbb} \\
\left[ \mathtt{T}_{m},\mathtt{T}_{n}\right] &=&\frac{c_{1}}{3}m\delta
_{m+n,0}\,,\quad \left[ \mathtt{T}_{m},\mathtt{B}_{n}\right] =\frac{c_{2}}{3}%
m\delta _{m+n,0}\,,  \notag \\
\left[ \mathtt{T}_{m},\mathtt{Z}_{n}\right] &=&\frac{c_{3}}{3}m\delta
_{m+n,0}\,,\quad \left[ \mathtt{B}_{m},\mathtt{B}_{n}\right] =\frac{c_{3}}{3}%
m\delta _{m+n,0}\,,  \notag
\end{eqnarray}%
\begin{eqnarray}
\left[ \mathcal{J}_{m},\mathcal{G}_{r}^{i}\right] &=&\left( \frac{m}{2}%
-r\right) \mathcal{G}_{m+r}^{i}\,,\quad \left[ \mathcal{P}_{m},\mathcal{G}%
_{r}^{i}\right] =\left( \frac{m}{2}-r\right) \mathcal{H}_{m+r}^{i}\,,  \notag
\\
\left[ \mathcal{J}_{m},\mathcal{H}_{r}^{i}\right] &=&\left( \frac{m}{2}%
-r\right) \mathcal{H}_{m+r}^{i}\,,  \notag \\
\left[ \mathcal{G}_{r}^{i},\mathtt{T}_{m}\right] &=&\epsilon ^{ij}\mathcal{G}%
_{m+r}^{j}\,,\qquad \left[ \mathcal{G}_{r}^{i},\mathtt{B}_{m}\right]
=\epsilon ^{ij}\mathcal{H}_{m+r}^{j}\,,  \notag \\
\left[ \mathcal{H}_{r}^{i},\mathtt{T}_{m}\right] &=&\epsilon ^{ij}\mathcal{H}%
_{m+r}^{j}\,,  \label{ccc} \\
\left\{ \mathcal{G}_{r}^{i},\mathcal{G}_{s}^{j}\right\} &=&\delta ^{ij}\left[
\mathcal{P}_{r+s}+\frac{c_{2}}{6}\,\left( r^{2}-\frac{1}{4}\right) \delta
_{r+s,0}\right] -2\epsilon ^{ij}\left( r-s\right) \mathtt{B}_{r+s}\,,  \notag
\\
\left\{ \mathcal{G}_{r}^{i},\mathcal{H}_{s}^{j}\right\} &=&\delta ^{ij}\left[
\mathcal{Z}_{r+s}+\frac{c_{3}}{6}\,\left( r^{2}-\frac{1}{4}\right) \delta
_{r+s,0}\right] -2\epsilon ^{ij}\left( r-s\right) \mathtt{Z}_{r+s}\,.  \notag
\end{eqnarray}%
Such $\mathcal{N}=2$ deformed super-$\widetilde{BMS}_{3}$ algebra differs
from the $\mathcal{N}=1$ one by the presence of additional bosonic
generators. In particular, one can see that the anticommutator of the
supercharges $\mathcal{G}_{r}^{i}$ closes to a combination of $\mathcal{P}$,
$\mathtt{B}$ and a central charge $c_{2}$. On the other hand, the
anticommutator of the supercharges $\mathcal{G}_{r}^{i}$ and $\mathcal{H}%
_{r}^{i}$ closes to a combination of $\mathcal{Z}$, $\mathtt{Z}$and a
central charge $c_{3}$. The present superalgebra can be seen as the $%
\mathcal{N}=2$ supersymmetric extension of deformed $\widetilde{BMS}_{3}$
algebra endowed with $\mathfrak{\hat{u}}\left( 1\right) \times \mathfrak{%
\hat{u}}\left( 1\right) \times \mathfrak{\hat{u}}\left( 1\right) $ current
algebra. This can be seen more clearly by the fact that the
infinite-dimensional superalgebra obtained here can alternatively be
recovered as an Inönü-Wigner contraction of three copies of a Virasoro
algebra, two of which augmented by supersymmetry, endowed with an affine $%
\mathfrak{\hat{u}}\left( 1\right) $ current algebra. In particular, the $%
\mathfrak{\hat{u}}\left( 1\right) $ current generators $\left\{ \mathfrak{k}%
_{n},\mathfrak{\bar{k}}_{n},\mathfrak{\tilde{k}}_{n}\right\} $ are related
to the $\mathcal{N}=2$ deformed super-$\widetilde{BMS}_{3}$ ones through the
following redefinitions%
\begin{equation}
\mathtt{T}_{m}=\mathfrak{k}_{m}+\mathfrak{\bar{k}}_{-m}+\mathfrak{\tilde{k}}%
_{-m}\,,\quad \mathtt{B}_{m}=\epsilon \left( \mathfrak{k}_{m}-\mathfrak{\bar{%
k}}_{-m}\right) \,,\quad \mathtt{Z}_{m}=\epsilon ^{2}\left( \mathfrak{k}_{m}+%
\mathfrak{\bar{k}}_{-m}\right) \,,
\end{equation}%
where the limit $\epsilon \rightarrow 0$ reproduces the $\mathcal{N}=2$
deformed super-$\widetilde{BMS}_{3}$ algebra presented here. The presence of
$\mathfrak{\hat{u}}\left( 1\right) $ current generators in asymptotic
symmetries is not new and has already be considered in $BMS_{3}$ algebra
\cite{DR, BDR, SA} and $\mathcal{N}=2$ super-$BMS_{3}$ algebra \cite{CCFR}.

One can note that the central extension of the $\mathcal{N}=2$ Maxwell
superalgebra endowed with $\mathfrak{so}\left( 2\right) $ internal symmetry
generators \cite{Concha} appears as a finite subalgebra of the $\mathcal{N}%
=2 $ deformed super-$\widetilde{BMS}_{3}$ algebra. Such subalgebra is
spanned by the generators $\mathcal{J}_{0}$, $\mathcal{J}_{\pm 1}$, $%
\mathcal{P}_{0}$, $\mathcal{P}_{\pm 1}$, $\mathcal{Z}_{0}$, $\mathcal{Z}%
_{\pm 1}$, $\mathtt{T}_{0}$, $\mathtt{B}_{0}$, $\mathtt{Z}_{0}$, $\mathcal{G}%
_{\pm \frac{1}{2}}$ and $\mathcal{H}_{\pm \frac{1}{2}}$ which are related to
the $\mathcal{N}=2$ Maxwell superalgebra $\left\{
J_{a},P_{a},Z_{a},T,B,Z,Q,\Sigma \right\} $ as follows%
\begin{equation}
\begin{tabular}{lllll}
$\mathcal{J}_{-1}=-\sqrt{2}J_{0}$ & , & $\mathcal{J}_{1}=\sqrt{2}J_{1}$ & ,
& $\mathcal{J}_{0}=J_{2}\,,$ \\
$\mathcal{P}_{-1}=-\sqrt{2}P_{0}$ & , & $\mathcal{P}_{1}=\sqrt{2}P_{1}$ & ,
& $\mathcal{P}_{0}=P_{2}\,,$ \\
$\mathcal{Z}_{-1}=-\sqrt{2}Z_{0}$ & , & $\mathcal{Z}_{1}=\sqrt{2}Z_{1}$ & ,
& $\mathcal{Z}_{0}=Z_{2}\,,$ \\
\texttt{T}$_{0}=-T$ & , & \texttt{B}$_{0}=-B$ & , & \texttt{Z}$_{0}=-Z\,\,,$
\\
$\mathcal{G}_{-\frac{1}{2}}=\sqrt{2}Q_{+}$ & , & $\mathcal{G}_{\frac{1}{2}}=%
\sqrt{2}Q_{-}$ & , &  \\
$\mathcal{H}_{-\frac{1}{2}}=\sqrt{2}\Sigma _{+}$ & , & $\mathcal{H}_{\frac{1%
}{2}}=\sqrt{2}\Sigma _{-}$ & . &
\end{tabular}
\label{redef}
\end{equation}%
Remarkably, the central extension of the $\mathcal{N}=2$ Maxwell
superalgebra endowed with $\mathfrak{so}\left( 2\right) $ internal symmetry
generators allows to define an invariant non-degenerate inner-product which
provides us with a consistent three-dimensional supergravity action \cite%
{Concha}. Interestingly, analogously to the results presented here, the $%
\mathcal{N}=2$ Maxwell supergravity theory can alternatively be obtained by
considering an $S$-expansion of the $\mathcal{N}=2$ super Lorentz algebra
using the same semigroup $S_{E}^{\left( 4\right) }$.

\subsubsection{$\mathcal{N}=4$ deformed super-$\widetilde{BMS}_{3}$ algebra}

Here we extend our construction to the obtention of the $\mathcal{N}=\left(
4,0\right) $ deformed super-$\widetilde{BMS_{3}}$ algebra by considering an $%
S_{E}^{\left( 4\right) }$-expansion of the $\mathcal{N}=4$ super-Virasoro
algebra. The infinite-dimensional superalgebra obtained corresponds to the
supersymmetric extension of the deformed $\widetilde{BMS}_{3}$ algebra (\ref%
{dbms3}) endowed with three $\mathfrak{su}\left( 2\right) $ current algebra.

Let us consider the $\mathcal{N}=4$ super-Virasoro algebra whose
(anti-)commutation relations are given by \cite{Ito}%
\begin{eqnarray}
\left[ \ell _{m},\ell _{n}\right] &=&\left( m-n\right) \ell _{m+n}+\frac{c}{%
12}\,m\left( m^{2}-1\right) \delta _{m+n,0}\,,  \notag \\
\left[ \ell _{m},\mathcal{Q}_{r}^{i,\pm }\right] &=&\left( \frac{m}{2}%
-r\right) \mathcal{Q}_{m+r}^{i,\pm }\,,  \notag \\
\left[ \ell _{m},\mathcal{R}_{n}^{a}\right] &=&-n\mathcal{R}_{m+n}^{a}\,,
\notag \\
\left[ \mathcal{R}_{m}^{a},\mathcal{R}_{n}^{b}\right] &=&i\epsilon ^{abc}%
\mathcal{R}_{m+n}^{c}+\frac{c}{12}\,m\delta ^{ab}\delta _{m+n,0}\,,
\label{N4SV} \\
\left[ \mathcal{R}_{m}^{a},\mathcal{Q}_{r}^{i,+}\right] &=&\,-\frac{1}{2}%
\left( \sigma ^{a}\right) _{j}^{i}\mathcal{Q}_{m+r}^{j,+}\,,\qquad \left[
\mathcal{R}_{m}^{a},\mathcal{Q}_{r}^{i,-}\right] =\,\frac{1}{2}\left( \bar{%
\sigma}^{a}\right) _{j}^{i}\mathcal{Q}_{m+r}^{j,-}\,,  \notag \\
\left\{ \mathcal{Q}_{r}^{i,+},\mathcal{Q}_{s}^{j,-}\right\} &=&\delta ^{ij}\,%
\left[ \ell _{r+s}+\frac{c}{6}\,\left( r^{2}-\frac{1}{4}\right) \delta
_{r+s,0}\,\right] -\left( r-s\right) \left( \sigma ^{a}\right) _{ij}\mathcal{%
R}_{r+s}^{a}\,,  \notag
\end{eqnarray}%
where $\sigma _{ij}^{a}=\sigma _{ji}^{a}$ are the Pauli matrices, $i,j=1,2$
and $a,b,c=1,2,3$. As the $\mathcal{N}=1$ and $\mathcal{N}=2$ version, the $%
\mathcal{N}=4$ super-Virasoro algebra can be written as the direct sum of a
bosonic subspace $V_{0}=\left\{ \ell _{m},\mathcal{R}_{m}^{a},c\right\} $
and a fermionic one $V_{1}=\left\{ Q_{r}^{i,\pm }\right\} $.

Let $S_{E}^{\left( 4\right) }=\left\{ \lambda _{0},\lambda _{1},\lambda
_{2},\lambda _{3},\lambda _{4},\lambda _{5}\right\} $ be the relevant
Abelian semigroup whose elements satisfy (\ref{ml3}). Then, by considering
the resonant decomposition (\ref{resonant}) and applying a resonant $0_{S}$%
-reduction $S_{E}^{\left( 4\right) }$-expansion of the $\mathcal{N}=4$
super-Virasoro algebra we find an expanded algebra spanned by the set of
generators:%
\begin{equation}
\left\{ \mathcal{J}_{m},\mathcal{P}_{m},\mathcal{Z}_{m},\mathtt{T}_{m}^{a},%
\mathtt{B}_{m}^{a},\mathtt{Z}_{m}^{a},\mathcal{G}_{r}^{i,\pm },\mathcal{H}%
_{r}^{i,\pm },c_{1},c_{2},c_{3}\right\} \,.
\end{equation}%
The expanded generators and central charges can be written in terms of the $%
\mathcal{N}=4$ super-Virasoro ones through the semigroup elements as%
\begin{equation}
\begin{tabular}{lll}
$\mathcal{J}_{m}=\lambda _{0}\ell _{m}$ & , & $c_{1}=\lambda _{0}c\,,$ \\
$\ell \,\mathcal{P}_{m}=\lambda _{2}\ell _{m}$ & , & $\ell \,c_{2}=\lambda
_{2}c\,,$ \\
$\ell ^{2}\,\mathcal{Z}_{m}=\lambda _{4}\ell _{m}\,$ & , & $\ell
^{2}\,c_{3}=\lambda _{4}c\,,$ \\
\texttt{T}$_{m}^{a}=\lambda _{0}\mathcal{R}_{m}^{a}\,$ & , & $\ell \,$%
\texttt{B}$_{m}^{a}=\lambda _{2}\mathcal{R}_{m}^{a}\,,$ \\
$\ell ^{2}\,$\texttt{Z}$_{m}^{a}=\lambda _{4}\mathcal{R}_{m}^{a}\,,$ & , &
\\
$\ell ^{1/2}\,\mathcal{G}_{r}^{i,\pm }=\lambda _{1}\mathcal{Q}_{r}^{i,\pm }$
& , & $\ell ^{3/2}\,\mathcal{H}_{r}^{i,\pm }=\lambda _{3}\mathcal{Q}%
_{r}^{i,\pm }\,.$%
\end{tabular}
\label{expgenb}
\end{equation}%
Using the semigroup multiplication law (\ref{ml3}) and the original
(anti-)commutators of the $\mathcal{N}=4$ super-Virasoro algebra (\ref{N4SV}%
) one can see that the expanded generators satisfy a $\mathcal{N}=4$
deformed super-$\widetilde{BMS}_{3}$ algebra whose (anti-)commutation
relations are given by (\ref{dbms3}) and%
\begin{eqnarray}
\left[ \mathcal{J}_{m},\mathtt{T}_{n}^{a}\right] &=&-n\mathtt{T}%
_{m+n}^{a}\,,\qquad \left[ \mathcal{P}_{m},\mathtt{T}_{n}^{a}\right] =-n%
\mathtt{B}_{m+n}^{a}\,,  \notag \\
\left[ \mathcal{Z}_{m},\mathtt{T}_{n}^{a}\right] &=&-n\mathtt{Z}%
_{m+n}^{a}\,,\qquad \left[ \mathcal{J}_{m},\mathtt{B}_{n}^{a}\right] =-n%
\mathtt{B}_{m+n}^{a}\,,  \notag \\
\left[ \mathcal{P}_{m},\mathtt{B}_{n}^{a}\right] &=&-n\mathtt{Z}%
_{m+n}^{a}\,,\qquad \left[ \mathcal{J}_{m},\mathtt{Z}_{n}^{a}\right] =-n%
\mathtt{Z}_{m+n}^{a}\,,  \notag \\
\left[ \mathtt{T}_{m}^{a},\mathtt{T}_{n}^{b}\right] &=&i\epsilon ^{abc}%
\mathtt{T}_{m+n}^{c}+\frac{c_{1}}{12}\,m\delta ^{ab}\delta _{m+n,0}\,,\,\quad
\label{dbms3a} \\
\left[ \mathtt{T}_{m}^{a},\mathtt{B}_{n}^{b}\right] &=&i\epsilon ^{abc}%
\mathtt{B}_{m+n}^{c}+\frac{c_{2}}{12}\,m\delta ^{ab}\delta _{m+n,0}\,,\,
\notag \\
\left[ \mathtt{T}_{m}^{a},\mathtt{Z}_{n}^{b}\right] &=&i\epsilon ^{abc}%
\mathtt{Z}_{m+n}^{c}+\frac{c_{3}}{12}\,m\delta ^{ab}\delta _{m+n,0}\,,\,
\notag \\
\left[ \mathtt{B}_{m}^{a},\mathtt{B}_{n}^{b}\right] &=&i\epsilon ^{abc}%
\mathtt{Z}_{m+n}^{c}+\frac{c_{3}}{12}\,m\delta ^{ab}\delta _{m+n,0}\,,\,
\notag
\end{eqnarray}%
\begin{eqnarray}
\left[ \mathcal{J}_{m},\mathcal{G}_{r}^{i,\pm }\right] &=&\left( \frac{m}{2}%
-r\right) \mathcal{G}_{m+r}^{i,\pm }\,,\quad \left[ \mathcal{P}_{m},\mathcal{%
G}_{r}^{i,\pm }\right] =\left( \frac{m}{2}-r\right) \mathcal{H}_{m+r}^{i,\pm
}\,,  \notag \\
\left[ \mathcal{J}_{m},\mathcal{H}_{r}^{i,\pm }\right] &=&\left( \frac{m}{2}%
-r\right) \mathcal{H}_{m+r}^{i,\pm }\,,  \notag \\
\left[ \mathtt{T}_{m}^{a},\mathcal{G}_{r}^{i,+}\right] &=&-\frac{1}{2}\left(
\sigma ^{a}\right) _{j}^{i}\mathcal{G}_{m+r}^{j,+}\,,\qquad \left[ \mathtt{T}%
_{m}^{a},\mathcal{G}_{r}^{i,-}\right] =\frac{1}{2}\left( \bar{\sigma}%
^{a}\right) _{j}^{i}\mathcal{G}_{m+r}^{j,-}\,,  \notag \\
\left[ \mathtt{B}_{m}^{a},\mathcal{G}_{r}^{i,+}\right] &=&-\frac{1}{2}\left(
\sigma ^{a}\right) _{j}^{i}\mathcal{H}_{m+r}^{j,+}\,,\qquad \left[ \mathtt{B}%
_{m}^{a},\mathcal{G}_{r}^{i,-}\right] =\frac{1}{2}\left( \bar{\sigma}%
^{a}\right) _{j}^{i}\mathcal{H}_{m+r}^{j,-}\,,  \label{dbms3b} \\
\left[ \mathtt{T}_{m}^{a},\mathcal{H}_{r}^{i,+}\right] &=&-\frac{1}{2}\left(
\sigma ^{a}\right) _{j}^{i}\mathcal{H}_{m+r}^{j,+}\,,\qquad \left[ \mathtt{T}%
_{m}^{a},\mathcal{H}_{r}^{i,-}\right] =\frac{1}{2}\left( \bar{\sigma}%
^{a}\right) _{j}^{i}\mathcal{H}_{m+r}^{j,-}\,,  \notag \\
\left\{ \mathcal{G}_{r}^{i,+},\mathcal{G}_{s}^{j,-}\right\} &=&\delta ^{ij}%
\left[ \mathcal{P}_{r+s}+\frac{c_{2}}{6}\,\left( r^{2}-\frac{1}{4}\right)
\delta _{r+s,0}\right] -\left( r-s\right) \left( \sigma ^{a}\right) _{ij}%
\mathtt{B}_{r+s}^{a},  \notag \\
\left\{ \mathcal{G}_{r}^{i,+},\mathcal{H}_{s}^{j,-}\right\} &=&\delta ^{ij}%
\left[ \mathcal{Z}_{r+s}+\frac{c_{3}}{6}\,\left( r^{2}-\frac{1}{4}\right)
\delta _{r+s,0}\right] -\left( r-s\right) \left( \sigma ^{a}\right) _{ij}%
\mathtt{Z}_{r+s}^{a}\,.  \notag
\end{eqnarray}%
Such infinite-dimensional superalgebra corresponds to the $\mathcal{N}=4$
supersymmetric extension of the deformed $\widetilde{BMS}_{3}$ algebra
endowed with $\mathfrak{su}\left( 2\right) $ current algebra spanned by $%
\mathfrak{k}_{m}^{a}$, $\mathfrak{\bar{k}}_{m}^{a}$ and $\mathfrak{\tilde{k}}%
_{m}^{a}$. The $\mathfrak{su}\left( 2\right) $ current generators are
related to $\mathtt{T}_{m}^{a}$, $\mathtt{B}_{m}^{a}$ and $\mathtt{Z}%
_{m}^{a} $ through the following redefinitions:%
\begin{equation}
\mathtt{T}_{m}^{a}=\mathfrak{k}_{m}^{a}+\mathfrak{\bar{k}}_{-m}^{a}+%
\mathfrak{\tilde{k}}_{-m}^{a}\,,\quad \mathtt{B}_{m}^{a}=\lim_{\epsilon
\rightarrow 0}\epsilon \left( \mathfrak{k}_{m}^{a}-\mathfrak{\bar{k}}%
_{-m}^{a}\right) \,,\quad \mathtt{Z}_{m}^{a}=\lim_{\epsilon \rightarrow
0}\epsilon ^{2}\left( \mathfrak{k}_{m}^{a}+\mathfrak{\bar{k}}%
_{-m}^{a}\right) \,,
\end{equation}%
One can note that the presence of the $\mathcal{Z}_{m}$, $\mathcal{H}%
_{r}^{i,\pm }$ and $\mathtt{Z}_{m}^{a}$ generators extends and deforms the $%
\mathcal{N}=4$ super-$BMS_{3}$ algebra presented in \cite{CCFR}.

On the other hand, such infinite-dimensional superalgebra contains a finite
subalgebra which corresponds to the central extension of the $\mathcal{N}%
=\left( 4,0\right) $ Maxwell superalgebra endowed with internal symmetry
generators \cite{Concha}. Indeed, the set of generators $\left\{ \mathcal{J}%
_{m},\mathcal{P}_{m},\emph{Z}_{m},\mathtt{T}_{0}^{a},\mathtt{B}_{0}^{a},%
\mathtt{Z}_{0}^{a},\mathcal{G}_{r}^{i,\pm },\mathcal{H}_{r}^{i,\pm }\right\}
$ with $m,n=0,\pm 1$ and $r=\pm \frac{1}{2}$ reproduces the $\mathcal{N}=4$
Maxwell superalgebra. It is worth it to mention that the $\mathcal{N}=4$
Maxwell superalgebra can also be obtained by applying an $S$-expansion to
the $\mathcal{N}=4$ super Lorentz algebra considering the same semigroup $%
S_{E}^{\left( 4\right) }$ used to obtain the $\mathcal{N}=4$ deformed super-$%
\widetilde{BMS}_{3}$ algebra.

\subsection{Supersymmetric extension of the asymptotic algebra of the $%
\mathfrak{so}\left( 2,2\right) \oplus \mathfrak{so}\left( 2,1\right) $
gravity theory}

Let us now explore the supersymmetric extension of the asymptotic
symmetry of the $\mathfrak{so}\left( 2,2\right) \oplus \mathfrak{so}\left(
2,1\right) $ CS gravity theory. As shown in \cite{CMRSV}, an explicit
realisation of the asymptotic symmetry at null infinity turned out to be a
semi-simple enlargement of the $BMS_{3}$ algebra (see Appendix B for further
details). Such infinite-dimensional algebra was first introduced in \cite%
{CCRS} as an $S$-expansion of the Virasoro algebra using the same semigroup $%
S_{\mathcal{M}}^{\left( 2\right) }$ used for obtaining its finite
subalgebra. Indeed, the AdS-Lorentz algebra can be obtained as a $S_{%
\mathcal{M}}^{\left( 2\right) }$-expansion of the Lorentz algebra.

Here, we present diverse supersymmetric extensions of the enlarged $BMS_{3}$
algebra by $S$-expanding the $\mathcal{N}$-extended super-Virasoro algebra
for $\mathcal{N}=1,2$ and $4$. To this purpose we shall use $S_{\mathcal{M}%
}^{\left( 4\right) }$ as the relevant semigroup. This semigroup is
characterized by the absence of zero elements which implies non-vanishing
commutators in the expanded algebra. Furthermore this election is due to the
fact that, as was discussed in \cite{CCRS}, the semigroup relating two
finite algebras can also be used to relate their respective
infinite-dimensional algebras. Since the AdS-Lorentz superalgebra can be
obtained as a S-expansion of the Lorentz superalgebra using $S_{\mathcal{M}%
}^{\left( 4\right) }$ as the semigroup, it seems then natural to apply the
same semigroup $S_{\mathcal{M}}^{\left( 4\right) }$ to the super Virasoro
algebra in order to obtain the infinite-dimensional enhancement of the
AdS-Lorentz superalgebra. One could argue that the new infinite-dimensional
superalgebras presented here would be the respective asymptotic symmetries
of three-dimensional AdS-Lorentz CS\ supergravity theories.

\subsubsection{Minimal enlarged super-$BMS_{3}$ algebra}

Let us consider the super-Virasoro algebra (\ref{svir}) as our starting
algebra. Let $S_{\mathcal{M}}^{\left( 4\right) }=\left\{ \lambda
_{0},\lambda _{1},\lambda _{2},\lambda _{3},\lambda _{4}\right\} $ be the
relevant semigroup whose elements satisfy the following multiplication law%
\begin{equation}
\begin{tabular}{l|lllll}
$\lambda _{4}$ & $\lambda _{4}$ & $\lambda _{1}$ & $\lambda _{2}$ & $\lambda
_{3}$ & $\lambda _{4}$ \\
$\lambda _{3}$ & $\lambda _{3}$ & $\lambda _{4}$ & $\lambda _{1}$ & $\lambda
_{2}$ & $\lambda _{3}$ \\
$\lambda _{2}$ & $\lambda _{2}$ & $\lambda _{3}$ & $\lambda _{4}$ & $\lambda
_{1}$ & $\lambda _{2}$ \\
$\lambda _{1}$ & $\lambda _{1}$ & $\lambda _{2}$ & $\lambda _{3}$ & $\lambda
_{4}$ & $\lambda _{1}$ \\
$\lambda _{0}$ & $\lambda _{0}$ & $\lambda _{1}$ & $\lambda _{2}$ & $\lambda
_{3}$ & $\lambda _{4}$ \\ \hline
& $\lambda _{0}$ & $\lambda _{1}$ & $\lambda _{2}$ & $\lambda _{3}$ & $%
\lambda _{4}$%
\end{tabular}
\label{mlsm}
\end{equation}%
One can notice that, unlike the $S_{E}$ semigroups, there is no zero element
in the $S_{\mathcal{M}}$ family. The absence of zero element implies that
there is no vanishing commutation relations in the expanded algebra.

Let us consider now a subset decomposition $S_{\mathcal{M}}^{\left( 4\right)
}=S_{0}\cup S_{1}$ with%
\begin{eqnarray}
S_{0} &=&\left\{ \lambda _{0},\lambda _{2},\lambda _{3}\right\} \,,  \notag
\\
S_{1} &=&\left\{ \lambda _{1},\lambda _{3}\right\} \,,
\end{eqnarray}%
which is resonant since it satisfies the same structure than the subspaces (%
\ref{subspace}).

A resonant subalgebra can be performed%
\begin{equation}
W_{R}=W_{0}\oplus W_{1}=S_{0}\times V_{0}\,\oplus S_{1}\times V_{1}\,,
\end{equation}%
with $V_{0}$ and $V_{1}$ being the subspaces of the super-Virasoro algebra.
Such resonant $S_{\mathcal{M}}^{\left( 4\right) }$-expansion of the
super-Virasoro reproduces a new infinite-dimensional algebra whose
generators are related to the super-Virasoro ones through the semigroup
elements as%
\begin{equation}
\begin{tabular}{lll}
$\mathcal{J}_{m}=\lambda _{0}\ell _{m}$ & , & $c_{1}=\lambda _{0}c\,,$ \\
$\ell \,\mathcal{P}_{m}=\lambda _{2}\ell _{m}$ & , & $\ell \,c_{2}=\lambda
_{2}c\,,$ \\
$\ell ^{2}\,\mathcal{Z}_{m}=\lambda _{4}\ell _{m}\,$ & , & $\ell
^{2}\,c_{3}=\lambda _{4}c\,,$ \\
$\ell ^{1/2}\,\mathcal{G}_{r}=\lambda _{1}\mathcal{Q}_{r}$ & , & $\ell
^{3/2}\,\mathcal{H}_{r}=\lambda _{3}\mathcal{Q}_{r}\,.$%
\end{tabular}%
\end{equation}%
Then, using the (anti-)commutation relations of the super-Virasoro algebra (%
\ref{svir}) and the multiplication law of the semigroup $S_{\mathcal{M}%
}^{\left( 4\right) }$ (\ref{mlsm}), one finds that the (anti-)commutators of
the expanded superalgebra are given by%
\begin{eqnarray}
\left[ \mathcal{J}_{m},\mathcal{J}_{n}\right] &=&\left( m-n\right) \mathcal{J%
}_{m+n}+\frac{c_{1}}{12}\,m\left( m^{2}-1\right) \delta _{m+n,0}\,,  \notag
\\
\left[ \mathcal{J}_{m},\mathcal{P}_{n}\right] &=&\left( m-n\right) \mathcal{P%
}_{m+n}+\frac{c_{2}}{12}\,m\left( m^{2}-1\right) \delta _{m+n,0}\,,  \notag
\\
\left[ \mathcal{P}_{m},\mathcal{P}_{n}\right] &=&\left( m-n\right) \mathcal{Z%
}_{m+n}+\frac{c_{3}}{12}\,m\left( m^{2}-1\right) \delta _{m+n,0}\,,  \notag
\\
\left[ \mathcal{J}_{m},\mathcal{Z}_{n}\right] &=&\left( m-n\right) \mathcal{Z%
}_{m+n}+\frac{c_{3}}{12}\,m\left( m^{2}-1\right) \delta _{m+n,0}\,,
\label{EBMS3a} \\
\left[ \mathcal{P}_{m},\mathcal{Z}_{n}\right] &=&\dfrac{1}{\ell ^{2}}\left(
m-n\right) \,\mathcal{P}_{m+n}+\dfrac{c_{2}}{12\ell ^{2}}\left(
m^{3}-m\right) \delta _{m+n,0}\,,  \notag \\
\left[ \mathcal{Z}_{m},\mathcal{Z}_{n}\right] &=&\dfrac{1}{\ell ^{2}}\left(
m-n\right) \,\mathcal{Z}_{m+n}+\dfrac{c_{3}}{12\ell ^{2}}\left(
m^{3}-m\right) \delta _{m+n,0}\,,  \notag
\end{eqnarray}%
\begin{eqnarray}
\left[ \mathcal{J}_{m},\mathcal{G}_{r}\right] &=&\left( \frac{m}{2}-r\right)
\mathcal{G}_{m+r}\,,\quad \left[ \mathcal{P}_{m},\mathcal{G}_{r}\right]
=\left( \frac{m}{2}-r\right) \mathcal{H}_{m+r}\,,  \notag \\
\left[ \mathcal{J}_{m},\mathcal{H}_{r}\right] &=&\left( \frac{m}{2}-r\right)
\mathcal{H}_{m+r}\,,\quad \left[ \mathcal{P}_{m},\mathcal{H}_{r}\right] =%
\dfrac{1}{\ell ^{2}}\left( \frac{m}{2}-r\right) \mathcal{G}_{m+r}\,,  \notag
\\
\left[ \mathcal{Z}_{m},\mathcal{G}_{r}\right] &=&\dfrac{1}{\ell ^{2}}\left(
\frac{m}{2}-r\right) \mathcal{G}_{m+r}\,,\quad \left[ \mathcal{Z}_{m},%
\mathcal{H}_{r}\right] =\dfrac{1}{\ell ^{2}}\left( \frac{m}{2}-r\right)
\mathcal{H}_{m+r}\,,  \notag \\
\left\{ \mathcal{G}_{r},\mathcal{G}_{s}\right\} &=&\mathcal{P}_{r+s}+\frac{%
c_{2}}{6}\,\left( r^{2}-\frac{1}{4}\right) \delta _{r+s,0}\,,  \label{EBMS3b}
\\
\left\{ \mathcal{G}_{r},\mathcal{H}_{s}\right\} &=&\mathcal{Z}_{r+s}+\frac{%
c_{3}}{6}\,\left( r^{2}-\frac{1}{4}\right) \delta _{r+s,0}\,,  \notag \\
\left\{ \mathcal{H}_{r},\mathcal{H}_{s}\right\} &=&\dfrac{1}{\ell ^{2}}%
\mathcal{P}_{r+s}+\frac{c_{2}}{6\ell ^{2}}\,\left( r^{2}-\frac{1}{4}\right)
\delta _{r+s,0}\,.  \notag
\end{eqnarray}%
Such infinite-dimensional superalgebra is a supersymmetric extension of the
enlarged $BMS_{3}$ algebra presented in \cite{CMRSV} and turns out to be the
infinite-dimensional lift of the minimal AdS-Lorentz superalgebra introduced
in \cite{CPR}. In fact, one can note that the set of generators $\left\{
\mathcal{J}_{m},\mathcal{P}_{m},\emph{Z}_{m},\mathcal{G}_{r},\mathcal{H}%
_{r}\right\} $ with $m=0,\pm 1$ and $r=\pm \frac{1}{2}$ defines a finite
subalgebra which reproduces the minimal AdS-Lorentz superalgebra spanned by $%
\left\{ J_{a},P_{a},Z_{a},Q,\Sigma \right\} $ through the change of basis (%
\ref{cob}) considered in the deformed super-$\widetilde{BMS}_{3}$ case. In
particular, the (anti-)commutators of the minimal AdS-Lorentz superalgebra
are given by (\ref{AdSL}) and%
\begin{eqnarray}
\left[ J_{a},Q_{\alpha }\right] &=&\frac{1}{2}\,\left( \Gamma _{a}\right) _{%
\text{ }\alpha }^{\beta }Q_{\beta }\,,\qquad \ \,\left[ J_{a},\Sigma
_{\alpha }\right] =\frac{1}{2}\,\left( \Gamma _{a}\right) _{\text{ }\alpha
}^{\beta }\Sigma _{\beta }\,,  \notag \\
\left[ P_{a},Q_{\alpha }\right] &=&\frac{1}{2}\,\,\left( \Gamma _{a}\right)
_{\text{ }\alpha }^{\beta }\Sigma _{\beta }\,,\qquad \ \,\left[ P_{a},\Sigma
_{\alpha }\right] =\frac{1}{2\ell ^{2}}\,\left( \Gamma _{a}\right) _{\text{ }%
\alpha }^{\beta }Q_{\beta }\,,  \notag \\
\left[ Z_{a},Q_{\alpha }\right] &=&\frac{1}{2\ell ^{2}}\,\left( \Gamma
_{a}\right) _{\text{ }\alpha }^{\beta }Q_{\beta }\,,\qquad \left[
Z_{a},\Sigma _{\alpha }\right] =\frac{1}{2\ell ^{2}}\,\left( \Gamma
_{a}\right) _{\text{ }\alpha }^{\beta }\Sigma _{\beta }\,,  \label{adsl} \\
\left\{ Q_{\alpha },Q_{\beta }\right\} &=&\frac{1}{2}\,\left( C\Gamma
^{a}\right) _{\alpha \beta }P_{a}\,,\text{ \ \ }\left\{ Q_{\alpha },\Sigma
_{\beta }\right\} =\frac{1}{2}\,\left( C\Gamma ^{a}\right) _{\alpha \beta
}Z_{a}\,,\,  \notag \\
\left\{ \Sigma _{\alpha },\Sigma _{\beta }\right\} &=&\frac{1}{2\ell ^{2}}%
\,\,\left( C\Gamma ^{a}\right) _{\alpha \beta }P_{a}\,.  \notag
\end{eqnarray}%
Further supersymmetric extensions of the AdS-Lorentz have also been explored
in four spacetime dimensions allowing to introduce a generalized
cosmological constant term in supergravity \cite{CRS, CIRR, BR, PR2}. The
(anti-)commutation relations (\ref{adsl}) can alternatively be obtained as
an $S$-expansion of the super-Lorentz algebra using the same semigroup $S_{%
\mathcal{M}}^{\left( 4\right) }$. On the other hand, as it bosonic version,
the limit $\ell \rightarrow \infty $ of the minimal AdS-Lorentz superalgebra
reproduces the minimal Maxwell superalgebra (\ref{sp1}). It is interesting
to notice that such limit can also be applied at the infinite-dimensional
level. Indeed the limit $\ell \rightarrow \infty $ applied to the minimal
enlarged super-$BMS_{3}$ algebra leads to the minimal deformed super-$%
\widetilde{BMS}_{3}$ algebra obtained previously. The following diagram
summarizes the expansion and limit relations:%
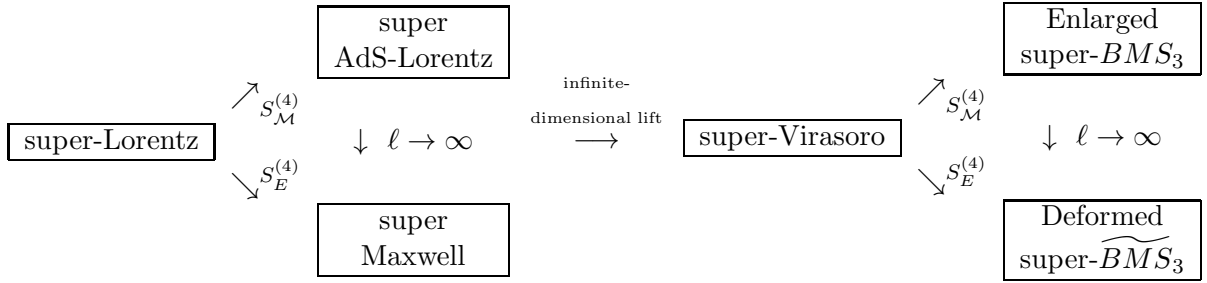
\begin{equation*}
\begin{tabular}{ccc}
\cline{3-3}
&  & \multicolumn{1}{|c|}{super} \\
&  & \multicolumn{1}{|c|}{AdS-Lorentz} \\ \cline{3-3}
& $\nearrow _{S_{\mathcal{M}}^{\left( 4\right) }}$ &  \\ \cline{1-1}
\multicolumn{1}{|c}{super-Lorentz} & \multicolumn{1}{|c}{} & $\downarrow $ \
$\ell \rightarrow \infty $ \\ \cline{1-1}
& $\searrow ^{S_{E}^{\left( 4\right) }}$ &  \\ \cline{3-3}
&  & \multicolumn{1}{|c|}{super} \\
&  & \multicolumn{1}{|c|}{Maxwell} \\ \cline{3-3}
\end{tabular}%
\overset{%
\begin{array}{c}
\text{{\tiny infinite-}} \\
\text{{\tiny dimensional lift}}%
\end{array}%
}{\longrightarrow }%
\begin{tabular}{ccc}
\cline{3-3}
&  & \multicolumn{1}{|c|}{Enlarged} \\
&  & \multicolumn{1}{|c|}{super-$BMS_{3}$} \\ \cline{3-3}
& $\nearrow _{S_{\mathcal{M}}^{\left( 4\right) }}$ &  \\ \cline{1-1}
\multicolumn{1}{|c}{super-Virasoro} & \multicolumn{1}{|c}{} & $\downarrow $
\ $\ell \rightarrow \infty $ \\ \cline{1-1}
& $\searrow ^{S_{E}^{\left( 4\right) }}$ &  \\ \cline{3-3}
&  & \multicolumn{1}{|c|}{Deformed} \\
&  & \multicolumn{1}{|c|}{super-$\widetilde{BMS}_{3}$} \\ \cline{3-3}
\end{tabular}%
\end{equation*}

There is an alternative basis in which the enlarged super-$BMS_{3}$ algebra
can be rewritten. Indeed three copies of the Virasoro algebra, two of which
augmented by supersymmetry, are revealed after the following redefinitions:%
\begin{eqnarray}
\mathcal{L}_{m}^{+} &=&\dfrac{1}{2}\left( \ell ^{2}\mathcal{Z}_{m}+\ell
\mathcal{P}_{m}\right) \,,\quad \quad \ \ \mathcal{L}_{m}^{-}=\dfrac{1}{2}%
\left( \ell ^{2}\mathcal{Z}_{-m}-\ell \mathcal{P}_{-m}\right) \,,\quad \quad
\hat{\mathcal{L}}_{m}=\mathcal{J}_{-m}-\ell ^{2}\mathcal{Z}_{-m}\,,  \notag
\\
Q_{r} &=&\dfrac{1}{2}\left( \ell ^{1/2}\mathcal{G}_{r}+\ell ^{3/2}\mathcal{H}%
_{r}\right) \,,\quad \bar{Q}_{r}=\dfrac{i}{2}\left( \ell ^{1/2}\mathcal{G}%
_{r}-\ell ^{3/2}\mathcal{H}_{r}\right) \,, \\
c^{\pm } &=&\dfrac{1}{2}\left( \ell ^{2}c_{3}\pm \ell c_{2}\right) \,,\quad
\qquad \quad \ \hat{c}=\left( c_{1}-\ell ^{2}c_{3}\right) \,.  \notag
\end{eqnarray}

Specifically, the (anti-)commutators are given by%
\begin{equation}
\begin{array}{lcl}
\left[ \mathcal{L}_{m}^{+},\mathcal{L}_{n}^{+}\right] & = & \left(
m-n\right) \mathcal{L}_{m+n}^{+}+\dfrac{c^{+}}{12}\,m\left( m^{2}-1\right)
\delta _{m+n,0}\,, \\[5pt]
\left[ \mathcal{L}_{m}^{-},\mathcal{L}_{n}^{-}\right] & = & \left(
m-n\right) \mathcal{L}_{m+n}^{-}+\dfrac{c^{-}}{12}\,m\left( m^{2}-1\right)
\delta _{m+n,0}\,, \\[5pt]
\left[ \hat{\mathcal{L}}_{m},\hat{\mathcal{L}}_{n}\right] & = & \left(
m-n\right) \mathcal{\hat{L}}_{m+n}+\dfrac{\hat{c}}{12}\,m\left(
m^{2}-1\right) \delta _{m+n,0}\,, \\
\left[ \mathcal{L}_{m}^{+},\mathcal{Q}_{r}\right] & = & \left( \dfrac{m}{2}%
-r\right) \mathcal{Q}_{m+r}\,\,, \\
\left[ \mathcal{L}_{m}^{-},\mathcal{\bar{Q}}_{r}\right] & = & \left( \dfrac{m%
}{2}-r\right) \mathcal{\bar{Q}}_{m+r}\,\,, \\
\left\{ \mathcal{Q}_{r},\mathcal{Q}_{s}\right\} & = & \mathcal{L}_{r+s}^{+}+%
\dfrac{c^{+}}{6}\,\left( r^{2}-\frac{1}{4}\right) \delta _{r+s,0}\,, \\
\left\{ \mathcal{\bar{Q}}_{r},\mathcal{\bar{Q}}_{s}\right\} & = & \mathcal{L}%
_{r+s}^{-}+\dfrac{c^{-}}{6}\,\left( r^{2}-\frac{1}{4}\right) \delta
_{r+s,0}\,.%
\end{array}%
\end{equation}%
This corresponds to the direct sum $\mathfrak{vir}\oplus \mathfrak{svir}%
\oplus \mathfrak{svir}$ and can be seen as the direct sum of the $\left(
1,1\right) $ superconformal algebra (\ref{scon}) and the Virasoro algebra.
Naturally the occurrence of such structure is due to the fact that the
finite AdS-Lorentz superalgebra is, in three spacetime dimensions,
isomorphic to three copies of the $\mathfrak{so}\left( 2,1\right) $ algebra,
two of which augmented by supersymmetry.

\subsubsection{$\mathcal{N}=2$ enlarged super-$BMS_{3}$ algebra}

A $\mathcal{N}=\left( 2,0\right) $ enlarged super-$BMS_{3}$ algebra can be
obtained considering the same semigroup $S_{\mathcal{M}}^{\left( 4\right) }$
but starting from the $\mathcal{N}=2$ super-Virasoro algebra (\ref{n2sv}).
Let us consider the subspace decomposition $\mathfrak{svir}_{2}=V_{0}\oplus
V_{1}$, with%
\begin{eqnarray}
V_{0} &=&\left\{ \ell _{m},\mathcal{R}_{m},c\right\} \,,  \notag \\
V_{1} &=&\left\{ Q_{r}^{i}\right\} \,,  \label{suba}
\end{eqnarray}%
which satisfies a graded Lie algebra (\ref{subspace}). Let $S_{\mathcal{M}%
}^{\left( 4\right) }=\left\{ \lambda _{0},\lambda _{1},\lambda _{2},\lambda
_{3},\lambda _{4}\right\} $ be the relevant semigroup whose elements satisfy
the multiplication law (\ref{mlsm}) and let $S_{\mathcal{M}}^{\left(
4\right) }=S_{0}\cup S_{1}$ be the resonant subset decomposition with%
\begin{eqnarray}
S_{0} &=&\left\{ \lambda _{0},\lambda _{2},\lambda _{4}\right\} \,,  \notag
\\
S_{1} &=&\left\{ \lambda _{1},\lambda _{3}\right\} \,,
\end{eqnarray}%
which satisfies the same structure as the subspaces (\ref{suba}).

After applying a resonant $S_{\mathcal{M}}^{\left( 4\right) }$-expansion to
the $\mathcal{N}=2$ super-Virasoro algebra one finds a new $\mathcal{N}=2$
infinite-dimensional superalgebra spanned by the set $\left\{ \mathcal{J}%
_{m},\mathcal{P}_{m},\mathcal{Z}_{m},\mathtt{T}_{m},\mathtt{B}_{m},\mathtt{Z}%
_{m},\mathcal{G}_{r}^{i},\mathcal{H}_{r}^{i},c_{1},c_{2},c_{3}\right\} $
whose generators are related to the super-Virasoro ones through the
semigroup elements as follows%
\begin{equation}
\begin{tabular}{lll}
$\mathcal{J}_{m}=\lambda _{0}\ell _{m}$ & , & $c_{1}=\lambda _{0}c\,,$ \\
$\ell \,\mathcal{P}_{m}=\lambda _{2}\ell _{m}$ & , & $\ell \,c_{2}=\lambda
_{2}c\,,$ \\
$\ell ^{2}\,\mathcal{Z}_{m}=\lambda _{4}\ell _{m}\,$ & , & $\ell
^{2}\,c_{3}=\lambda _{4}c\,,$ \\
\texttt{T}$_{m}=\lambda _{0}\mathcal{R}_{m}\,$ & , & $\ell \,$\texttt{B}$%
_{m}=\lambda _{2}\mathcal{R}_{m}\,,$ \\
$\ell ^{2}\,$\texttt{Z}$_{m}=\lambda _{4}\mathcal{R}_{m}\,,$ & , &  \\
$\ell ^{1/2}\,\mathcal{G}_{r}=\lambda _{1}\mathcal{Q}_{r}$ & , & $\ell
^{3/2}\,\mathcal{H}_{r}=\lambda _{3}\mathcal{Q}_{r}\,.$%
\end{tabular}%
\end{equation}%
One can show that the expanded generators satisfy an $\mathcal{N}=2$
supersymmetric extension of the enlarged $BMS_{3}$ algebra (\ref{EBMS3})
endowed with internal symmetry algebra. In particular, the
(anti-)commutation relations can directly be obtained by combining the
original (anti-)commutation relations of the $\mathcal{N}=2$ super-Virasoro
algebra (\ref{n2sv}) and the multiplication law of the semigroup (\ref{mlsm}%
). Indeed, the $\mathcal{N}=2$ enlarged super-$BMS_{3}$ algebra is given by
its bosonic subalgebra (\ref{EBMS3}) and%
\begin{eqnarray}
\left[ \mathcal{J}_{m},\mathtt{T}_{n}\right] &=&-n\mathtt{T}_{m+n}\,,\qquad %
\left[ \mathcal{P}_{m},\mathtt{T}_{n}\right] =-n\mathtt{B}_{m+n}\,,  \notag
\\
\left[ \mathcal{Z}_{m},\mathtt{T}_{n}\right] &=&-n\mathtt{Z}_{m+n}\,,\qquad %
\left[ \mathcal{J}_{m},\mathtt{B}_{n}\right] =-n\mathtt{B}_{m+n}\,,  \notag
\\
\left[ \mathcal{P}_{m},\mathtt{B}_{n}\right] &=&-n\mathtt{Z}_{m+n}\,,\qquad %
\left[ \mathcal{Z}_{m},\mathtt{B}_{n}\right] =-\frac{1}{\ell ^{2}}n\mathtt{B}%
_{m+n}\,,  \notag \\
\left[ \mathcal{J}_{m},\mathtt{Z}_{n}\right] &=&-n\mathtt{Z}_{m+n}\,,\qquad %
\left[ \mathcal{P}_{m},\mathtt{Z}_{n}\right] =-\frac{1}{\ell ^{2}}n\mathtt{B}%
_{m+n}\,,  \notag \\
\left[ \mathcal{Z}_{m},\mathtt{Z}_{n}\right] &=&-\frac{1}{\ell ^{2}}n\mathtt{%
Z}_{m+n}\,,  \label{EBMS3c} \\
\left[ \mathtt{T}_{m},\mathtt{T}_{n}\right] &=&\frac{c_{1}}{3}m\delta
_{m+n,0}\,,\quad \left[ \mathtt{T}_{m},\mathtt{B}_{n}\right] =\frac{c_{2}}{3}%
m\delta _{m+n,0}\,,  \notag \\
\left[ \mathtt{T}_{m},\mathtt{Z}_{n}\right] &=&\frac{c_{3}}{3}m\delta
_{m+n,0}\,,\quad \left[ \mathtt{B}_{m},\mathtt{B}_{n}\right] =\frac{c_{3}}{3}%
m\delta _{m+n,0}\,,  \notag \\
\left[ \mathtt{B}_{m},\mathtt{Z}_{n}\right] &=&\frac{c_{2}}{3\ell ^{2}}%
m\delta _{m+n,0}\,,\text{ \ }\left[ \mathtt{Z}_{m},\mathtt{Z}_{n}\right] =%
\frac{c_{3}}{3\ell ^{2}}m\delta _{m+n,0}\,,  \notag
\end{eqnarray}%
\begin{eqnarray}
\left[ \mathcal{J}_{m},\mathcal{G}_{r}^{i}\right] &=&\left( \frac{m}{2}%
-r\right) \mathcal{G}_{m+r}^{i}\,,\qquad \left[ \mathcal{P}_{m},\mathcal{G}%
_{r}^{i}\right] =\left( \frac{m}{2}-r\right) \mathcal{H}_{m+r}^{i}\,,  \notag
\\
\left[ \mathcal{J}_{m},\mathcal{H}_{r}^{i}\right] &=&\left( \frac{m}{2}%
-r\right) \mathcal{H}_{m+r}^{i}\,,\qquad \left[ \mathcal{P}_{m},\mathcal{H}%
_{r}^{i}\right] =\frac{1}{\ell ^{2}}\left( \frac{m}{2}-r\right) \mathcal{G}%
_{m+r}^{i}\,,  \notag \\
\left[ \mathcal{Z}_{m},\mathcal{G}_{r}^{i}\right] &=&\frac{1}{\ell ^{2}}%
\left( \frac{m}{2}-r\right) \mathcal{G}_{m+r}^{i}\,,\text{\quad }\left[
\mathcal{Z}_{m},\mathcal{H}_{r}^{i}\right] =\frac{1}{\ell ^{2}}\left( \frac{m%
}{2}-r\right) \mathcal{H}_{m+r}^{i}\,,  \notag \\
\left[ \mathcal{G}_{r}^{i},\mathtt{T}_{m}\right] &=&\epsilon ^{ij}\mathcal{G}%
_{m+r}^{j}\,,\qquad \left[ \mathcal{G}_{r}^{i},\mathtt{B}_{m}\right]
=\epsilon ^{ij}\mathcal{H}_{m+r}^{j}\,,  \notag \\
\left[ \mathcal{H}_{r}^{i},\mathtt{T}_{m}\right] &=&\epsilon ^{ij}\mathcal{H}%
_{m+r}^{j}\,,\quad \ \ \left[ \mathcal{G}_{r}^{i},\mathtt{Z}_{m}\right] =%
\frac{1}{\ell ^{2}}\epsilon ^{ij}\mathcal{G}_{m+r}^{j}\,,  \label{EBMS3d} \\
\left[ \mathcal{H}_{r}^{i},\mathtt{B}_{m}\right] &=&\frac{1}{\ell ^{2}}%
\epsilon ^{ij}\mathcal{G}_{m+r}^{j}\,,\quad \left[ \mathcal{H}_{r}^{i},%
\mathtt{Z}_{m}\right] =\frac{1}{\ell ^{2}}\epsilon ^{ij}\mathcal{H}%
_{m+r}^{j}\,,  \notag \\
\left\{ \mathcal{G}_{r}^{i},\mathcal{G}_{s}^{j}\right\} &=&\delta ^{ij}\left[
\mathcal{P}_{r+s}+\frac{c_{2}}{6}\,\left( r^{2}-\frac{1}{4}\right) \delta
_{r+s,0}\right] -2\epsilon ^{ij}\left( r-s\right) \mathtt{B}_{r+s}\,,  \notag
\\
\left\{ \mathcal{G}_{r}^{i},\mathcal{H}_{s}^{j}\right\} &=&\delta ^{ij}\left[
\mathcal{Z}_{r+s}+\frac{c_{3}}{6}\,\left( r^{2}-\frac{1}{4}\right) \delta
_{r+s,0}\right] -2\epsilon ^{ij}\left( r-s\right) \mathtt{Z}_{r+s}\,,  \notag
\\
\left\{ \mathcal{H}_{r}^{i},\mathcal{H}_{s}^{j}\right\} &=&\frac{\delta ^{ij}%
}{\ell ^{2}}\left[ \mathcal{P}_{r+s}+\frac{c_{2}}{6}\,\left( r^{2}-\frac{1}{4%
}\right) \delta _{r+s,0}\right] -\frac{2}{\ell ^{2}}\epsilon ^{ij}\left(
r-s\right) \mathtt{B}_{r+s}\,.  \notag
\end{eqnarray}%
The $\mathcal{N}=2$ supersymmetric extension of the enlarged $BMS_{3}$
algebra requires the introduction of new bosonic generators $\left\{ \mathtt{%
T}_{m},\mathtt{B}_{m},\mathtt{Z}_{m}\right\} $ which satisfy internal
symmetry algebras. Their presence is due to the R-symmetry generator
appearing in the $\mathcal{N}=2$ super-Virasoro algebra. Interestingly, a
particular redefinition of the generators allows us to rewrite the present
infinite-dimensional superalgebra as three copies of the Virasoro algebra,
two of which are augmented by supersymmetry, endowed with a $\mathfrak{\hat{u%
}}\left( 1\right) $ current algebra. In particular, the structure in
presence of $\mathfrak{\hat{u}}\left( 1\right) $ current generators is
apparent by considering the following redefinitions%
\begin{eqnarray}
\mathcal{L}_{m}^{+} &=&\frac{1}{2}\left( \ell ^{2}\mathcal{Z}_{m}+\ell
\mathcal{P}_{m}\right) \,,\qquad \ \mathcal{L}_{m}^{-}=\frac{1}{2}\left(
\ell ^{2}\mathcal{Z}_{-m}-\ell \mathcal{P}_{-m}\right) \,,\qquad \hat{%
\mathcal{L}}_{m}=\mathcal{J}_{-m}-\ell ^{2}\mathcal{Z}_{-m}\,,  \notag \\
\mathfrak{k}_{m}^{+} &=&\frac{1}{2}\left( \ell \mathtt{B}_{m}+\ell ^{2}%
\mathtt{Z}_{m}\right) \,,\qquad \ \ \ \mathfrak{k}_{m}^{-}=\frac{1}{2}\left(
\ell \mathtt{B}_{m}-\ell ^{2}\mathtt{Z}_{m}\right) \,,\qquad \ \ \ \
\mathfrak{\hat{k}}_{m}=\frac{1}{2}\left( \mathtt{T}_{m}-\ell ^{2}\mathtt{Z}%
_{m}\right) \,,  \notag \\
Q_{r}^{i} &=&\frac{1}{2}\left( \ell ^{1/2}\mathcal{G}_{r}^{i}+\ell ^{3/2}%
\mathcal{H}_{r}^{i}\right) \,,\quad \bar{Q}_{r}^{i}=\frac{i}{2}\left( \ell
^{1/2}\mathcal{G}_{r}^{i}-\ell \mathcal{H}_{r}^{i}\right) \,, \\
c^{\pm } &=&\frac{1}{2}\left( \ell ^{2}c_{3}\pm \ell c_{2}\right) \,,\qquad
\quad \ \ \ \ \ \hat{c}=\left( c_{1}-\ell ^{2}c_{3}\right) \,.  \notag
\end{eqnarray}%
With these redefinitions the (anti-)commutation relations (\ref{EBMS3}), (%
\ref{EBMS3c}) and (\ref{EBMS3d}) change into%
\begin{eqnarray}
\left[ \mathcal{L}_{m}^{\pm },\mathcal{L}_{n}^{\pm }\right] &=&\left(
m-n\right) \mathcal{L}_{m+n}^{\pm }+\dfrac{c^{\pm }}{12}\,m\left(
m^{2}-1\right) \delta _{m+n,0}\,,  \notag \\
\left[ \hat{\mathcal{L}}_{m},\hat{\mathcal{L}}_{n}\right] &=&\left(
m-n\right) \mathcal{\hat{L}}_{m+n}+\dfrac{\hat{c}}{12}\,m\left(
m^{2}-1\right) \delta _{m+n,0}\,,  \notag \\
\left[ \mathcal{L}_{m}^{\pm },\mathfrak{k}_{n}^{\pm }\right] &=&-n\mathfrak{k%
}_{m+n}^{\pm }\,,\qquad \qquad \left[ \mathcal{\hat{L}}_{m},\mathfrak{\hat{k}%
}_{n}\right] =-n\mathfrak{\hat{k}}_{m+n}\,,  \notag \\
\left[ \mathfrak{k}_{m}^{\pm },\mathfrak{k}_{n}^{\pm }\right] &=&\dfrac{%
c^{\pm }}{3}m\delta _{m+n,0}\,,\quad \quad \ \ \ \left[ \mathfrak{\hat{k}}%
_{m},\mathfrak{\hat{k}}_{n}\right] =\dfrac{\hat{c}}{3}m\delta _{m+n,0}\,,
\notag \\
\left[ \mathcal{L}_{m}^{+},\mathcal{Q}_{r}^{i}\right] &=&\left( \dfrac{m}{2}%
-r\right) \mathcal{Q}_{m+r}^{i}\,\,,\quad \left[ \mathcal{L}_{m}^{-},%
\mathcal{\bar{Q}}_{r}^{i}\right] =\left( \dfrac{m}{2}-r\right) \mathcal{\bar{%
Q}}_{m+r}^{i}\,\,,  \label{V2SV} \\
\left[ \mathcal{Q}_{r}^{i},\mathfrak{k}_{m}^{+}\right] &=&\epsilon ^{ij}%
\mathcal{Q}_{m+r}^{j}\,,\quad \ \ \ \ \ \qquad \left[ \mathcal{\bar{Q}}%
_{r}^{i},\mathfrak{k}_{m}^{-}\right] =\epsilon ^{ij}\mathcal{\bar{Q}}%
_{m+r}^{j}\,,  \notag \\
\left\{ \mathcal{Q}_{r}^{i},\mathcal{Q}_{s}^{j}\right\} &=&\delta ^{ij}\,%
\left[ \mathcal{L}_{r+s}^{+}+\dfrac{c^{+}}{6}\,\left( r^{2}-\frac{1}{4}%
\right) \delta _{r+s,0}\,\right] -2\epsilon ^{ij}\left( r-s\right) \mathfrak{%
k}_{r+s}^{+}\,,  \notag \\
\left\{ \mathcal{\bar{Q}}_{r}^{i},\mathcal{\bar{Q}}_{s}^{j}\right\}
&=&\delta ^{ij}\left[ \mathcal{L}_{r+s}^{-}+\dfrac{c^{-}}{6}\,\left( r^{2}-%
\frac{1}{4}\right) \delta _{r+s,0}\right] -2\epsilon ^{ij}\left( r-s\right)
\mathfrak{k}_{r+s}^{-}\,,\,.  \notag
\end{eqnarray}%
The infinite-dimensional superalgebra (\ref{V2SV}) corresponds to the direct
sum of the $\left( 2,2\right) $ superconformal algebra and the Virasoro
algebra endowed with a $\mathfrak{\hat{u}}\left( 1\right) $ current algebra.
In particular the $\mathfrak{\hat{u}}\left( 1\right) $ current generators $%
\mathfrak{k}_{m}^{+}$ and $\mathfrak{k}_{m}^{-}$ are R-symmetry generators
each one belonging to a $\mathcal{N}=2$ super-Virasoro algebra. Although
such structure seems more natural, the $\mathcal{N}=2$ enlarged super-$%
BMS_{3}$ algebra reproduces the $\mathcal{N}=2$ deformed super-$\widetilde{%
BMS}_{3}$ algebra (\ref{BosDefBMS3})-(\ref{ccc}) in the limit $\ell
\rightarrow \infty $ considering the basis $\left\{ \mathcal{J}_{m},\mathcal{%
P}_{m},\mathcal{Z}_{m},\mathtt{T}_{m},\mathtt{B}_{m},\mathtt{Z}_{m},\mathcal{%
G}_{r}^{i},\mathcal{H}_{r}^{i}\right\} $. Naturally, the diagram summarizing
the expansion and limit relations appearing in the minimal case can also be
reproduced in the $\mathcal{N}=2$ case showing that the expansions and flat
limit appearing at the infinite-dimensional level are also present in their
finite subalgebras.

On the other hand, one can note that the $\mathcal{N}=2$ AdS-Lorentz
superalgebra endowed with $\mathfrak{so}\left( 2\right) $ internal symmetry
generators \cite{Concha} is as a finite subalgebra of the $\mathcal{N}=2$
enlarged super-$BMS_{3}$ algebra. Indeed, the subalgebra spanned by the
generators $\mathcal{J}_{0}$, $\mathcal{J}_{\pm 1}$, $\mathcal{P}_{0}$, $%
\mathcal{P}_{\pm 1}$, $\mathcal{Z}_{0}$, $\mathcal{Z}_{\pm 1}$, $\mathtt{T}%
_{0}$, $\mathtt{B}_{0}$, $\mathtt{Z}_{0}$, $\mathcal{G}_{\pm \frac{1}{2}}$
and $\mathcal{H}_{\pm \frac{1}{2}}$ are related to the $\mathcal{N}=2$
AdS-Lorentz superalgebra $\left\{ J_{a},P_{a},Z_{a},T,B,Z,Q,\Sigma \right\} $
by considering the redefinitions (\ref{redef}).

\subsubsection{$\mathcal{N}=4$ enlarged super-$BMS_{3}$ algebra}

For completeness we provide with the $\mathcal{N}=4$ enlarged super-$BMS_{3}$
algebra which corresponds to the infinite-dimensional lift of the $\mathcal{N%
}=4$ AdS-Lorentz superalgebra endowed with internal symmetry algebra. The
new infinite-dimensional superalgebra can be obtained by applying an $S$%
-expansion of the $\mathcal{N}=4$ super-Virasoro algebra (\ref{N4SV}). In
particular, considering $S_{\mathcal{M}}^{\left( 4\right) }$ as the relevant
finite semigroup, whose elements satisfy the multiplication law (\ref{mlsm}%
), and applying a resonant $S_{\mathcal{M}}^{\left( 4\right) }$-expansion of
the $\mathcal{N}=4$ super-Virasoro algebra we find an expanded $\mathcal{N}%
=4 $ infinite-dimensional superalgebra spanned by the generators%
\begin{equation}
\left\{ \mathcal{J}_{m},\mathcal{P}_{m},\mathcal{Z}_{m},\mathtt{T}_{m}^{a},%
\mathtt{B}_{m}^{a},\mathtt{Z}_{m}^{a},\mathcal{G}_{r}^{i,\pm },\mathcal{H}%
_{r}^{i,\pm },c_{1},c_{2},c_{3}\right\} \,,
\end{equation}%
which are related to the $\mathcal{N}=4$ super-Virasoro ones through (\ref%
{expgenb}). One can show that, using the multiplication law (\ref{mlsm}) of
the $S_{\mathcal{M}}^{\left( 4\right) }$ semigroup and the original
(anti-)commutators (\ref{N4SV}) of the $\mathcal{N}=4$ super-Virasoro
algebra, the expanded generators satisfy a $\mathcal{N}=4$ enlarged super-$%
BMS_{3}$ algebra. In particular the new infinite-dimensional superalgebra
contains the enlarged $BMS_{3}$ algebra (\ref{EBMS3}) as a bosonic
subalgebra. On the other hand, the set of R-symmetry generators $\left\{
\mathtt{T}_{m}^{a},\mathtt{B}_{m}^{a},\mathtt{Z}_{m}^{a}\right\} $ with $%
a=1,2,3$ obeys the following commutators:
\begin{eqnarray}
\left[ \mathcal{J}_{m},\mathtt{T}_{n}^{a}\right] &=&-n\mathtt{T}%
_{m+n}^{a}\,,\qquad \left[ \mathcal{P}_{m},\mathtt{T}_{n}^{a}\right] =-n%
\mathtt{B}_{m+n}^{a}\,,  \notag \\
\left[ \mathcal{Z}_{m},\mathtt{T}_{n}^{a}\right] &=&-n\mathtt{Z}%
_{m+n}^{a}\,,\qquad \left[ \mathcal{J}_{m},\mathtt{B}_{n}^{a}\right] =-n%
\mathtt{B}_{m+n}^{a}\,,  \notag \\
\left[ \mathcal{P}_{m},\mathtt{B}_{n}^{a}\right] &=&-n\mathtt{Z}%
_{m+n}^{a}\,,\qquad \left[ \mathcal{Z}_{m},\mathtt{B}_{n}^{a}\right] =-\frac{%
1}{\ell ^{2}}n\mathtt{B}_{m+n}^{a}\,,  \label{ebms3a} \\
\left[ \mathcal{J}_{m},\mathtt{Z}_{n}^{a}\right] &=&-n\mathtt{Z}%
_{m+n}^{a}\,,\qquad \left[ \mathcal{P}_{m},\mathtt{Z}_{n}^{a}\right] =-\frac{%
1}{\ell ^{2}}n\mathtt{B}_{m+n}^{a}\,,  \notag \\
\left[ \mathcal{Z}_{m},\mathtt{Z}_{n}^{a}\right] &=&-\frac{1}{\ell ^{2}}n%
\mathtt{Z}_{m+n}^{a}\,,  \notag
\end{eqnarray}%
\begin{eqnarray}
\left[ \mathtt{T}_{m}^{a},\mathtt{T}_{n}^{b}\right] &=&i\epsilon ^{abc}%
\mathtt{T}_{m+n}^{c}+\frac{c_{1}}{12}\,m\delta ^{ab}\delta _{m+n,0}\,,
\notag \\
\left[ \mathtt{T}_{m}^{a},\mathtt{B}_{n}^{b}\right] &=&i\epsilon ^{abc}%
\mathtt{B}_{m+n}^{c}+\frac{c_{2}}{12}\,m\delta ^{ab}\delta _{m+n,0}\,,
\notag \\
\left[ \mathtt{T}_{m}^{a},\mathtt{Z}_{n}^{b}\right] &=&i\epsilon ^{abc}%
\mathtt{Z}_{m+n}^{c}+\frac{c_{3}}{12}\,m\delta ^{ab}\delta _{m+n,0}\,,
\notag \\
\left[ \mathtt{B}_{m}^{a},\mathtt{B}_{n}^{b}\right] &=&i\epsilon ^{abc}%
\mathtt{Z}_{m+n}^{c}+\frac{c_{3}}{12}\,m\delta ^{ab}\delta _{m+n,0}\,,
\label{ebms3b} \\
\left[ \mathtt{B}_{m}^{a},\mathtt{Z}_{n}^{b}\right] &=&\frac{1}{\ell ^{2}}%
\left( i\epsilon ^{abc}\mathtt{B}_{m+n}^{c}+\frac{c_{2}}{12}\,m\delta
^{ab}\delta _{m+n,0}\right) \,,  \notag \\
\left[ \mathtt{Z}_{m}^{a},\mathtt{Z}_{n}^{b}\right] &=&\frac{1}{\ell ^{2}}%
\left( i\epsilon ^{abc}\mathtt{Z}_{m+n}^{c}+\frac{c_{3}}{12}\,m\delta
^{ab}\delta _{m+n,0}\right) \,,  \notag
\end{eqnarray}%
\begin{eqnarray}
\left[ \mathtt{T}_{m}^{a},\mathcal{G}_{r}^{i,+}\right] &=&-\frac{1}{2}\left(
\sigma ^{a}\right) _{j}^{i}\mathcal{G}_{m+r}^{j,+}\,,\qquad \left[ \mathtt{T}%
_{m}^{a},\mathcal{G}_{r}^{i,-}\right] =\frac{1}{2}\left( \bar{\sigma}%
^{a}\right) _{j}^{i}\mathcal{G}_{m+r}^{j,-}\,,  \notag \\
\left[ \mathtt{B}_{m}^{a},\mathcal{G}_{r}^{i,+}\right] &=&-\frac{1}{2}\left(
\sigma ^{a}\right) _{j}^{i}\mathcal{H}_{m+r}^{j,+}\,,\qquad \left[ \mathtt{B}%
_{m}^{a},\mathcal{G}_{r}^{i,-}\right] =\frac{1}{2}\left( \bar{\sigma}%
^{a}\right) _{j}^{i}\mathcal{H}_{m+r}^{j,-}\,,  \notag \\
\left[ \mathtt{T}_{m}^{a},\mathcal{H}_{r}^{i,+}\right] &=&-\frac{1}{2}\left(
\sigma ^{a}\right) _{j}^{i}\mathcal{H}_{m+r}^{j,+}\,,\qquad \left[ \mathtt{T}%
_{m}^{a},\mathcal{H}_{r}^{i,-}\right] =\frac{1}{2}\left( \bar{\sigma}%
^{a}\right) _{j}^{i}\mathcal{H}_{m+r}^{j,-}\,,  \notag \\
\left[ \mathtt{B}_{m}^{a},\mathcal{H}_{r}^{i,+}\right] &=&-\frac{1}{2\ell
^{2}}\left( \sigma ^{a}\right) _{j}^{i}\mathcal{G}_{m+r}^{j,+}\,,\qquad %
\left[ \mathtt{B}_{m}^{a},\mathcal{H}_{r}^{i,-}\right] =\frac{1}{2\ell ^{2}}%
\left( \bar{\sigma}^{a}\right) _{j}^{i}\mathcal{G}_{m+r}^{j,-}\,,
\label{ebms3c} \\
\left[ \mathtt{Z}_{m}^{a},\mathcal{G}_{r}^{i,+}\right] &=&-\frac{1}{2\ell
^{2}}\left( \sigma ^{a}\right) _{j}^{i}\mathcal{G}_{m+r}^{j,+}\,,\qquad %
\left[ \mathtt{Z}_{m}^{a},\mathcal{G}_{r}^{i,-}\right] =\frac{1}{2\ell ^{2}}%
\left( \bar{\sigma}^{a}\right) _{j}^{i}\mathcal{G}_{m+r}^{j,-}\,,  \notag \\
\left[ \mathtt{Z}_{m}^{a},\mathcal{H}_{r}^{i,+}\right] &=&-\frac{1}{2\ell
^{2}}\left( \sigma ^{a}\right) _{j}^{i}\mathcal{H}_{m+r}^{j,+}\,,\qquad %
\left[ \mathtt{Z}_{m}^{a},\mathcal{H}_{r}^{i,-}\right] =\frac{1}{2\ell ^{2}}%
\left( \bar{\sigma}^{a}\right) _{j}^{i}\mathcal{H}_{m+r}^{j,-}\,,  \notag
\end{eqnarray}%
Here, $\bar{\sigma}_{ij}^{a}=\sigma _{ji}^{a}$ with $\sigma ^{a}$ being the
Pauli matrices. Furthermore, the fermionic generators $\mathcal{G}%
_{r}^{i,\pm }$ and $\mathcal{H}_{r}^{i,\pm }$, with $r=\pm \frac{1}{2}$
satisfy the following (anti-)commutation relations:%
\begin{eqnarray}
\left[ \mathcal{J}_{m},\mathcal{G}_{r}^{i,\pm }\right] &=&\left( \frac{m}{2}%
-r\right) \mathcal{G}_{m+r}^{i,\pm }\,,\qquad \left[ \mathcal{P}_{m},%
\mathcal{G}_{r}^{i,\pm }\right] =\left( \frac{m}{2}-r\right) \mathcal{H}%
_{m+r}^{i,\pm }\,,  \notag \\
\left[ \mathcal{J}_{m},\mathcal{H}_{r}^{i,\pm }\right] &=&\left( \frac{m}{2}%
-r\right) \mathcal{H}_{m+r}^{i,\pm }\,,\qquad \left[ \mathcal{Z}_{m},%
\mathcal{G}_{r}^{i,\pm }\right] =\frac{1}{\ell ^{2}}\left( \frac{m}{2}%
-r\right) \mathcal{G}_{m+r}^{i,\pm }\,,  \notag \\
\left[ \mathcal{P}_{m},\mathcal{H}_{r}^{i,\pm }\right] &=&\frac{1}{\ell ^{2}}%
\left( \frac{m}{2}-r\right) \mathcal{G}_{m+r}^{i,\pm }\,,\quad \left[
\mathcal{Z}_{m},\mathcal{H}_{r}^{i,\pm }\right] =\frac{1}{\ell ^{2}}\left(
\frac{m}{2}-r\right) \mathcal{H}_{m+r}^{i,\pm }\,,  \notag \\
\left\{ \mathcal{G}_{r}^{i,+},\mathcal{G}_{s}^{j,-}\right\} &=&\delta ^{ij}%
\left[ \mathcal{P}_{r+s}+\frac{c_{2}}{6}\,\left( r^{2}-\frac{1}{4}\right)
\delta _{r+s,0}\right] -\left( r-s\right) \left( \sigma ^{a}\right) _{ij}%
\mathtt{B}_{r+s}^{a}, \\
\left\{ \mathcal{G}_{r}^{i,+},\mathcal{H}_{s}^{j,-}\right\} &=&\delta ^{ij}%
\left[ \mathcal{Z}_{r+s}+\frac{c_{3}}{6}\,\left( r^{2}-\frac{1}{4}\right)
\delta _{r+s,0}\right] -\left( r-s\right) \left( \sigma ^{a}\right) _{ij}%
\mathtt{Z}_{r+s}^{a}\,,  \notag \\
\left\{ \mathcal{H}_{r}^{i,+},\mathcal{H}_{s}^{j,-}\right\} &=&\frac{\delta
^{ij}}{\ell ^{2}}\left[ \mathcal{P}_{r+s}+\frac{c_{2}}{6}\,\left( r^{2}-%
\frac{1}{4}\right) \delta _{r+s,0}\right] -\frac{1}{\ell ^{2}}\left(
r-s\right) \left( \sigma ^{a}\right) _{ij}\mathtt{B}_{r+s}^{a}\,.  \notag
\end{eqnarray}%
It is interesting to note that the $\mathcal{N}=4$ deformed super-$%
\widetilde{BMS}_{3}$ algebra given by (\ref{dbms3}), (\ref{dbms3a}) and (\ref%
{dbms3b}) can alternatively be recovered by applying a flat limit $\ell
\rightarrow \infty $ to the present $\mathcal{N}=4$ enlarged super-$BMS_{3}$
algebra. In particular, the internal symmetry generator $\mathtt{Z}_{r}^{a}$
is no more a R-symmetry generator after the limit $\ell \rightarrow \infty $%
. \ Such flat limit can also be reproduced at the finite subalgebra level.
Indeed, the finite set of generators $\left\{ \mathcal{J}_{m},\mathcal{P}%
_{m},\mathcal{Z}_{m},\mathtt{T}_{m}^{a},\mathtt{B}_{m}^{a},\mathtt{Z}%
_{m}^{a},\mathcal{G}_{r}^{i,\pm },\mathcal{H}_{r}^{i,\pm }\right\} $ with $%
m=0,\pm 1$ and $r=\pm \frac{1}{2}$ reproduces the $\mathcal{N}=4$
AdS-Lorentz superalgebra \cite{Concha}. As was discussed in \cite{Concha},
the $\mathcal{N}$-extended AdS-Lorentz superalgebra leads to the $\mathcal{N}
$-extended Maxwell superalgebra in the limit $\ell \rightarrow \infty $
which results to be the finite subalgebra of the $\mathcal{N}$-extended
deformed super-$\widetilde{BMS}_{3}$ algebra. In particular, the internal
symmetry generator $\mathtt{Z}_{m}^{a}$ with $m=0,\pm 1$ becomes a central
charge after the flat limit.

Note that the semigroup used to obtain the $\mathcal{N}$-extended enlarged
super-$BMS_{3}$ algebra from the $\mathcal{N}$-extended super-Virasoro
algebra is the same used to recovered the $\mathcal{N}$-extended AdS-Lorentz
superalgebra from an $\mathcal{N}$-extended super-Lorentz algebra. The
following diagram summarizes the limit and expansion relations present in
the new infinite-dimensional algebras and their finite subalgebra:%
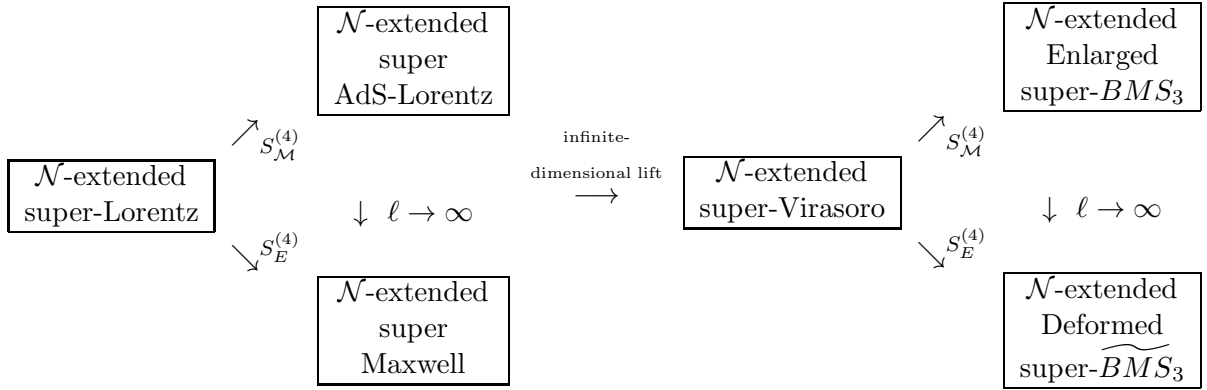
\begin{equation*}
\begin{tabular}{ccc}
\cline{3-3}
&  & \multicolumn{1}{|c|}{$\mathcal{N}$-extended} \\
&  & \multicolumn{1}{|c|}{super} \\
&  & \multicolumn{1}{|c|}{AdS-Lorentz} \\ \cline{3-3}
& $\nearrow _{S_{\mathcal{M}}^{\left( 4\right) }}$ &  \\ \cline{1-1}
\multicolumn{1}{|c}{$\mathcal{N}$-extended} & \multicolumn{1}{|c}{} &  \\
\multicolumn{1}{|c}{super-Lorentz} & \multicolumn{1}{|c}{} & $\downarrow $ \
$\ell \rightarrow \infty $ \\ \cline{1-1}
& $\searrow ^{S_{E}^{\left( 4\right) }}$ &  \\ \cline{3-3}
&  & \multicolumn{1}{|c|}{$\mathcal{N}$-extended} \\
&  & \multicolumn{1}{|c|}{super} \\
&  & \multicolumn{1}{|c|}{Maxwell} \\ \cline{3-3}
\end{tabular}%
\overset{%
\begin{array}{c}
\text{{\tiny infinite-}} \\
\text{{\tiny dimensional lift}}%
\end{array}%
}{\longrightarrow }%
\begin{tabular}{ccc}
\cline{3-3}
&  & \multicolumn{1}{|c|}{$\mathcal{N}$-extended} \\
&  & \multicolumn{1}{|c|}{Enlarged} \\
&  & \multicolumn{1}{|c|}{super-$BMS_{3}$} \\ \cline{3-3}
& $\nearrow _{S_{\mathcal{M}}^{\left( 4\right) }}$ &  \\ \cline{1-1}
\multicolumn{1}{|c}{$\mathcal{N}$-extended} & \multicolumn{1}{|c}{} &  \\
\multicolumn{1}{|c}{super-Virasoro} & \multicolumn{1}{|c}{} & $\downarrow $
\ $\ell \rightarrow \infty $ \\ \cline{1-1}
& $\searrow ^{S_{E}^{\left( 4\right) }}$ &  \\ \cline{3-3}
&  & \multicolumn{1}{|c|}{$\mathcal{N}$-extended} \\
&  & \multicolumn{1}{|c|}{Deformed} \\
&  & \multicolumn{1}{|c|}{super-$\widetilde{BMS}_{3}$} \\ \cline{3-3}
\end{tabular}%
\end{equation*}

Let us note that the $\mathcal{N}=4$ enlarged super-$BMS_{3}$ can be
rewritten in an alternative basis. Indeed a particular redefinition of the
generators allows us to rewrite the infinite-dimensional superalgebra as
three copies of the Virasoro algebra, two of which augmented by
supersymmetry, endowed with $\mathfrak{su}\left( 2\right) $ current
generators:%
\begin{eqnarray}
\left[ \mathcal{L}_{m},\mathcal{L}_{n}\right] &=&\left( m-n\right) \mathcal{L%
}_{m+n}+\frac{c}{12}\,m\left( m^{2}-1\right) \delta _{m+n,0}\,,  \notag \\
\left[ \mathcal{\bar{L}}_{m},\mathcal{\bar{L}}_{n}\right] &=&\left(
m-n\right) \mathcal{\bar{L}}_{m+n}+\frac{\bar{c}}{12}\,m\left(
m^{2}-1\right) \delta _{m+n,0}\,,  \notag \\
\left[ \mathcal{\hat{L}}_{m},\mathcal{\hat{L}}_{n}\right] &=&\left(
m-n\right) \mathcal{\hat{L}}_{m+n}+\frac{\hat{c}}{12}\,m\left(
m^{2}-1\right) \delta _{m+n,0}\,,  \notag \\
\left[ \mathcal{L}_{m},\mathfrak{k}_{n}^{a}\right] &=&-n\mathfrak{k}%
_{m+n}^{a}\,,\quad \left[ \mathcal{\bar{L}}_{m},\mathfrak{\bar{k}}_{n}^{a}%
\right] =-n\mathfrak{\bar{k}}_{m+n}^{a}\,,  \notag \\
\left[ \mathcal{\hat{L}}_{m},\mathfrak{\hat{k}}_{n}^{a}\right] &=&-n%
\mathfrak{\hat{k}}_{m+n}^{a}\,, \\
\left[ \mathfrak{k}_{m}^{a},\mathfrak{k}_{n}^{b}\right] &=&i\epsilon ^{abc}%
\mathfrak{k}_{m+n}^{c}+\frac{c}{12}\,m\delta ^{ab}\delta _{m+n,0}\,,  \notag
\\
\left[ \mathfrak{\bar{k}}_{m}^{a},\mathfrak{\bar{k}}_{n}^{b}\right]
&=&i\epsilon ^{abc}\mathfrak{\bar{k}}_{m+n}^{c}+\frac{\bar{c}}{12}\,m\delta
^{ab}\delta _{m+n,0}\,,  \notag \\
\left[ \mathfrak{\hat{k}}_{m}^{a},\mathfrak{\hat{k}}_{n}^{b}\right]
&=&i\epsilon ^{abc}\mathfrak{\hat{k}}_{m+n}^{c}+\frac{\hat{c}}{12}\,m\delta
^{ab}\delta _{m+n,0}\,,  \notag
\end{eqnarray}%
\begin{eqnarray}
\left[ \mathcal{L}_{m},\mathcal{Q}_{r}^{i,\pm }\right] &=&\left( \frac{m}{2}%
-r\right) \mathcal{Q}_{m+r}^{i,\pm }\,,\qquad \left[ \mathcal{\bar{L}}_{m},%
\mathcal{\bar{Q}}_{r}^{i,\pm }\right] =\left( \frac{m}{2}-r\right) \mathcal{%
\bar{Q}}_{m+r}^{i,\pm }\,,  \notag \\
\left[ \mathfrak{k}_{m}^{a},\mathcal{Q}_{r}^{i,+}\right] &=&\,-\frac{1}{2}%
\left( \sigma ^{a}\right) _{j}^{i}\mathcal{Q}_{m+r}^{j,+}\,,\qquad \left[
\mathfrak{k}_{m}^{a},\mathcal{Q}_{r}^{i,-}\right] =\,\frac{1}{2}\left( \bar{%
\sigma}^{a}\right) _{j}^{i}\mathcal{Q}_{m+r}^{j,-}\,,  \notag \\
\left[ \mathfrak{\bar{k}}_{m}^{a},\mathcal{\bar{Q}}_{r}^{i,+}\right] &=&\,-%
\frac{1}{2}\left( \sigma ^{a}\right) _{j}^{i}\mathcal{\bar{Q}}%
_{m+r}^{j,+}\,,\qquad \left[ \mathfrak{\bar{k}}_{m}^{a},\mathcal{\bar{Q}}%
_{r}^{i,-}\right] =\,\frac{1}{2}\left( \bar{\sigma}^{a}\right) _{j}^{i}%
\mathcal{\bar{Q}}_{m+r}^{j,-}\,, \\
\left\{ \mathcal{Q}_{r}^{i,+},\mathcal{Q}_{s}^{j,-}\right\} &=&\delta ^{ij}\,%
\left[ \mathcal{L}_{r+s}+\frac{c}{6}\,\left( r^{2}-\frac{1}{4}\right) \delta
_{r+s,0}\,\right] -\left( r-s\right) \left( \sigma ^{a}\right) _{ij}%
\mathfrak{k}_{r+s}^{a}\,,  \notag \\
\left\{ \mathcal{\bar{Q}}_{r}^{i,+},\mathcal{\bar{Q}}_{s}^{j,-}\right\}
&=&\delta ^{ij}\,\left[ \mathcal{\bar{L}}_{r+s}+\frac{\bar{c}}{6}\,\left(
r^{2}-\frac{1}{4}\right) \delta _{r+s,0}\,\right] -\left( r-s\right) \left(
\sigma ^{a}\right) _{ij}\mathfrak{\bar{k}}_{r+s}^{a}\,.  \notag
\end{eqnarray}%
Such structure is revealed by considering the following redefinitions%
\begin{equation}
\begin{tabular}{lll}
$\mathcal{L}_{m}=\dfrac{1}{2}\left( \ell ^{2}\mathcal{Z}_{m}+\ell \mathcal{P}%
_{m}\right) \,,$ & $\mathcal{\bar{L}}_{m}=\dfrac{1}{2}\left( \ell ^{2}%
\mathcal{Z}_{-m}-\ell \mathcal{P}_{-m}\right) \,,$ & $\hat{\mathcal{L}}_{m}=%
\mathcal{J}_{-m}-\ell ^{2}\mathcal{Z}_{-m}\,,$ \\
$\mathfrak{k}_{m}^{a}=\dfrac{1}{2}\left( \ell \mathtt{B}_{m}^{a}+\ell ^{2}%
\mathtt{Z}_{m}^{a}\right) \,,$ & $\mathfrak{\bar{k}}_{m}^{a}=\dfrac{1}{2}%
\left( \ell \mathtt{B}_{m}^{a}-\ell ^{2}\mathtt{Z}_{m}^{a}\right) \,,$ & $%
\mathfrak{\hat{k}}_{m}^{a}=\dfrac{1}{2}\left( \mathtt{T}_{m}^{a}-\ell ^{2}%
\mathtt{Z}_{m}^{a}\right) \,,$ \\
$Q_{r}^{i,\pm }=\dfrac{1}{2}\left( \ell ^{1/2}\mathcal{G}_{r}^{i,\pm }+\ell
^{3/2}\mathcal{H}_{r}^{i,\pm }\right) \,,$ & $\bar{Q}_{r}^{i,\pm }=\dfrac{i}{%
2}\left( \ell ^{1/2}\mathcal{G}_{r}^{i,\pm }-\ell \mathcal{H}_{r}^{i,\pm
}\right) \,,$ &  \\
$c=\dfrac{1}{2}\left( \ell ^{2}c_{3}+\ell c_{2}\right) \,,$ & $\bar{c}=%
\dfrac{1}{2}\left( \ell ^{2}c_{3}-\ell c_{2}\right) \,,$ & $\hat{c}=\left(
c_{1}-\ell ^{2}c_{3}\right) \,.$%
\end{tabular}%
\end{equation}%
Thus the $\mathcal{N}=4$ enlarged super-$BMS_{3}$ algebra given by (\ref%
{EBMS3}), (\ref{ebms3a}), (\ref{ebms3b}) and (\ref{ebms3c}) can be seen as
the direct sum of the $\left( 4,4\right) $ superconformal algebra and the
Virasoro algebra endowed with a $\mathfrak{su}\left( 2\right) $ current
algebra. In particular, the $\mathfrak{su}\left( 2\right) $ current
generators $\mathfrak{k}_{m}^{a}$ and $\mathfrak{\bar{k}}_{m}^{a}$ present
in the $\left( 4,4\right) $ superconformal algebra correspond to $\mathfrak{%
su}\left( 2\right) $ R-symmetry generators.

\subsubsection{Non-standard enlarged super-$BMS_{3}$ algebra}

An alternative supersymmetric extension of the so-called enlarged $BMS_{3}$
algebra (\ref{EBMS3}) can be obtained considering a different semigroup and
a different starting infinite-dimensional algebra.

Let us consider the superconformal algebra:%
\begin{eqnarray}
\left[ \mathcal{J}_{m},\mathcal{J}_{n}\right] &=&\left( m-n\right) \mathcal{J%
}_{m+n}+\frac{c_{1}}{12}\,m\left( m^{2}-1\right) \delta _{m+n,0}\,,  \notag
\\
\left[ \mathcal{J}_{m},\mathcal{P}_{n}\right] &=&\left( m-n\right) \mathcal{P%
}_{m+n}+\frac{c_{2}}{12}\,m\left( m^{2}-1\right) \delta _{m+n,0}\,,  \notag
\\
\left[ \mathcal{P}_{m},\mathcal{P}_{n}\right] &=&\left( m-n\right) \mathcal{J%
}_{m+n}+\frac{c_{1}}{12}\,m\left( m^{2}-1\right) \delta _{m+n,0}\,,  \notag
\\
\left[ \mathcal{J}_{m},\mathcal{G}_{r}\right] &=&\left( \frac{m}{2}-r\right)
\mathcal{G}_{m+r}\,,  \label{sconf} \\
\left[ \mathcal{P}_{m},\mathcal{G}_{r}\right] &=&\left( \frac{m}{2}-r\right)
\mathcal{G}_{m+r}\,,  \notag \\
\left\{ \mathcal{G}_{r},\mathcal{G}_{s}\right\} &=&\mathcal{J}_{r+s}+%
\mathcal{P}_{r+s}+\frac{\left( c_{1}+c_{2}\right) }{6}\,\left( r^{2}-\frac{1%
}{4}\right) \delta _{r+s,0}\,.  \notag
\end{eqnarray}%
which, as was discussed in section 3.2, can be written as two copies of the
Virasoro algebra, one of which is augmented by supersymmetry.

Let \thinspace $\mathfrak{g}=V_{0}\oplus V_{1}\oplus V_{2}$ be a subspace
decomposition where $V_{0}$ corresponds to the Virasoro subalgebra which is
generated by $\left\{ \mathcal{J}_{m},c_{1}\right\} $, $V_{1}$ is the
fermionic subspace spanned by $\mathcal{G}_{r}$ and $V_{2}$ is the bosonic
subspace generated by the set $\left\{ \mathcal{P}_{m},c_{2}\right\} $. One
can notice that the subspaces satisfy%
\begin{eqnarray}
\left[ V_{0},V_{0}\right] &\subset &V_{0}\,,\quad \left[ V_{0},V_{1}\right]
\subset V_{1}\,,\quad \left[ V_{0},V_{2}\right] \subset V_{2}\,,  \notag \\
\left[ V_{1},V_{2}\right] &\subset &V_{1}\,,\quad \left[ V_{2},V_{2}\right]
\subset V_{0}\,,\quad \left[ V_{1},V_{1}\right] \subset V_{0}\oplus V_{2}\,.
\end{eqnarray}%
Let $S_{\mathcal{M}}^{\left( 2\right) }=\left\{ \lambda _{0},\lambda
_{1},\lambda _{2}\right\} $ be the relevant semigroup whose elements satisfy%
\begin{equation}
\begin{tabular}{l|lll}
$\lambda _{2}$ & $\lambda _{2}$ & $\lambda _{1}$ & $\lambda _{2}$ \\
$\lambda _{1}$ & $\lambda _{1}$ & $\lambda _{2}$ & $\lambda _{1}$ \\
$\lambda _{0}$ & $\lambda _{0}$ & $\lambda _{1}$ & $\lambda _{2}$ \\ \hline
& $\lambda _{0}$ & $\lambda _{1}$ & $\lambda _{2}$%
\end{tabular}
\label{mlf}
\end{equation}%
and let us consider the following subset decomposition $S_{\mathcal{M}%
}^{\left( 2\right) }=S_{0}\cup S_{1}\cup S_{2}$ with%
\begin{eqnarray}
S_{0} &=&\left\{ \lambda _{0},\lambda _{2}\right\} \,,  \notag \\
S_{1} &=&\left\{ \lambda _{1}\right\} \,, \\
S_{2} &=&\left\{ \lambda _{2}\right\} \,,  \notag
\end{eqnarray}%
which is resonant since they satisfy the same structure than the subspaces $%
V_{0}$, $V_{1}$ and $V_{2}$. Then, the resonant subalgebra is given by%
\begin{equation}
W_{R}=W_{0}\oplus W_{1}\oplus W_{2}=S_{0}\times V_{0}\,\oplus S_{1}\times
V_{1}\oplus S_{2}\times V_{2}\,.
\end{equation}%
A new supersymmetric extension of the enlarged $BMS_{3}$ algebra is obtained
after performing a resonant $S_{\mathcal{M}}^{\left( 2\right) }$-expansion
whose generators are related to the superconformal ones through%
\begin{equation}
\begin{tabular}{ll}
$\mathcal{\tilde{J}}_{m}=\lambda _{0}\mathcal{J}_{m}\,,$ & $\tilde{c}%
_{1}=\lambda _{0}c_{1}\,,$ \\
$\mathcal{\tilde{Z}}_{m}=\lambda _{2}\mathcal{J}_{m}\,,$ & $\tilde{c}%
_{3}=\lambda _{2}c_{1}\,,$ \\
$\mathcal{\tilde{P}}_{m}=\lambda _{2}\mathcal{P}_{m}\,,$ & $\tilde{c}%
_{2}=\lambda _{2}c_{2}\,,$ \\
$\mathcal{\tilde{G}}_{r}=\lambda _{1}\mathcal{G}_{r}\,.$ &
\end{tabular}%
\end{equation}%
Then, using the (anti-)commutation relations of the superconformal algebra (%
\ref{sconf}) and the multiplication law of the semigroup (\ref{mlf}), one
find the non-standard enlarged super-$BMS_{3}$ algebra:%
\begin{eqnarray}
\left[ \mathcal{\tilde{J}}_{m},\mathcal{\tilde{J}}_{n}\right] &=&\left(
m-n\right) \mathcal{\tilde{J}}_{m+n}+\frac{\tilde{c}_{1}}{12}\,m\left(
m^{2}-1\right) \delta _{m+n,0}\,,  \notag \\
\left[ \mathcal{\tilde{J}}_{m},\mathcal{\tilde{P}}_{n}\right] &=&\left(
m-n\right) \mathcal{\tilde{P}}_{m+n}+\frac{\tilde{c}_{2}}{12}\,m\left(
m^{2}-1\right) \delta _{m+n,0}\,,  \notag \\
\left[ \mathcal{\tilde{P}}_{m},\mathcal{\tilde{P}}_{n}\right] &=&\left(
m-n\right) \mathcal{\tilde{Z}}_{m+n}+\frac{\tilde{c}_{3}}{12}\,m\left(
m^{2}-1\right) \delta _{m+n,0}\,,  \notag \\
\left[ \mathcal{\tilde{J}}_{m},\mathcal{\tilde{Z}}_{n}\right] &=&\left(
m-n\right) \mathcal{\tilde{Z}}_{m+n}+\frac{\tilde{c}_{3}}{12}\,m\left(
m^{2}-1\right) \delta _{m+n,0}\,,  \notag \\
\left[ \mathcal{\tilde{P}}_{m},\mathcal{\tilde{Z}}_{n}\right] &=&\left(
m-n\right) \mathcal{\tilde{P}}_{m+n}+\frac{\tilde{c}_{2}}{12}\,m\left(
m^{2}-1\right) \delta _{m+n,0}\,,  \label{NSEBMS3} \\
\left[ \mathcal{\tilde{Z}}_{m},\mathcal{\tilde{Z}}_{n}\right] &=&\left(
m-n\right) \mathcal{\tilde{Z}}_{m+n}+\frac{\tilde{c}_{3}}{12}\,m\left(
m^{2}-1\right) \delta _{m+n,0}\,,  \notag \\
\left[ \mathcal{\tilde{J}}_{m},\mathcal{\tilde{G}}_{r}\right] &=&\left(
\frac{m}{2}-r\right) \mathcal{\tilde{G}}_{m+r}\,,\quad \left[ \mathcal{%
\tilde{P}}_{m},\mathcal{\tilde{G}}_{r}\right] =\left( \frac{m}{2}-r\right)
\mathcal{\tilde{G}}_{m+r}\,,  \notag \\
\left[ \mathcal{\tilde{Z}}_{m},\mathcal{\tilde{G}}_{r}\right] &=&\left(
\frac{m}{2}-r\right) \mathcal{\tilde{G}}_{m+r}\,,  \notag \\
\left\{ \mathcal{\tilde{G}}_{r},\mathcal{\tilde{G}}_{s}\right\} &=&\mathcal{%
\tilde{Z}}_{r+s}+\mathcal{\tilde{P}}_{r+s}+\frac{\left( \tilde{c}_{3}+\tilde{%
c}_{2}\right) }{6}\,\left( r^{2}-\frac{1}{4}\right) \delta _{r+s,0}\,.
\notag
\end{eqnarray}%
The new superalgebra obtained corresponds to the infinite-dimensional lift
of a particular AdS-Lorentz superalgebra introduced in \cite{FISV}. One can
see that the finite subalgebra spanned by $\mathcal{\tilde{J}}_{0}$, $%
\mathcal{\tilde{J}}_{1}$, $\mathcal{\tilde{J}}_{-1}$, $\mathcal{\tilde{P}}%
_{0}$, $\mathcal{\tilde{P}}_{1}$, $\mathcal{\tilde{P}}_{-1}$, $\mathcal{%
\tilde{Z}}_{0}$, $\mathcal{\tilde{Z}}_{1}$, $\mathcal{\tilde{Z}}_{-1}$, $%
\mathcal{\tilde{G}}_{\frac{1}{2}}$ and$\ \mathcal{\tilde{G}}_{-\frac{1}{2}}$
are related to the super AdS-Lorentz ones through%
\begin{equation}
\begin{tabular}{lllll}
$\mathcal{\tilde{J}}_{-1}=-\sqrt{2}\tilde{J}_{0}$ & , & $\mathcal{\tilde{J}}%
_{1}=\sqrt{2}\tilde{J}_{1}$ & , & $\mathcal{\tilde{J}}_{0}=\tilde{J}_{2}\,,$
\\
$\mathcal{\tilde{P}}_{-1}=-\sqrt{2}\tilde{P}_{0}$ & , & $\mathcal{\tilde{P}}%
_{1}=\sqrt{2}\tilde{P}_{1}$ & , & $\mathcal{\tilde{P}}_{0}=\tilde{P}_{2}\,,$
\\
$\mathcal{\tilde{Z}}_{-1}=-\sqrt{2}\tilde{Z}_{0}$ & , & $\mathcal{\tilde{Z}}%
_{1}=\sqrt{2}\tilde{Z}_{1}$ & , & $\mathcal{\tilde{Z}}_{0}=\tilde{Z}_{2}\,,$
\\
$\mathcal{\tilde{G}}_{-\frac{1}{2}}=\sqrt{2}\tilde{Q}_{+}$ & , & $\mathcal{%
\tilde{G}}_{\frac{1}{2}}=\sqrt{2}\tilde{Q}_{-}$ & . &
\end{tabular}%
\end{equation}%
As the minimal enlarged super-$BMS_{3}$ algebra (\ref{EBMS3a})-(\ref{EBMS3b}%
), the non-standard one obtained here can be rewritten as three copies of
the Virasoro algebra but only one augmented by supersymmetry,%
\begin{equation*}
\begin{array}{lcl}
\left[ \mathcal{L}_{m}^{+},\mathcal{L}_{n}^{+}\right] & = & \left(
m-n\right) \mathcal{L}_{m+n}^{+}+\dfrac{c^{+}}{12}\,m\left( m^{2}-1\right)
\delta _{m+n,0}\,, \\[5pt]
\left[ \mathcal{L}_{m}^{-},\mathcal{L}_{n}^{-}\right] & = & \left(
m-n\right) \mathcal{L}_{m+n}^{-}+\dfrac{c^{-}}{12}\,m\left( m^{2}-1\right)
\delta _{m+n,0}\,, \\[5pt]
\left[ \hat{\mathcal{L}}_{m},\hat{\mathcal{L}}_{n}\right] & = & \left(
m-n\right) \mathcal{\hat{L}}_{m+n}+\dfrac{\hat{c}}{12}\,m\left(
m^{2}-1\right) \delta _{m+n,0}\,, \\
\left[ \mathcal{L}_{m}^{+},\mathcal{Q}_{r}\right] & = & \left( \dfrac{m}{2}%
-r\right) \mathcal{Q}_{m+r}\,\,, \\
\left\{ \mathcal{Q}_{r},\mathcal{Q}_{s}\right\} & = & \mathcal{L}_{r+s}^{+}+%
\dfrac{c^{+}}{6}\,\left( r^{2}-\frac{1}{4}\right) \delta _{r+s,0}\,,%
\end{array}%
\end{equation*}%
This corresponds to the direct sum of the $\left( 1,0\right) $
superconformal algebra and the Virasoro algebra. Such structure appears
after the following redefinitions:%
\begin{equation*}
\begin{tabular}{ll}
$\mathcal{L}_{m}^{+}=\frac{1}{2}\left( \mathcal{\tilde{Z}}_{m}+\mathcal{%
\tilde{P}}_{m}\right) \,,$ & $\mathcal{L}_{m}^{-}=\frac{1}{2}\left( \mathcal{%
\tilde{Z}}_{-m}-\mathcal{\tilde{P}}_{-m}\right) \,,$ \\
$\hat{\mathcal{L}}_{m}=\mathcal{\tilde{J}}_{-m}-\mathcal{\tilde{Z}}_{-m}\,,$
& $Q_{r}=\frac{1}{\sqrt{2}}\mathcal{\tilde{G}}_{r}\,,$ \\
$c^{\pm }=\dfrac{1}{2}\left( \tilde{c}_{3}\pm \tilde{c}_{2}\right) \,,$ & $%
\hat{c}=\left( \tilde{c}_{1}-\tilde{c}_{3}\right) \,.$%
\end{tabular}%
\end{equation*}%
The main difference with the enlarged super-$BMS_{3}$ algebra (\ref{EBMS3a}%
)-(\ref{EBMS3b}) introduced previously is the absence of a second spinor
charge. Then, it seems that the connection with the deformed super-$%
\widetilde{BMS}_{3}$ algebra, which possesses two spinor charges $\mathcal{G}%
_{r}$ and $\mathcal{H}_{r}$, cannot be done from this non-standard enlarged
super-$BMS_{3}$ algebra. Nevertheless, a non-standard deformed super-$%
\widetilde{BMS_{3}}$ algebra can be obtained considering an Inönü-Wigner
contraction to (\ref{NSEBMS3}). Indeed, the rescaling of the generators of (%
\ref{NSEBMS3})%
\begin{eqnarray*}
\mathcal{\tilde{J}}_{m} &\rightarrow &\mathcal{\tilde{J}}_{m}\,,\quad
\mathcal{\tilde{P}}_{m}\rightarrow \sigma \mathcal{\tilde{P}}_{m}\,,\quad
\mathcal{\tilde{Z}}_{m}\rightarrow \sigma ^{2}\mathcal{\tilde{Z}}%
_{m}\,,\quad \mathcal{\tilde{G}}_{r}\rightarrow \sigma \mathcal{\tilde{G}}%
_{r}\,, \\
\tilde{c}_{1} &\rightarrow &\tilde{c}_{1}\,,\quad \tilde{c}_{2}\rightarrow
\sigma \tilde{c}_{2}\,,\quad \tilde{c}_{3}\rightarrow \sigma ^{2}\tilde{c}%
_{3}\,,
\end{eqnarray*}%
leads to a non-standard deformed super-$\widetilde{BMS}_{3}$ algebra in the
limit $\sigma \rightarrow \infty $:%
\begin{eqnarray}
\left[ \mathcal{\tilde{J}}_{m},\mathcal{\tilde{J}}_{n}\right] &=&\left(
m-n\right) \mathcal{\tilde{J}}_{m+n}+\frac{\tilde{c}_{1}}{12}\,m\left(
m^{2}-1\right) \delta _{m+n,0}\,,  \notag \\
\left[ \mathcal{\tilde{J}}_{m},\mathcal{\tilde{P}}_{n}\right] &=&\left(
m-n\right) \mathcal{\tilde{P}}_{m+n}+\frac{\tilde{c}_{2}}{12}\,m\left(
m^{2}-1\right) \delta _{m+n,0}\,,  \notag \\
\left[ \mathcal{\tilde{P}}_{m},\mathcal{\tilde{P}}_{n}\right] &=&\left(
m-n\right) \mathcal{\tilde{Z}}_{m+n}+\frac{\tilde{c}_{3}}{12}\,m\left(
m^{2}-1\right) \delta _{m+n,0}\,,  \notag \\
\left[ \mathcal{\tilde{J}}_{m},\mathcal{\tilde{Z}}_{n}\right] &=&\left(
m-n\right) \mathcal{\tilde{Z}}_{m+n}+\frac{\tilde{c}_{3}}{12}\,m\left(
m^{2}-1\right) \delta _{m+n,0}\,, \\
\left[ \mathcal{\tilde{J}}_{m},\mathcal{\tilde{G}}_{r}\right] &=&\left(
\frac{m}{2}-r\right) \mathcal{\tilde{G}}_{m+r}\,,  \notag \\
\left\{ \mathcal{\tilde{G}}_{r},\mathcal{\tilde{G}}_{s}\right\} &=&\mathcal{%
\tilde{Z}}_{r+s}+\frac{\tilde{c}_{3}}{6}\,\left( r^{2}-\frac{1}{4}\right)
\delta _{r+s,0}\,.  \notag
\end{eqnarray}%
The name "non-standard" is due to its finite subalgebra which corresponds to
the so-called non-standard Maxwell superalgebra \cite{Lukierski, Sorokas2}.
Such supersymmetric extension of the Maxwell algebra has the particularity
that the four-momentum generators $\tilde{P}_{a}$ are no more expressed as
bilinear expressions of the fermionic ones leading to an exotic
three-dimensional supersymmetric action. As was discussed in \cite{CPR}, a
well-defined supergravity action invariant under a supersymmetric extension
of the Maxwell algebra requires the introduction of a second spinorial
charge.

\section{Conclusions}

The $S$-expansion procedure has been useful to obtain new (super)symmetries
and novel (super)gravity theories. Here, based on the recent applications of
the $S$-expansion method in the asymptotic symmetries context \cite{CCRS,
CCFR}, we have obtained known and new supersymmetric extensions of
asymptotic symmetries of diverse three-dimensional gravity theories. By
expanding the super-Virasoro algebra we have found the minimal
supersymmetric extensions of the asymptotic algebras of the Maxwell gravity
\cite{CMMRSV} and $\mathfrak{so}\left( 2,2\right) \oplus \mathfrak{so}\left(
2,1\right) $ gravity \cite{CMRSV} theories defined in three spacetime
dimensions. The new infinite-dimensional superalgebras obtained can be seen
as enlargement, extension and deformation of the super-$BMS_{3}$ algebra and
corresponds to infinite-dimensional lift of the AdS-Lorentz and Maxwell
superalgebras.

The $\mathcal{N}$-extensions of our results have also been considered with $%
\mathcal{N}=2$ and $\mathcal{N}=4$. We have found that the new $\mathcal{N}$%
-extended infinite-dimensional superalgebras involve new features. Indeed,
in the $\mathcal{N}=2$ case we have shown that $\mathfrak{\hat{u}}\left(
1\right) $ R-symmetry generators are required. On the other hand, in the $%
\mathcal{N}=4$ case, we have $\mathfrak{su}\left( 2\right) $ R-symmetry
generators. Interestingly, we have shown that the $\mathcal{N}$-extended
enlarged super-$BMS_{3}$ and the $\mathcal{N}$-extended deformed super-$%
\widetilde{BMS}_{3}$ algebras are related through a flat limit $\ell
\rightarrow \infty $. \ Such limit is not a particularity of the
infinite-dimensional superalgebras obtained but is already present at the
finite level \cite{CPR, Concha}.

On the other hand, it is important to clarify that the election of the
semigroups to obtain the diverse infinite-dimensional superalgebras is not
arbitrary. Indeed, following the results obtained in the bosonic case in
\cite{CCRS} and subsequently at the supersymmetric level \cite{CCFR}, the
semigroups chosen are those used at the finite-dimensional level.

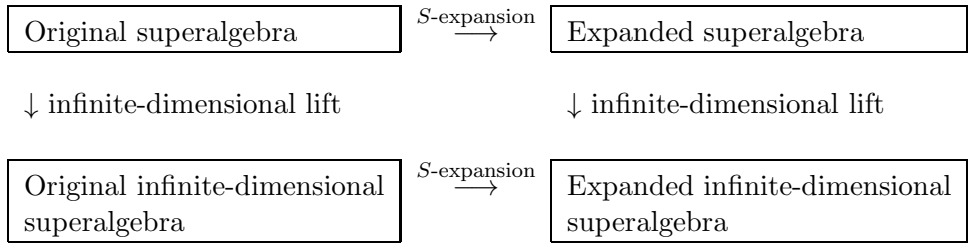
\begin{equation*}
\begin{tabular}{lll}
\cline{1-1}\cline{3-3}
\multicolumn{1}{|l}{Original superalgebra} & \multicolumn{1}{|l}{$\overset{S%
\text{-expansion}}{\longrightarrow }$} & \multicolumn{1}{|l|}{Expanded
superalgebra} \\ \cline{1-1}\cline{3-3}
&  &  \\
$\downarrow $ infinite-dimensional lift &  & $\downarrow $
infinite-dimensional lift \\
&  &  \\ \cline{1-1}\cline{3-3}
\multicolumn{1}{|l}{Original infinite-dimensional} & \multicolumn{1}{|l}{$%
\overset{S\text{-expansion}}{\longrightarrow }$} & \multicolumn{1}{|l|}{
Expanded infinite-dimensional} \\
\multicolumn{1}{|l}{superalgebra} & \multicolumn{1}{|l}{} &
\multicolumn{1}{|l|}{superalgebra} \\ \cline{1-1}\cline{3-3}
\end{tabular}%
\end{equation*}

The new results obtained here could have important consequences in further
studies of the Maxwell and AdS-Lorentz supergravity theories. In particular,
we conjecture that the new infinite-dimensional structures introduced here
are the respective asymptotic supersymmetries of the three-dimensional
Maxwell and AdS-Lorentz supergravities presented in \cite{CPR, Concha}. If
our conjecture is true, it would mean that the $S$-expansion method not only
allows us to construct new and consistent (super)algebras and (super)gravity
theories but also provides us with their asymptotic symmetry. At the bosonic
level, we have recently shown that the infinite-dimensional lifts of the
Maxwell and AdS-Lorentz algebra obtained as $S$-expansion \cite{CCRS} result
to be the respective asymptotic symmetries of the CS\ gravity theory based
on the Maxwell \cite{CMMRSV} and AdS-Lorentz gravity theories \cite{CMRSV},
respectively. More recently in \cite{CCFR}, we have shown that the semigroup
allowing to obtain $\mathcal{N}$-extended Poincaré superalgebras can also be
used to obtain $\mathcal{N}$-extended super-$BMS_{3}$ algebras. It is then
natural to expect that the novel infinite-dimensional superalgebras obtained
here are the respective asymptotic supersymmetries of the Maxwell and AdS-Lorentz supergravity theories. It would be interesting to explore the
explicit obtention of our infinite-dimensional superalgebras by imposing
suitable boundary conditions. A direct asymptotic symmetry analysis for the
Maxwell and $\mathfrak{so}\left( 2,2\right) \oplus \mathfrak{so}\left(
2,1\right) $ supergravity theory would be approached in a future work.

Another aspect that it would be worth exploring is the connection of our new
infinite-dimensional superalgebras with the two-dimensional super Galilean
conformal algebra (GCA) \cite{BM, M, MR}. In particular, one could expect to
obtain a deformed super-GCA algebra by contracting the relativistic $\left(
1,1\right) $ Superconformal $\oplus $ Virasoro algebra obtained here. It
would be interesting to evaluate if the deformed super-$\widetilde{BMS}_{3}$
algebra introduced here is isomorphic to a deformed super-GCA [work in
progress].

As an ending remark: It would be worth it to generalize the recent results
obtained with the algebraic expansion method at the non-relativistic level
\cite{BIOR, AGI, CR4, PSR, Romano}. It would be interesting to study the
non-relativistic version of the Maxwell and AdS-Lorentz superalgebra using
the $S$-expansion method.

\section{Acknowledgment}

This work was supported by the CONICYT - PAI grant N$^{\circ }$77190078 (P.C.) and FONDECYT Projects N$^{\circ }$11191175 (O.F.) and N$^{\circ }$3170438 (E.R.) . This work was supported by the Research
project Code DIN 11/2012 (R.C.) and DINREG 19/2018 (O.F.) of the Universidad
Católica de la Santisima Concepción, Chile. R.C., P.C. and O.F. would like to
thank to the Dirección de Investigación and Vice-rectoría de Investigación
of the Universidad Católica de la Santísima Concepción, Chile, for their
constant support.

{\small \appendix}

\section{Appendix\label{App}}

In three spacetime dimensions, the Maxwell algebra has the following
non-vanishing commutation relations:%
\begin{eqnarray}
\left[ J_{a},J_{b}\right] &=&\epsilon _{abc}J^{c}\,,\qquad \left[ J_{a},P_{b}%
\right] =\epsilon _{abc}P^{c}\,,  \notag \\
\left[ J_{a},Z_{b}\right] &=&\epsilon _{abc}Z^{c}\,,\qquad \left[ P_{a},P_{b}%
\right] =\epsilon _{abc}Z^{c}\,,  \label{Maxwell}
\end{eqnarray}%
where $J_{a}$ is the Lorentz generator, $P_{a}$ is the translation generator
and $Z_{a}$ is the so-called gravitational Maxwell generator. Such symmetry
has been introduced in \cite{BCR, Schrader, GK} in order to describe a
particle moving in a four-dimensional background in presence of a constant
electromagnetic field.

As shown in \cite{SSV, HR, AFGHZ, CMMRSV}, a three-dimensional CS\ action
invariant under the Maxwell symmetry reads%
\begin{eqnarray}
I &=&\frac{k}{4\pi }\int \left[ \alpha _{0}\left( \,\omega ^{a}d\omega _{a}+%
\frac{1}{3}\,\epsilon _{abc}\omega ^{a}\omega ^{b}\omega ^{c}\right)
+2\alpha _{1}e^{a}R_{a}\,\right.  \notag  \label{sMCS} \\
&&\left. +\alpha _{2}\left( 2R^{a}\sigma _{a}+e^{a}T_{a}\right) -d\left(
\alpha _{1}\omega ^{e}e_{a}+\alpha _{2}\omega ^{a}\sigma _{a}\right) \right]
\,,  \label{MCS}
\end{eqnarray}%
where $e^{a}$ is the vielbein, $\omega ^{a}$ corresponds to the spin
connection, $\sigma ^{a}$ is the gravitational Maxwell gauge field, $%
R^{a}=d\omega ^{a}+\frac{1}{2}\epsilon ^{abc}\omega _{b}\omega _{c}$ is the
Lorentz curvature and $T^{a}=D_{\omega }e^{a}~$is the torsion two-form. As
was discussed in \cite{CMMRSV}, the vacuum angular momentum and vacuum
energy of the stationary configuration are influenced by the gravitational
Maxwell field. More recent results have been presented in \cite{BS, CS} in
the dual version of the Maxwell algebra known as Hietarinta algebra \cite%
{Hietarinta}. The Hietarinta symmetry appears by interchanging the role of
the Maxwell gravitational generator $Z_{a}$ with the momentum generator $%
P_{a}$.

Interestingly, the boundary dynamics results to be described by an extension
and deformation of the $BMS_{3}$ algebra with three independent central
charges \cite{CMMRSV}. Such analysis was based on the charge algebra of the
theory in the BMS gauge which includes the solutions of standard
asymptotically flat case.

As the Maxwell algebra, the asymptotic symmetry contains an additional
Abelian generator $\mathcal{Z}_{m}$ which modifies the $BMS_{3}$ symmetry as
follows%
\begin{equation}
\begin{array}{lcl}
\left[ \mathcal{J}_{m},\mathcal{J}_{n}\right] & = & \left( m-n\right)
\mathcal{J}_{m+n}+\dfrac{c_{1}}{12}\left( m^{3}-m\right) \delta _{m+n,0}\,,
\\[6pt]
\left[ \mathcal{J}_{m},\mathcal{P}_{n}\right] & = & \left( m-n\right)
\mathcal{P}_{m+n}+\dfrac{c_{2}}{12}\left( m^{3}-m\right) \delta _{m+n,0}\,,
\\[6pt]
\left[ \mathcal{P}_{m},\mathcal{P}_{n}\right] & = & \left( m-n\right)
\mathcal{Z}_{m+n}+\dfrac{c_{3}}{12}\left( m^{3}-m\right) \delta _{m+n,0}\,,
\\[6pt]
\left[ \mathcal{J}_{m},\mathcal{Z}_{n}\right] & = & \left( m-n\right)
\mathcal{Z}_{m+n}+\dfrac{c_{3}}{12}\left( m^{3}-m\right) \delta _{m+n,0}\,,
\\[6pt]
\left[ \mathcal{P}_{m},\mathcal{Z}_{n}\right] & = & 0\,, \\[6pt]
\left[ \mathcal{Z}_{m},\mathcal{Z}_{n}\right] & = & 0\,.%
\end{array}
\label{dbms3}
\end{equation}%
Here, the central charges $c_{1},c_{2}$ and $c_{3}$ are related to the three
terms of the CS action (\ref{MCS}) with%
\begin{equation}
c_{1}=12k\alpha _{0}\,,\qquad c_{2}=12k\alpha _{1}\,,\qquad c_{3}=12k\alpha
_{2}\,.
\end{equation}%
Let us note that the Maxwell algebra (\ref{Maxwell}) is a finite subalgebra
of the deformed $\widetilde{BMS}_{3}$ algebra (\ref{dbms3}) spanned by $%
\mathcal{J}_{0}$, $\mathcal{J}_{1}$, $\mathcal{J}_{-1}$, $\mathcal{P}_{0}$, $%
\mathcal{P}_{1}$, $\mathcal{P}_{-1}$ and $\mathcal{Z}_{0}$, $\mathcal{Z}_{1}$%
, $\mathcal{Z}_{-1}$ with%
\begin{equation}
\begin{tabular}{lllll}
$\mathcal{J}_{-1}=-\sqrt{2}J_{0}$ & , & $\mathcal{J}_{1}=\sqrt{2}J_{1}$ & ,
& $\mathcal{J}_{0}=J_{2}\,,$ \\
$\mathcal{P}_{-1}=-\sqrt{2}P_{0}$ & , & $\mathcal{P}_{1}=\sqrt{2}P_{1}$ & ,
& $\mathcal{P}_{0}=P_{2}\,,$ \\
$\mathcal{Z}_{-1}=-\sqrt{2}Z_{0}$ & , & $\mathcal{Z}_{1}=\sqrt{2}Z_{1}$ & ,
& $\mathcal{Z}_{0}=Z_{2}\,.$%
\end{tabular}%
\end{equation}

\section{Appendix\label{App2}}

A semi-simple enlargement of the Poincaré symmetry has been introduced in
\cite{Sorokas, GKL, DFIMRSV, SS} which can be seen as the direct sum of the
Lorentz and AdS algebra. In three spacetime dimensions, the so-called
AdS-Lorentz algebra can be written in the basis $\left\{
J_{a},P_{a},Z_{a}\right\} $ whose generators satisfy%
\begin{equation}
\begin{tabular}{lll}
$\left[ J_{a},J_{b}\right] =\epsilon _{abc}J^{c}\,,$ & \medskip & $\left[
P_{a},P_{b}\right] =\epsilon _{abc}Z^{c}\,$, \\
$\left[ J_{a},Z_{b}\right] =\epsilon _{abc}Z^{c}\,,$ & \medskip & $\left[
Z_{a},Z_{b}\right] =\dfrac{1}{\ell ^{2}}\epsilon _{abc}Z^{c}$\thinspace , \\
$\left[ J_{a},P_{b}\right] =\epsilon _{abc}P^{c}$\thinspace $,$ & \medskip &
$\left[ Z_{a},P_{b}\right] =\dfrac{1}{\ell ^{2}}\epsilon _{abc}P^{c}$%
\thinspace ,%
\end{tabular}
\label{AdSL}
\end{equation}%
where $Z_{a}$ are non-Abelian generators. Interestingly, in such basis, the
Maxwell algebra (\ref{Maxwell}) can also be obtained as a flat limit $\ell
\rightarrow \infty $ of the AdS-Lorentz algebra. Such limit can be
reproduced not only at the bosonic level but also at the supersymmetric \cite%
{CPR, CRR, Concha}, non-relativistic \cite{CR4, PSR}, higher-spin \cite%
{CCFRS} and asymptotic level \cite{CMRSV}.

The three-dimensional CS gravity action invariant under the AdS-Lorentz
algebra (\ref{AdSL}) reads \cite{DFIMRSV, HR, CMRSV, CR4, Durka2}%
\begin{align}
I_{R}=& \int \left[ \alpha _{0}\left( \omega ^{a}d\omega _{a}+\frac{1}{3}%
\,\epsilon ^{abc}\omega _{a}\omega _{b}\omega _{c}\right) +\alpha _{1}\left(
2e_{a}R^{a}+\frac{2}{\ell ^{2}}\,e_{a}F^{a}+\frac{1}{3\ell ^{2}}\,\epsilon
^{abc}e_{a}e_{b}e_{c}\right) \right.  \notag \\
& +\left. \alpha _{2}\left( T^{a}e_{a}+\frac{1}{\ell ^{2}}\,\epsilon
^{abc}e_{a}\sigma _{b}e_{c}+2\sigma _{a}R^{a}+\frac{2}{\ell ^{2}}\sigma
_{a}\,D_{\omega }\sigma ^{a}+\frac{1}{3\ell ^{4}}\epsilon ^{abc}\sigma
_{a}\sigma _{b}\sigma _{c}\right) \rule{0pt}{15pt}\right] \,\,,  \label{cs}
\end{align}%
where, $R^{a}=d\omega ^{a}+\frac{1}{2}\epsilon ^{abc}\omega _{b}\omega _{c}$
is the Lorentz curvature two-form, $T^{a}=D_{\omega }e^{a}$ is the torsion
two-form and $F^{a}=D_{\omega }\sigma ^{a}+\frac{1}{2\ell ^{2}}\epsilon
^{abc}\sigma _{b}\sigma _{c}$ is the curvature two-form related to $\sigma
^{a}$.

An explicit realisation of the asymptotic symmetry at null infinity was
presented in \cite{CMRSV} and turned out to be a semi-simple enlargement of
the $BMS_{3}$ algebra:%
\begin{equation}
\begin{array}{lcl}
\left[ \mathcal{J}_{m},\mathcal{J}_{n}\right] & = & \left( m-n\right)
\mathcal{J}_{m+n}+\dfrac{c_{1}}{12}\left( m^{3}-m\right) \delta _{m+n,0}\,,
\\[6pt]
\left[ \mathcal{J}_{m},\mathcal{P}_{n}\right] & = & \left( m-n\right)
\mathcal{P}_{m+n}+\dfrac{c_{2}}{12}\left( m^{3}-m\right) \delta _{m+n,0}\,,
\\[6pt]
\left[ \mathcal{P}_{m},\mathcal{P}_{n}\right] & = & \left( m-n\right)
\mathcal{Z}_{m+n}+\dfrac{c_{3}}{12}\left( m^{3}-m\right) \delta _{m+n,0}\,,
\\[6pt]
\left[ \mathcal{J}_{m},\mathcal{Z}_{n}\right] & = & \left( m-n\right)
\mathcal{Z}_{m+n}+\dfrac{c_{3}}{12}\left( m^{3}-m\right) \delta _{m+n,0}\,,
\\[6pt]
\left[ \mathcal{P}_{m},\mathcal{Z}_{n}\right] & = & \dfrac{1}{\ell ^{2}}%
\left( m-n\right) \,\mathcal{P}_{m+n}+\dfrac{c_{2}}{12\ell ^{2}}\left(
m^{3}-m\right) \delta _{m+n,0}\,, \\[6pt]
\left[ \mathcal{Z}_{m},\mathcal{Z}_{n}\right] & = & \dfrac{1}{\ell ^{2}}%
\left( m-n\right) \,\mathcal{Z}_{m+n}+\dfrac{c_{3}}{12\ell ^{2}}\left(
m^{3}-m\right) \delta _{m+n,0}\,,%
\end{array}
\label{EBMS3}
\end{equation}%
where the central charges $c_{1}$, $c_{2}$ and $c_{3}$ are related to the
coupling constant of the CS action (\ref{cs}) as follows%
\begin{equation}
c_{1}=12k\alpha _{0}\,,\qquad c_{2}=12k\alpha _{1}\,,\qquad c_{3}=12k\alpha
_{2}\,.
\end{equation}%
Such enlarged $BMS_{3}$ algebra results to be isomorphic to three copies of
the Virasoro algebra. Indeed, by considering the following change of basis%
\begin{equation}
\begin{tabular}{lll}
$\mathcal{L}_{m}^{+}=\dfrac{1}{2}\left( \ell ^{2}\mathcal{Z}_{m}+\ell
\mathcal{P}_{m}\right) \,,$ & $\mathcal{L}_{m}^{-}=\dfrac{1}{2}\left( \ell
^{2}\mathcal{Z}_{-m}-\ell \mathcal{P}_{-m}\right) \,,$ & $\hat{\mathcal{L}}%
_{m}=\mathcal{J}_{-m}-\ell ^{2}\mathcal{Z}_{-m}\,,$%
\end{tabular}%
\end{equation}%
one can rewrite the enlarged $BMS_{3}$ algebra as three copies of the
Virasoro algebra,%
\begin{equation}
\begin{array}{lcl}
i\left\{ \mathcal{L}_{m}^{+},\mathcal{L}_{n}^{+}\right\} & = & \left(
m-n\right) \mathcal{L}_{m+n}^{+}+\dfrac{c^{+}}{12}m^{3}\delta _{m+n,0}\,, \\%
[5pt]
i\left\{ \mathcal{L}_{m}^{-},\mathcal{L}_{n}^{-}\right\} & = & \left(
m-n\right) \mathcal{L}_{m+n}^{-}+\dfrac{c^{-}}{12}m^{3}\delta _{m+n,0}\,, \\%
[5pt]
i\left\{ \hat{\mathcal{L}}_{m},\hat{\mathcal{L}}_{n}\right\} & = & \left(
m-n\right) \hat{\mathcal{L}}_{m+n}+\dfrac{\hat{c}}{12}m^{3}\delta _{m+n,0}\,,%
\end{array}
\label{3virasoro}
\end{equation}%
with the following central charges%
\begin{equation}
c^{\pm }=\dfrac{1}{2}\left( \ell ^{2}c_{3}\pm \ell c_{2}\right) \;,\quad
\hat{c}=\left( c_{1}-\ell ^{2}c_{3}\right) \,.  \label{cpmandhatc}
\end{equation}%
It is interesting to note that the AdS-Lorentz algebra (\ref{AdSL}) is a
finite subalgebra of the enlarged $BMS_{3}$ algebra (\ref{EBMS3}). In fact,
the finite set of generators $\left\{ \mathcal{J}_{m},\mathcal{P}_{m},%
\mathcal{Z}_{m}\right\} $ with $m=0,\pm 1$ is related to the AdS-Lorentz
ones through the following redefinitions%
\begin{equation}
\begin{tabular}{lllll}
$\mathcal{J}_{-1}=-\sqrt{2}J_{0}$ & , & $\mathcal{J}_{1}=\sqrt{2}J_{1}$ & ,
& $\mathcal{J}_{0}=J_{2}\,,$ \\
$\mathcal{P}_{-1}=-\sqrt{2}P_{0}$ & , & $\mathcal{P}_{1}=\sqrt{2}P_{1}$ & ,
& $\mathcal{P}_{0}=P_{2}\,,$ \\
$\mathcal{Z}_{-1}=-\sqrt{2}Z_{0}$ & , & $\mathcal{Z}_{1}=\sqrt{2}Z_{1}$ & ,
& $\mathcal{Z}_{0}=J_{2}\,.$%
\end{tabular}%
\end{equation}


\begin{thebibliography}{999}
\bibitem{BH} J.D. Brown, M. Henneaux, \textit{Central charges in the
canonical realization of asymptotic symmetries: an example from
three-dimensional gravity}. Commun. Math. Phys. \textbf{104} (1986) 207.

\bibitem{ABS} A. Ashtekar, J. Bicak, B.G. Schmidt, \textit{Asymptotic
structure of symmetry reduced general relativity}, Phys. Rev. D\textbf{55}
(1997) 669. [gr-qc/9608042].

\bibitem{BC} G. Barnich, G. Compere, \textit{Classical central extension for
asymptotic symmetries at null infinity in three spacetime dimensions},
Class. Quant. Grav. \textbf{24} (2007) F15. [gr-qc/0610130].

\bibitem{BT1} G. Barnich, C. Troessaert, \textit{Aspects of the BMS/CFT
correspondence}, JHEP \textbf{1005} (2010) 062. arXiv:1001.1541 [hep-th].

\bibitem{BBM} H. Bondi, M.G.J. van der Burg, A.W.K. Metzner, \textit{%
Gravitational waves in general relativity. 7. Waves from axisymmetric
isolated systems}, Proc. Roy. Soc. Lond. A \textbf{269} (1962) 21.

\bibitem{Sachs} R.K. Sachs, \textit{Gravitational waves in general
relativity. 8. Waves in asymptotically flat space-times}, Proc. Roy. Soc.
Lond. A \textbf{270} (1962) 103.

\bibitem{GMPT} H.A. Gonzalez, J. Matulich, M. Pino, R. Troncoso, \textit{%
Asymptotically flat spacetimes in three-dimensional higher spin gravity},
JHEP \textbf{1309} (2013) 016. arXiv:1307.5651 [hep-th].

\bibitem{ABFGR} H. Afshar, A. Bagchi, R. Fareghbal, D. Grumiller, J.
Rosseel, \textit{Spin-3 Gravity in Three-Dimensional Flat Space}, Phys. Rev.
Lett. \textbf{111} (2013) no.12, 121603. arXiv:1307.4768 [hep-th].

\bibitem{GP} H.A. Gonzalez, M. Pino, \textit{Boundary dynamics of
asymptotically flat 3D gravity coupled to higher spin fields}, JHEP \textbf{%
05} (2014) 127. arXiv:1403.4898 [hep-th].

\bibitem{MPTT} J. Matulich, A. Perez, D. Tempo, R. Troncoso, \textit{Higher
spin extension of cosmological spacetimes in 3D: asymptotically flat
behavior with chemical potentials and thermodynamics}, JHEP \textbf{05}
(2015) 025. arXiv:1412.1464 [hep-th].

\bibitem{FMT1} O. Fuentealba, J. Matulich, R. Troncoso, \textit{%
Asymptotically flat structure of hypergravity in three spacetime dimensions}%
, JHEP \textbf{10} (2015) 009. arXiv:1508.04663 [hep-th].

\bibitem{BJMN} N. Banerjee, D.P. Jatkar, S. Mukhi, T. Neogi, \textit{%
Free-field realisations of the BMS}$_{3}\mathit{\ }$\textit{algebra and its
extensions}, JHEP \textbf{06} (2016) 024. arXiv:1512.06240 [hep-th].

\bibitem{DR} S. Detournay, M. Riegler, \textit{Enhanced Asymptotic Symmetry
Algebra of 2+1 Dimensional Flat Space}, Phys. Rev. D\textbf{95} (2017)
046008. arXiv:1612.00278 [hep-th].

\bibitem{SA} M.R. Setare, H. Adami, \textit{Enhanced asymptotic BMS}$_{3}$
\textit{algebra of the flat spacetime solutions of generalized minimal
massive gravity}, Nucl. Phys. B \textbf{926} (2018) 70. arXiv:1703.00936
[hep-th].

\bibitem{PSSJ} A. Farmhand Parsa, H.R. Safari, M.M. Sheikh-Jabbari, \textit{%
On Rigidity of 3d Asymptotic Symmetry Algebras}, arXiv:1809.08209 [hep-th].

\bibitem{SSJ} H.R. Safari, M.M. Sheikh-Jabbari, \textit{BMS}$_{4}$\textit{\
algebra, its stability and deformations}, JHEP \textbf{1904} (2019) 068.
arXiv:1902.03260 [hep-th].

\bibitem{CMMRSV} P. Concha, N. Merino, O. Miskovic, E. Rodríguez, P.
Salgado-Rebolledo, O. Valdivia, \textit{Asymptotic symmetries of
three-dimensional Chern-Simons gravity for the Maxwell algebra}. JHEP
\textbf{10} (2018) 079. arXiv:1805.08834 [hep-th].

\bibitem{BCR} H. Bacry, P. Combe, J.L. Richard, \textit{Group-theoretical
analysis of elementary particles in an external electromagnetic fields. 1.
The relativistic particle in a constant and uniform field}, Nuovo Cim. A
\textbf{67} (1970) 267.

\bibitem{Schrader} R. Schrader, \textit{The Maxwell group and the quantum
theory of particles in classical homogeneous electromagnetic fields},
Fortsch. Phys. \textbf{20} (1972) 701.

\bibitem{GK} J. Gomis, A. Kleinschmidt, \textit{On free Lie algebras and
particles in electro-magnetic fields}, JHEP \textbf{07} (2017) 085.
arXiv:1705.05854 [hep-th].

\bibitem{AKL} J.A. de Azcarraga, K. Kamimura, J. Lukierski, \textit{%
Generalized cosmological term from Maxwell symmetries}, Phys. Rev. D\textbf{%
83} (2011) 124036. arXiv:1012.4402 [hep-th].

\bibitem{DKGS} R. Durka, J. Kowalski-Glikman, M. Szczachor, \textit{Gauges
AdS-Maxwell algebra and gravity}, Mod. Phys. Lett. A \textbf{26} (2011)
2689. arXiv:1107.4728 [hep-th].

\bibitem{AKL2} J.A. de Azcarraga, K. Kamimura, J. Lukierski, \textit{Maxwell
symmetries and some applications}, Int. J. Mod. Phys. Conf. Ser. \textbf{23}
(2013) 01160. arXiv:1201.2850 [hep-th].

\bibitem{CPRS1} P.K. Concha, D.M. Peñafiel, E.K. Rodríguez, P. Salgado,
\textit{Even-dimensional General Relativity from Born-Infeld gravity}, Phys.
Lett. B \textbf{725} (2013) 419. arXiv:1309.0062 [hep-th].

\bibitem{CPRS2} P.K. Concha, D.M. Peñafiel, E.K. Rodríguez, P. Salgado,
\textit{Chern-Simons and Born-Infeld gravity theories and Maxwell algebras
type}, Eur. Phys. J. C \textbf{74} (2014) 2741. arXiv:1402.0023 [hep-th].

\bibitem{CPRS3} P.K. Concha, D.M. Peñafiel, E.K. Rodríguez, P. Salgado,
\textit{Generalized Poincaré algebras and Lovelock-Cartan gravity theory},
Phys. Lett. B \textbf{742} (2015) 310. arXiv:1405.7078 [hep.th].

\bibitem{SSV} P. Salgado, R.J. Szabo, O. Valdivia, \textit{Topological
gravity and transgression holography}, Phys. Rev. D\textbf{89} (2014)
084077. arXiv:1401.3653 [hep-th].

\bibitem{HR} S. Hoseinzadeh, A. Rezaei-Aghdam, \textit{(2+1)-dimensional
gravity from Maxwell and semisimple extension of the Poincaré gauge
symmetric models}, Phys. Rev. D\textbf{90} (2014) 084008. arXiv:1402.0320
[hep-th].

\bibitem{CK} O. Cebecio\u{g}lu, S. Kibaro\u{g}lu,\textit{\ Maxwell-affine
gauge theory of gravity}, Phys. Lett. B \textbf{751} (2015) 131.
arXiv:1503.09003 [hep-th].

\bibitem{AFGHZ} L. Avilés, E. Frodden, J. Gomis, D. Hidalgo, J. Zanelli,
\textit{Non-Relativistic Maxwell Chern-Simons Gravity}, JHEP \textbf{1805 }%
(2018) 047. arXiv:1802.08453 [hep-th].

\bibitem{GKP} J. Gomis, A. Kleinschmidt, J. Palmkvist, \textit{Symmetries of
M-theory and free Lie superalgebras}, JHEP \textbf{03} (2019) 160.
arXiv:1809.09171 [hep-th].

\bibitem{KSC} S. Kibaro\u{g}lu, M. \c{S}enay, O. Cebecio\u{g}lu, $D=4\mathit{%
\ }$\textit{topological gravity from gauging the Maxwell-special-affine group%
}, Mod. Phys. Lett. A\textbf{34} (2019) 1950016. arXiv:1810.01635 [hep-th].

\bibitem{SR} P. Salgado-Rebolledo, \textit{The Maxwell group in 2+1
dimensions and its infinite-dimensional enhancements}, arXiv:1905.09421
[hep-th].

\bibitem{CMRSV} P. Concha, N. Merino, E. Rodríguez, P. Salgado-Rebolledo, O.
Valdivia, \textit{Semi-simple enlargement of the} $\mathfrak{bms}_{3}$
\textit{algebra from a} $\mathfrak{so}(2,2)\oplus \mathfrak{so}(2,1)$
\textit{Chern-Simons theory}, JHEP \textbf{1902} (2019) 002.
arXiv:1810.12256 [hep-th].

\bibitem{Sorokas} D.V. Soroka, V.A. Soroka, \textit{Semi-simple extension of
the (super)Poincaré algebra}, Adv. High Energy Phys. \textbf{2009} (2009)
[hep-th/0605251].

\bibitem{GKL} J.\ Gomis, K. Kamimura, J. Lukierski, \textit{Deformations of
Maxwell algebra and their dynamical realizations}, JHEP \textbf{0908 }(2009)
039. arXiv:0906.4464 [hep-th].

\bibitem{DFIMRSV} J. Díaz, O. Fierro, F. Izaurieta, N. Merino, E. Rodriguez,
P. Salgado, O. Valdivia, \textit{A generalized action for }$\mathit{(2+1)}$%
\textit{-dimensional Chern-Simons gravity}, J. Phys. A. Math. Theor.\textbf{%
\ 45} (2012) 255207, arXiv:1311.2215 [gr-qc].

\bibitem{SS} P. Salgado, S. Salgado, $\mathit{so}\left( D-1,1\right) \otimes
so\left( D-1,2\right) $\textit{\ algebras and gravity}, Phys. Lett. B
\textbf{728} (2014) 5.

\bibitem{CDIMR} P.K. Concha, R. Durka, C. Inostroza, N. Merino, E.K. Rodrí%
guez, \textit{Pure Lovelock gravity and Chern-Simons theory}, Phys. Rev. D
\textbf{94} (2016) 024055. arXiv:1603.09424 [hep-th],

\bibitem{CMR} P.K. Concha, N. Merino, E.K. Rodríguez, \textit{Lovelock
gravity from Born-Infeld gravity theory}, Phys. Lett. B \textbf{765} (2017)
395. arXiv:1606.07083 [hep-th].

\bibitem{CR3} P. Concha, E. Rodríguez, \textit{Generalized Pure Lovelock
Gravity}, Phys. Lett. B \textbf{774} (2017) 616. arXiv:1708.08827 [hep-th].

\bibitem{CR4} P. Concha, E. Rodríguez, \textit{Non-Relativistic Gravity
Theory based on an Enlargement of the Extended Bargmann Algebra}, JHEP
\textbf{07} (2019) 085. arXiv:1906.00086 [hep-th].

\bibitem{BDMT} G. Barnich, L. Donnay, J. Matulich, R. Troncoso, \textit{%
Asymptotic symmetries and dynamics of three-dimensional flat supergravity},
JHEP \textbf{1408} (2014) 071. arXiv:1407.4275 [hep-th].

\bibitem{LM} I. Lodato, W. Merbis, \textit{Super-BMS}$_{3}$ algebras from $%
\mathcal{N}=2$ \textit{flat supergravities}, JHEP \textbf{1611} (2016) 150.
arXiv:1610.07506 [hep-th].

\bibitem{FMT} O. Fuentealba, J. Matulich, R. Troncoso, \textit{Asymptotic
structure of }$\mathcal{N}=2$ \textit{supergravity in 3D: extended super-BMS}%
$_{3}$\textit{\ and nonlinear energy bounds}, JHEP \textbf{1709} (2017) 030.
arXiv:1706.07542 [hep-th].

\bibitem{BBNN} N. Banerjee, A. Bhattacharjee, Neetu, T. Neogi, \textit{New
N=2 SuperBMS}$_{3}$\textit{\ algebra and Invariant Dual Theory for 3D
Supergravity}, arXiv:1905.10239 [hep-th].

\bibitem{BLN} N. Banerjee, I. Lodato, T. Neogi, \textit{N=4 Supersymmetric
BMS}$_{3}$\textit{\ algebras from asymptotic symmetry analysis}, Phys. Rev.
D \textbf{96} (2017) 066029. arXiv:1706.02922 [hep-th].

\bibitem{BBLN} N. Banerjee, A. Bhattacharjee, I. Lodato, T. Neogi, \textit{%
Maximmaly }$\mathcal{N}$\textit{-extended super-BMS}$_{3}\mathit{\ }$\textit{%
algebras and Generalized 3D Gravity Solutions}, JHEP \textbf{1901} (2019)
115. arXiv:1807.06768 [hep-th].

\bibitem{Sexp} F. Izaurieta, E. Rodríguez, P. Salgado, \textit{Expanding Lie
(super)algebras through Abelian semigroups}, J. Math. Phys. \textbf{47}
(2006) 123512. [hep-th/0606215].

\bibitem{CCRS} R. Caroca, P. Concha, E. Rodríguez, P. Salgado-Rebolledo,
\textit{Generalizing the }$bms_{3}$ \textit{and 2D-conformal algebra by
expanding the Virasoro algebra}, Eur. Phys. J. C \textbf{78} (2018) 262.
arXiv:1707.07209 [hep-th].

\bibitem{CCFR} R. Caroca, P. Concha, O. Fierro, E. Rodríguez, \textit{%
Three-dimensional Poincaré supergravity and }$\mathcal{N}$\textit{-extended
supersymmetric BMS}$_{3}$\textit{\ algebra}, Phys. Lett. B \textbf{792}
(2019) 93. arXiv:1812.05065 [hep-th].

\bibitem{BGKL} S. Bonanos, J. Gomis, K. Kamimura, J. Lukierski, \textit{%
Maxwell superalgebra and superparticle in constant Gauge background}. Phys.
Rev. Lett. \textbf{104} (2010) 090401. arXiv:0911.5072 [hep-th].

\bibitem{CPR} P. Concha, D.M. Peñafiel, E. Rodríguez, \textit{On the Maxwell
supergravity and flat limit in 2+1 dimensions}, Phys. Lett. B \textbf{785}
(2018) 247. arXiv:1807.00194 [hep-th].

\bibitem{HS} M. Hastsuda, M. Sakaguchi, \textit{Wess-Zumino term for the AdS
superstring and generalized Inönü-Wigner contraction}. Prog. Theor. Phys.
\textbf{109} (2003) 853. [hep-th/0106114].

\bibitem{AIPV} J.A. de Azcarraga, J.M. Izquierdo, M. Picon, O. Varela,
\textit{Generating Lie and gauge free differential (super)algebras by
expanding Maurer-Cartan forms and Chern-Simons supergravity}, Nucl. Phys. B
\textbf{662} (2003) 185. [hep-th/0212347].

\bibitem{AIPV2} J.A. de Azcarraga, J.M. Izquierdo, M. Picon, O. Varela,
\textit{Extensions, expansions, Lie algebra cohomology and enlarged
superspaces}. Class. Quant. Grav. \textbf{21} (2004) S1375-1384.
[hep-th/0401033].

\bibitem{AIPV3} J.A. de Azcarraga, J.M. Izquierdo, M. Picon, O. Varela,
\textit{Expansions of algebras and superalgebras and some applications},
Int. J. Theor. Phys. \textbf{46} (2007) 2734. [hep-th/0401033].

\bibitem{Caroca2010a} R. Caroca, N. Merino, P. Salgado, \textit{%
{}{}S-Expansion of Higher-Order Lie Algebras{}{}}, J. Math. Phys. {}{}%
\textbf{50} {}{}(2009) 013503. arXiv:1004.5213 [math-ph].

\bibitem{Caroca2010b} R. Caroca, N. Merino, A. Perez, P. Salgado, {}{}%
\textit{Generating Higher-Order Lie Algebras by Expanding Maurer Cartan Forms%
}{}{}, J. Math. Phys. {}{}\textbf{50{}{}} (2009) 123527. arXiv:1004.5503
[hep-th].

\bibitem{Caroca2011} R. Caroca, N. Merino, P. Salgado, O. Valdivia, {}{}%
\textit{Generating infinite-dimensional algebras from loop algebras by
expanding Maurer-Cartan forms}{}{}, J. Math. Phys. {}{}\textbf{52}{}{}
(2011) 043519. arXiv:1311.2623 [math-ph].

\bibitem{CKMN} R. Caroca, I. Kondrashuk, N. Merino, F. Nadal, {}{}\textit{%
Bianchi spaces and their three-dimensional isometries as S-expansions of
two-dimensional isometries}{}{}, J. Phys. A{}{}\textbf{46{}{}} (2013)
225201. arXiv:1104.3541 [math-ph].

\bibitem{AMNT} L. Andrianopoli, N. Merino, F. Nadal, M. Trigiante, {}{}%
\textit{General properties of the expansion methods of Lie algebras}{}{}, J.
Phys. A {}{}\textbf{46}{}{} (2013) 365204. arXiv:1308.4832 [gr-qc].

\bibitem{ACCSP} M. Artebani, R. Caroca. M.C. Ipinza, D.M. Peñafiel, P.
Salgado, {}{}\textit{Geometrical aspects of the Lie Algebra S-Expansion
Procedure{}{}}, J. Math. Phys. {}{}\textbf{57}{}{} (2016) 023516.
arXiv:1602.04525 [math-ph].

\bibitem{ILPR} M.C. Ipinza, F. Lingua, D.M. Peñafiel, L. Ravera, {}{}\textit{%
An Analytic Method for }$\mathit{S}$\textit{-expansion involving Resonance
and Reduction}{}{}, Fortschr. Phys. \textbf{64} (2016) 854. arXiv:1609.05042
[hep-th].

\bibitem{IKMN} C. Inostroza, I. Kondrashuk, N. Merino, F. Nadal, {}{}\textit{%
On a Java library to perform S-expansions of Lie algebras}{}{}, J. Phys.
Conf. Ser. \textbf{1085} (2018) 052010. arXiv:1802.04468 [math-ph].{}

\bibitem{IKMN2} C. Inostroza, I. Kondrashuk, N. Merino, F. Nadal,\textit{\
On the algorithm to find S-related Lie algebras}, J. Phys. Conf. Ser.
\textbf{1085} (2018) 052011. arXiv:1802.05765 [physics.comp-ph].

\bibitem{LuNo} J. Lukierski, A. Nowicki, \textit{Superspinors and Graded
Lorentz Groups in Three, Four and Five Dimensions}, Fortsch. Phys. \textbf{30%
} (1982) 75.

\bibitem{BJLMN} N. Banerjee, D.P. Jatkar, I. Lodato, S. Mukhi, T. Neogi,
\textit{Extended Supersymmetric BMS}$_{3}$ \textit{algebras and Their Free
Field Realisations}, JHEP \textbf{11} (2016) 059. arXiv:1609.09210 [hep-th].

\bibitem{BDMT2} G. Barnich, L. Donnay, J. Matulich, R. Troncoso, \textit{%
Super-BMS}$_{3}$ \textit{invariant boundary theory from three-dimensional
flat supergravity}, JHEP \textbf{1701} (2017) 029. arXiv:1510.08824 [hep-th].

\bibitem{BM} A. Bagchi, I. Mandal, \textit{Supersymmetric Extension of
Galilean Conformal Algebras}, Phys. Rev. D \textbf{80} (2009) 086011.
arXiv:0905.0580 [hep-th].

\bibitem{M} I. Mandal,\textit{\ Supersymmetric Extension of GCA in 2d}, JHEP
\textbf{1011} (2010) 018. arXiv:1003.0209 [hep-th].

\bibitem{KRR} C. Krishnan, A. Raju, S. Roy, \textit{A Grassmann path from AdS%
}$_{3}$\textit{\ to flat space, J. High. Energy Phys. }\textbf{1403 }(2014)
036. arXiv:1312.2941 [hep-th].

\bibitem{Concha} P. Concha, $\mathcal{N}$\textit{-extended Maxwell
supergravities as Chern-Simons theories in three spacetime dimensions},
Phys. Lett. B \textbf{792} (2019) 290. arXiv:1903.03081 [hep-th].

\bibitem{GRCS} F. Izaurieta, E. Rodríguez, P. Minning, P. Salgado, A. Perez,
\textit{Standard General Relativity from Chern-Simons Gravity}, Phys. Lett.
B \textbf{678} (2009) 213. arXiv:0905.2187 [hep-th].

\bibitem{BGKL2} S. Bonanos, J. Gomis, K. Kamimura, J. Lukierski, \textit{%
Deformations of Maxwell Superalgebras and Their Applications}, J. Math.
Phys. \textbf{51} (2010) 102301. arXiv:1005.3714 [hep-th].

\bibitem{Lukierski} J. Lukierski, \textit{Generalized Wigner-Inönü
Contractions and Maxwell (Super)Algebras}, Proc. Steklov Inst. Math. \textbf{%
272} (2011) no.1 183. arXiv:1007.3405 [hep-th].

\bibitem{FL} S. Fedoruk, J. Lukierski, \textit{New spinorial particle model
in tensorial space-time and interacting higher spin fields}, JHEP \textbf{%
1302} (2013) 128. arXiv:1210.1506 [hep-th].

\bibitem{AILW} J.A. de Azcarraga, J.M. Izquierdo, J. Lukierski, M.
Woronowicz, \textit{Generalizations of Maxwell (super)algebras by the
expansion method}, Nucl. Phys. B \textbf{869} (2013) 303. arXiv:1210.1117
[hep-th].

\bibitem{AI} J.A. de Azcarraga, J.M. Izquierdo, \textit{Minimal D=4
supergravity from superMaxwell algebra}, Nucl. Phys. B \textbf{885} (2014)
34. arXiv:1403.4128 [hep-th].

\bibitem{CR1} P.K. Concha, E.K. Rodríguez, \textit{Maxwell superalgebras and
Abelian semigroup expansion}, Nucl. Phys. B \textbf{886} (2014) 1128.
arXiv:1405.1334 [hep-th].

\bibitem{CR2} P.K. Concha, E.K. Rodríguez, \textit{N=1 Supergravity and
Maxwell superalgebras}, JHEP \textbf{1409} (2014) 090. arXiv:1407.4635
[hep-th].

\bibitem{CFRS} P.K. Concha, O. Fierro, E.K. Rodríguez, P. Salgado, \textit{%
Chern-Simons supergravity in D=3 and Maxwell superalgebra}, Phys. Lett. B
\textbf{750} (2015) 117. arXiv:1507.02335 [hep-th].

\bibitem{CFR} P.K. Concha, O. Fierro, E.K. Rodríguez, \textit{Inönü-Wigner
contraction and D=2+1 supergravity}, Eur. Phys. J. C \textbf{77} (2017) 48.
arXiv:1611.05018 [hep-th].

\bibitem{PR} D.M. Peñafiel, L. Ravera, \textit{On the Hidden Maxwell
Superalgebra underlying D=4 Supergravity}, Fortsch. Phys. \textbf{65} (2017)
1700005. arXiv:1701.04234 [hep-th].

\bibitem{Ravera} L. Ravera, \textit{Hidden role of Maxwell superalgebras in
the free differential algebras of }$D=4$\textit{\ and }$D=11$\textit{\
supergravity}, Eur. Phys. J. C \textbf{78} (2018) 211. arXiv:1801.08860
[hep-th].

\bibitem{KC} S. Kibaro\u{g}lu, O. Cebecio\u{g}lu, $D=4$\textit{\
supergravity from the Maxwell-Weyl superalgebra}, arXiv:1812.09861 [hep-th].

\bibitem{Green} M.B. Green, \textit{Supertranslations, superstrings and
Chern-Simons forms}. Phys. Lett. B \textbf{223} (1989) 157.

\bibitem{AF} R. D'Auria, P. Fré, \textit{Geometric supergravity in d=11 and
its hidden supergroup}, Nucl. Phys. B \textbf{201} (1982) 101.

\bibitem{BDR} R. Basu, S. Detournay, M. Riegler, \textit{Spectral Flow in 3D
Flat Spacetimes}, JHEP \textbf{12} (2017) 134. arXiv:1706.07438 [hep-th].

\bibitem{Ito} K. Ito, \textit{Extended superconformal algebras on AdS(3)},
Phys. Lett. B \textbf{449} (1999) 48. [hep-th/9811002].

\bibitem{CRS} P.K. Concha, E.K. Rodríguez, P. Salgado, \textit{Generalized
supersymmetric cosmological term in N=1 Supergravity}, JHEP 08 (2015) 009.
arXiv:1504.01898 [hep-th].

\bibitem{CIRR} P.K. Concha, M.C. Ipinza, L. Ravera, E.K. Rodríguez, \textit{%
On the supersymmetric extension of Gauss-Bonnet like gravity}, JHEP \textbf{%
09} (2016) 007. arXiv:1607.00373 [hep-th].

\bibitem{BR} A. Baunadi, L. Ravera, \textit{Generalized AdS-Lorentz deformed
supergravity on a manifold with boundary}, Eur. Phys. J. Plus \textbf{133} (2018) 514. arXiv:1803.08738 [hep-th].

\bibitem{PR2} D.M. Peñafiel, L. Ravera, \textit{Generalized cosmological
term in }$D=4$\textit{\ supergravity from a new AdS-Lorentz superalgebra},
Eur. Phys. J. C \textbf{78} (2018) 945. arXiv:1807.07673 [hep-th].

\bibitem{FISV} O. Fierro, F. Izaurieta, P. Salgado, O. Valdivia, \textit{%
Minimal AdS-Lorentz supergravity in three-dimensions}, Phys. Lett. B \textbf{%
788} (2019) 198. arXiv:1401.3697 [hep-th].

\bibitem{Sorokas2} D.V. Soroka, V.A. Soroka, \textit{Tensor extension of the
Poincaré algebra}, Phys. Lett. B \textbf{607} (2005) 302. [hep-th/0410012].

\bibitem{MR} I. Manda, A. Rayyan, \textit{Super-GCA from }$\mathcal{N}%
=\left( 2,2\right) $\textit{\ Super-Virasoro}, Phys. Lett. B \textbf{754}
(2016) 195. arXiv:1601.04723 [hep-th].

\bibitem{BIOR} E. Bergshoeff, J. Izquierdo, T. Ortín, L. Romano, \textit{Lie
Algebra Expansions and Actions for Non-Relativistic Gravity}, JHEP \textbf{08} (2019) 048.
arXiv:1904.08304 [hep-th].

\bibitem{AGI} J.A. de Azcárraga, D. Gútiez, J.M. Izquierdo, \textit{Extended
}$D=3$\textit{\ Bargmann supergravity from a Lie algebra expansion}, Nucl. Phys. B \textbf{946} (2019) 114706.
arXiv:1904.12786 [hep-th].

\bibitem{Romano} L. Romano, \textit{Non-Relativistic Four Dimensional
p-Brane Supersymmetric Theories and Lie Algebra Expansion}, arXiv:1906.08220
[hep-th].

\bibitem{PSR} D.M. Peñafiel, P. Salgado-Rebolledo, \textit{Non-relativistic
symmetries in three space-time dimensions and the Nappi-Witten algebra}, Phys. Lett. B \textbf{798} (2019) 135005.
arXiv:1906.02161 [hep-th].

\bibitem{BS} S. Bansal, D. Sorokin, \textit{Can Chern-Simons or
Rarita-Schwinger be a Volkov-Akulov Goldstone?}, JHEP \textbf{07} (2018)
106. arXiv:1806.05945 [hep-th].

\bibitem{CS} D. Chernyavsky, D. Sorokin, \textit{Three-dimensional
(higher-spin) gravities with extended Schrödinger and l-conformal Galilean
symmetries}, JHEP \textbf{07} (2019) 156. arXiv:1905.13154 [hep-th].

\bibitem{Hietarinta} J. Hietarinta, \textit{Supersymmetry Generators of
Arbitrary Spin}, Phys. Rev. D\textbf{13} (1976) 838.

\bibitem{CRR} P. Concha, L. Ravera, E. Rodríguez, \textit{On the
supersymmetry invariance of flat supergravity with boundary}, JHEP \textbf{01%
} (2019) 192. arXiv:1809.07871 [hep-th].

\bibitem{CCFRS} R. Caroca, P. Concha, O. Fierro, E. Rodríguez, P.
Salgado-Rebolledo,\textit{\ Generalized Chern-Simons higher-spin gravity
theories in three dimensions}, Nucl. Phys. B \textbf{934} (2018) 240.
arXiv:1712.09975 [hep-th].

\bibitem{Durka2} R. Durka, J. Kowalski-Glikman, \textit{Resonant algebras in
Chern-Simons model of topological insulators}, Phys. Lett. B \textbf{795} (2019) 516. arXiv:1906.02356 [hep-th].
\end{thebibliography}
\end{document}